\documentclass[format=acmsmall, review=false, screen=true]{acmart}

\usepackage{booktabs} 

\usepackage[ruled]{algorithm2e} 

\SetAlFnt{\small}
\SetAlCapFnt{\small}
\SetAlCapNameFnt{\small}
\SetAlCapHSkip{0pt}
\IncMargin{-\parindent}

\acmJournal{TOIS}
\acmVolume{9}
\acmNumber{4}
\acmArticle{39}
\acmYear{2017}
\acmMonth{10}
\copyrightyear{2009}

\setcopyright{acmlicensed}

\acmDOI{}

\usepackage{amsmath,amsthm,amssymb}
\usepackage[FIGTOPCAP]{subfigure}

\begin{document}
\title[PNNs for User Response Prediction]{Product-based Neural Networks for User Response Prediction over Multi-field Categorical Data}  


\author{Yanru Qu}
\author{Bohui Fang}
\author{Weinan Zhang}
\affiliation{%
  \institution{Shanghai Jiao Tong University}
  \country{China}}
\email{kevinqu@apex.sjtu.edu.cn}
\email{fangbohui@sjtu.edu.cn}
\email{wnzhang@sjtu.edu.cn}
\author{Ruiming Tang}
\affiliation{%
	\institution{Noah's Ark Lab, Huawei}
	\country{China}}
\email{tangruiming@huawei.com}
\author{Minzhe Niu}
\affiliation{%
\institution{Shanghai Jiao Tong University}
\country{China}}
\email{nmzfrank@apex.sjtu.edu.cn}
\author{Huifeng Guo}
\authornote{This work is done when Huifeng Guo was an intern at Noah's Ark Lab, Huawei}
\affiliation{%
\institution{Shenzhen Graduate School, Harbin Institute of Technology}
\country{China}}
\email{huifengguo@yeah.net}
\author{Yong Yu}
\affiliation{%
	\institution{Shanghai Jiao Tong University}
	\country{China}}
\email{yyu@apex.sjtu.edu.cn}
\author{Xiuqiang He}
\authornote{This work is done when Xiuqiang He worked at Noah's Ark Lab, Huawei}
\affiliation{%
\institution{Data service center, MIG, Tencent}
\country{China}}
\email{xiuqianghe@tencent.com}

\newcommand{\qu}[1]{{\color{red} [qu: ``#1'']}}
\newcommand{\change}[1]{{\color{black} #1}}
\newcommand{\revise}[1]{{\color{black} #1}}

\begin{abstract}
User response prediction is a crucial component for personalized information retrieval and filtering scenarios\revise{, such as recommender system and web search}.
The data in user response prediction is mostly in a multi-field categorical format and transformed into sparse representations via one-hot encoding.
\change{Due to the sparsity problems in representation and optimization, most research focuses on feature engineering and shallow modeling.}
Recently, deep neural networks have attracted research attention on such a problem for their high capacity and end-to-end training scheme. 
In this paper, we study user response prediction in the scenario of click prediction. 
We first analyze a coupled gradient issue in latent vector-based models and propose kernel product \revise{to learn field-aware feature interactions}.
Then we discuss an insensitive gradient issue in DNN-based models and propose Product-based Neural Network (PNN) \revise{which adopts a feature extractor to explore feature interactions}.
Generalizing the kernel product to a net-in-net architecture, we further propose Product-network In Network (PIN) which can generalize previous models.
Extensive experiments on 4 industrial datasets \revise{and 1 contest dataset demonstrate that our models consistently outperform 8 baselines on both AUC and log loss}.
\revise{Besides, PIN makes great CTR improvement (relatively 34.67\%) in online A/B test.}
\end{abstract}

%
%


%
%


\maketitle

\renewcommand{\shortauthors}{Y. Qu et al.}

\newcommand{\tsf}{\textsf}
\newcommand{\ttt}{\textsc}
\newcommand{\bv}{\mathbf{v}}
\newcommand{\bV}{\mathbf{V}}
\newcommand{\bk}{\mathbf{k}}
\newcommand{\bR}{\mathbb{R}}
\newcommand{\bN}{\mathbb{N}}
\newcommand{\bx}{\mathbf{x}}
\newcommand{\bp}{\mathbf{p}}
\newcommand{\bq}{\mathbf{q}}
\newcommand{\bE}{\mathbb{E}}
\newcommand{\bz}{\mathbf{z}}
\newcommand{\Bf}{\mathbf{f}}
\newcommand{\bw}{\mathbf{w}}
\newcommand{\bh}{\mathbf{v}}
\newcommand{\bH}{\mathbf{H}}
\newcommand{\bb}{\mathbf{b}}
\newcommand{\bL}{\mathcal{L}}
\newcommand{\bfm}{\text{Male}}
\newcommand{\bft}{\text{Tue}}
\newcommand{\bfl}{\text{London}}
\newcommand{\embed}{\text{embed}}
\newcommand{\inter}{\text{interact}}
\newcommand{\net}{\text{net}}
\newcommand{\tabincell}[2]{\begin{tabular}{@{}#1@{}}#2\end{tabular}}  

\renewcommand{\qu}[1]{{\bf \color{red} [yanru says ``#1'']}}
\newcommand{\tang}[1]{{\bf \color{blue} [ruiming says ``#1'']}}
\newcommand{\weinan}[1]{{\bf \color{cyan} [weinan says ``#1'']}}
\newcommand{\niu}[1]{{\bf \color{purple} [minzhe says ``#1'']}}

\section{Introduction}\label{sec:intro}
Predicting a user's response to some item (e.g., movie, news article, advertising post) under certain context (e.g., website) is a crucial component for personalized information retrieval (IR) and filtering scenarios, such as online advertising \cite{mcmahan2013ad,ren2016user}, recommender system \cite{koren2009matrix,rendle2010factorization}, and web search \cite{agichtein2006learning,chapelle2009dynamic}.

The core of personalized 
\revise{service} 
is to estimate the probability that a user will ``like'', ``click'', or ``purchase'' an item, given features about the user, the item, and the context \cite{menon2011response}.
This probability indicates the user's interest in the specific item and influences the subsequent decision-making such as item ranking \cite{xue2004optimizing} and ad bidding \cite{zhang2014optimal}.
Taking online advertising as an example, the estimated Click-Through Rate (CTR) will be utilized to calculate a bid price in an ad auction to improve the advertisers' budget efficiency and the platform revenue \cite{zhang2014optimal,perlich2012bid,ren2016user}.
Hence, it is much desirable to gain accurate prediction to not only improve the user experience, but also boost the volume and profit for the service providers.

\begin{table}[htbp]
\caption{An example of multi-field categorical data.}
\label{tab:example}
\centering
\small
\begin{tabular}{c|ccc}
\ttt{TARGET} & \ttt{WEEKDAY} & \ttt{GENDER} & \ttt{CITY}\\ \hline
\ttt{1} & \ttt{Tuesday} & \ttt{Male} & \ttt{London}\\
\ttt{0} & \ttt{Monday} & \ttt{Female} & \ttt{New York}\\
\ttt{1} & \ttt{Tuesday} & \ttt{Female} & \ttt{Hong Kong}\\
\ttt{0} & \ttt{Tuesday} & \ttt{Male} & \ttt{Tokyo}\\ \hline
\ttt{Number} & \ttt{7} & \ttt{2} & \ttt{1000}\\
\end{tabular}
\end{table}

The data collection in these tasks is mostly in a multi-field categorical form.
And each data instance is normally transformed into a high-dimensional sparse (binary) vector via one-hot encoding \cite{he2014practical}.
Taking Table~\ref{tab:example} as an example, the 3 fields are one-hot encoded and concatenated as
\[\underbrace{[0,1,0,0,0,0,0]}_{\ttt{WEEKDAY=Tuesday}}\underbrace{[1,0]}_{\ttt{GENDER=Male}} \underbrace{[0,0,1,0,\ldots,0,0]}_{\ttt{CITY=London}}~.\]
Each field is represented as a binary vector, of which only 1 entry corresponding to the input is set as 1 while others are 0.
The dimension of a vector is determined by its field size, i.e., the number of unique categories\footnote{
For clarity, we use ``category'' instead of ``feature'' to represent a certain value in a categorical field. For consistency with previous literature, we preserve  ``feature'' in some terminologies, e.g., feature combination, feature interaction, and feature representation.} in that field.
The one-hot vectors of these fields are then concatenated together in a predefined order. 

\begin{figure}[htbp]
	\centering
	\includegraphics[width=0.6\textwidth]{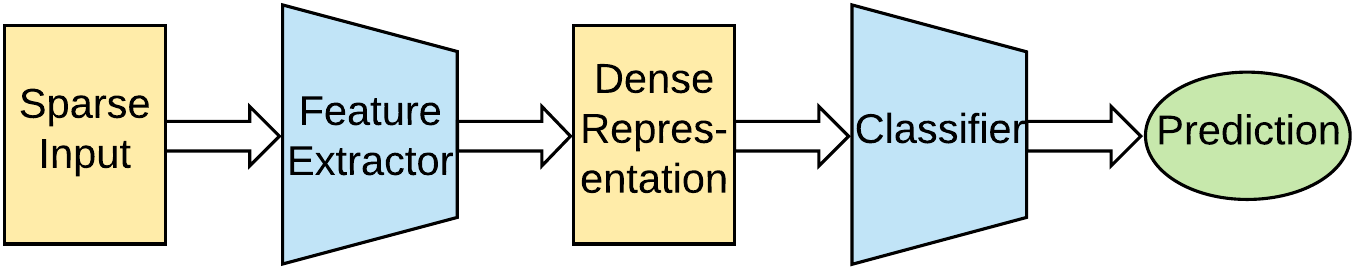}
	\caption{Model decomposition.}
	\label{fig:decom}
\end{figure}

Without loss of generality, user response prediction can be regarded as a binary classification problem, and 1/0 are used to denote positive/negative responses respectively \cite{richardson2007predicting, graepel2010web}.
\revise{A main challenge of such a problem is sparsity.
For parametric models, they usually convert the sparse (binary) input into dense representations (e.g., weights, latent vectors), and then search for a separable hyperplane.
Fig.~\ref{fig:decom} shows the model decomposition.
In this paper, we mainly focus on modeling and training,
thus we exclude preliminary feature engineering \cite{cui2011bid}.}

\change{
Many machine learning models are leveraged or proposed to work on such a problem, 
including linear models, latent vector-based models, tree models, and DNN-based models.
Linear models, such as Logistic Regression (LR) \cite{lee2012estimating} and Bayesian Probit Regression \cite{graepel2010web}, are easy to implement and with high efficiency. 
A typical latent vector-based model is Factorization Machine (FM) \cite{rendle2010factorization}. 
FM uses weights and latent vectors to represent categories. 
\revise{According to their parametric representations, LR has a linear feature extractor, and FM has a bi-linear\footnote{Although FM has higher-order formulations \cite{rendle2010factorization}, due to the efficiency and practical performance, FM is usually implemented with second-order interactions.} feature extractor.
The prediction of LR and FM are simply based on the sum over weights, thus their classifiers are linear.}
}
\revise{FM works well on sparse data, and inspires a lot of extensions, including Field-aware FM (FFM) \cite{juan2016field}. 
FFM introduces field-aware latent vectors, which gain FFM higher capacity and better performance.
However, FFM is restricted by space complexity.}
Inspired by FFM, we find a \textbf{coupled gradient issue} of latent vector-based models and refine 
feature interactions\footnote{In \cite{rendle2010factorization}, the cross features learned by FM are called feature interactions.} 
as field-aware feature interactions. 
\revise{ 
To solve this issue as well as saving memory}, we propose kernel product methods and derive \textit{Kernel FM} (KFM) and \textit{Network in FM} (NIFM).

\revise{Trees and DNNs are potent function approximators.}
\change{
Tree models, such as Gradient Boosting Decision Tree (GBDT) \cite{chen2016xgboost}, are popular in various data science contests as well as industrial applications. 
\revise{GBDT explores very high order feature combinations in a non-parametric way}, yet its exploration ability is restricted when feature space becomes extremely high-dimensional and sparse.
}
DNN has also been preliminarily studied in information system literature \cite{zhang2016deep, covington2016deep, shan2016deep, qu2016product}.
\revise{In \cite{zhang2016deep}, FM supported Neural Network (FNN) is proposed. 
FNN has a pre-trained embedding\footnote{We use ``latent vector'' in shallow models, and ``embedding vector'' in DNN models.} layer and several fully connected layers.}
\revise{Since the embedding layer indeed performs a linear transformation, FNN mainly extracts linear information from the input.}
\revise{Inspired by \cite{shalev2017failures}, we find an \textbf{insensitive gradient issue} that fully connected layers cannot fit such target functions perfectly.}

\revise{From the model decomposition perspective, the above models are restricted by inferior feature extractors or weak classifiers.
Incorporating product operations in DNN, we propose \textit{Product-based Neural Network} (PNN).
PNN consists of an embedding layer, a product layer, and a DNN classifier. 
The product layer serves as the feature extractor which can make up for the deficiency of DNN in modeling feature interactions.}
We take FM, KFM, and NIFM as feature extractors in PNN, leading to \textit{Inner Product-based Neural Network} (IPNN), \textit{Kernel Product-based Neural Network} (KPNN), and \textit{Product-network In Network} (PIN).  

CTR estimation is a fundamental task in personalized advertising and recommender systems, and we take CTR estimation as the working example to evaluate our models. 
Extensive experiments on 4 large-scale real-world datasets and 1 contest dataset demonstrate the consistent superiority of our models over 8 baselines \cite{lee2012estimating, rendle2010factorization, liu2015convolutional, zhang2016deep, juan2016field, guo2017deepfm, chen2016xgboost, xiao2017attentional} on both AUC and log loss. 
\revise{Besides, PIN makes great CTR improvement (34.67\%) in online A/B test.}
To sum up, our contributions can be highlighted as follows:
\begin{itemize}	
\item \change{We 
analyze a coupled gradient issue of latent vector-based models. We refine feature interactions as field-aware feature interactions and extend FM with kernel product methods. 
Our experiments on KFM and NIFM successfully verify our assumption.}

\item \change{We analyze an insensitive gradient issue of DNN-based models and propose PNNs to tackle this issue. }
\revise{PNN has a flexible architecture which can generalize previous models.}

\item We study optimization, regularization, and other practical issues in training and generalization. 
In our extensive experiments, \revise{our models achieve consistently good results in 4 offline datasets, 1 contest, and 1 online A/B test.} 
\end{itemize}

The rest of this paper is organized as follows. 
In Section~\ref{sec:related}, we \revise{introduce related work in user response prediction and other involved techniques.} 
In Section~\ref{sec:method} and \ref{sec:issue}, we present our PNN models in detail and discuss several practical issues. 
In Section~\ref{sec:exp}, \revise{we show offline evaluation, parameter study, online A/B test, and synthetic experiments respectively.} 
We finally conclude this paper and discuss future work in Section~\ref{sec:con}.

\section{Background and Related Work}\label{sec:related}
\subsection{User Response Prediction}\label{sec:related-problem}
User response prediction is normally formulated as a binary classification problem with prediction log-likelihood or cross-entropy as the training objective \cite{richardson2007predicting, graepel2010web, agarwal2010estimating}.
area under ROC curve (AUC), log loss and relative information gain are common evaluation metrics \cite{graepel2010web}.
Due to the one-hot encoding of categorical data, sparsity is a big challenge in user response prediction.

\begin{figure}[tbp]
\subfigure[LR]{
	\includegraphics[width=.31\textwidth]{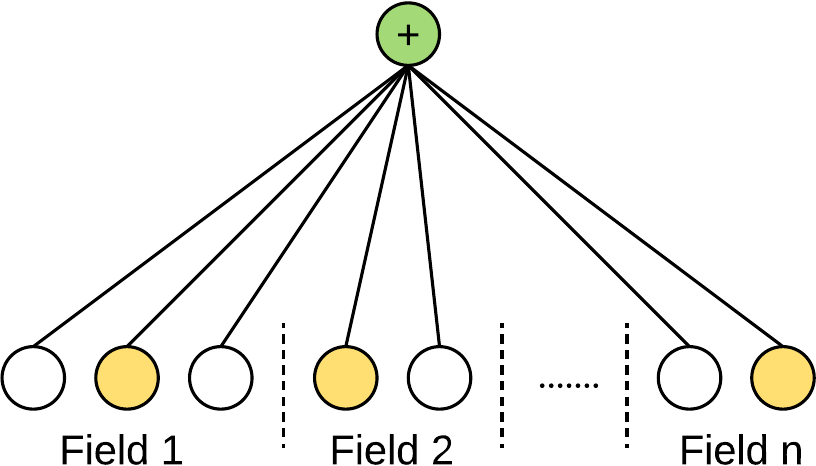}
	\label{fig:lr}
	}
\subfigure[FM (FFM)]{
	\includegraphics[width=.31\textwidth]{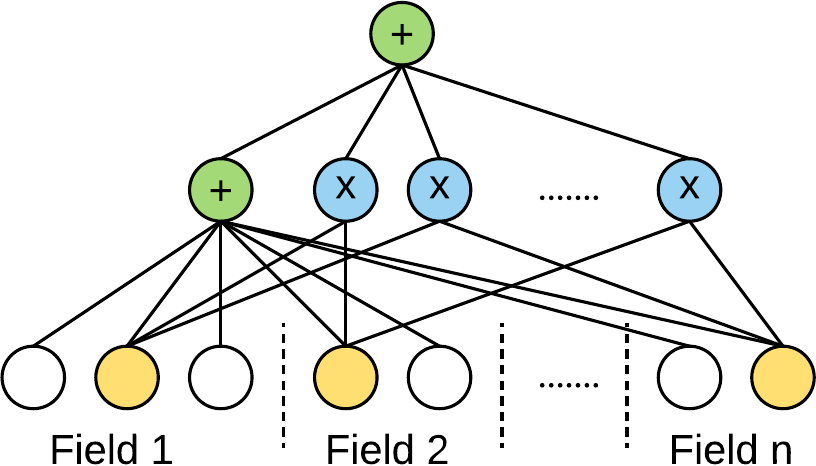}
	\label{fig:fm}
	}
\subfigure[GBDT]{
	\includegraphics[width=.31\textwidth]{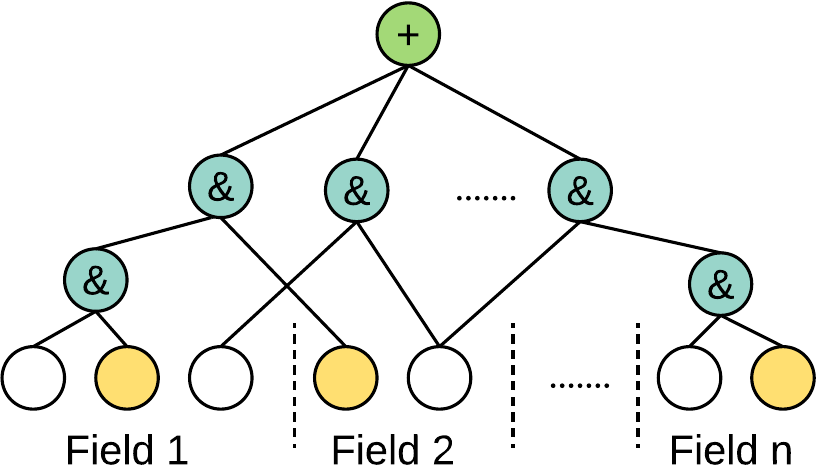}
	\label{fig:gbdt}
}
\caption{Illustration of LR, FM, FFM, and GBDT. \emph{Note:} Yellow node means one-hot input of a field; green `$+$' node means addition operation; blue `$\times$' node means multiplication operation; cyan `$\&$' node means feature combination.}
\label{fig:shallow}
\end{figure}

From the modeling perspective, linear Logistic Regression (LR) \cite{lee2012estimating, ren2016user}, \revise{bi-linear} Factorization Machine (FM) \cite{rendle2010factorization} and Gradient Boosting Decision Tree (GBDT) \cite{he2014practical} are widely used in industrial applications. 
As illustrated in Fig.~\ref{fig:shallow}, LR extracts linear information from the input, FM further extracts bi-linear information, while GBDT explores feature combinations in a non-parametric way. 
From the training perspective, \revise{many adaptive optimization algorithms can speed up training of sparse data}, including Follow the Regularized Leader (FTRL) \cite{mcmahan2013ad}, Adaptive Moment Estimation (Adam) \cite{kingma2014adam}, etc. 
\revise{These algorithms follow a per-coordinate learning rate scheme, 
making them converge much faster than stochastic gradient descent (SGD).}

\revise{From the representation perspective, latent vectors are expressive in representing categorical data.}
In FM, the side information and user/item identifiers are represented by low-dimensional latent vectors, and the feature interactions are modeled as the inner product of latent vectors.
As an extension of FM, Field-aware FM (FFM) \cite{juan2016field} enables each category to have multiple latent vectors. 
\revise{From the classification perspective, powerful function approximators like GBDT and DNN are more suitable for continuous input.
Therefore, in many contests, the winning solutions take FM/FFM as feature extractors to process discrete data, and use the latent vectors or interactions as the input of successive classifiers (e.g., GBDT, DNN).}
\revise{According to model decomposition (Fig.~\ref{fig:decom}), latent vector-based models make predictions simply based on the sum of interactions.
This weakness motivates the DNN variants of latent vector-based models.
}

\begin{figure}[tbp]
\subfigure[FNN]{
	\label{fig:fnn}
	\includegraphics[width=.3\textwidth]{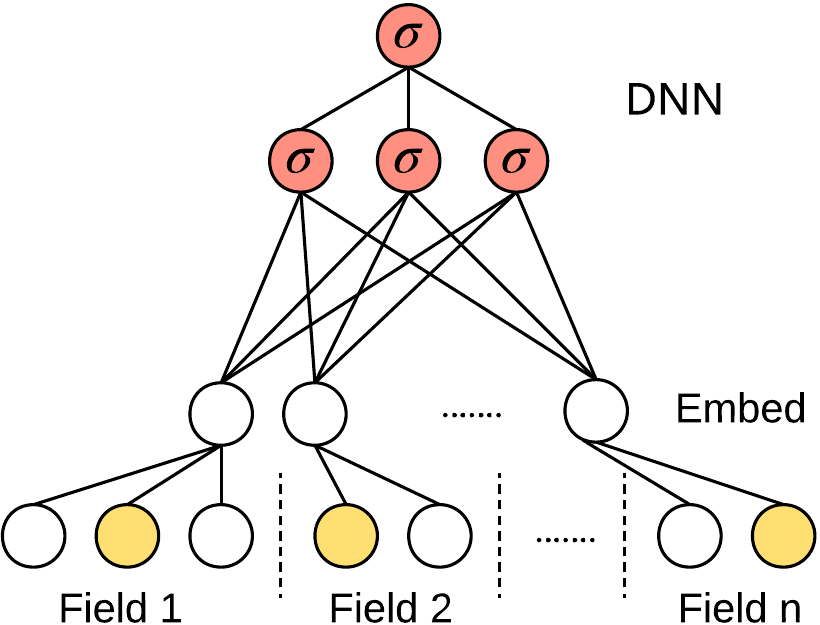}
}
\subfigure[CCPM]{
	\label{fig:ccpm}
	\includegraphics[width=.3\textwidth]{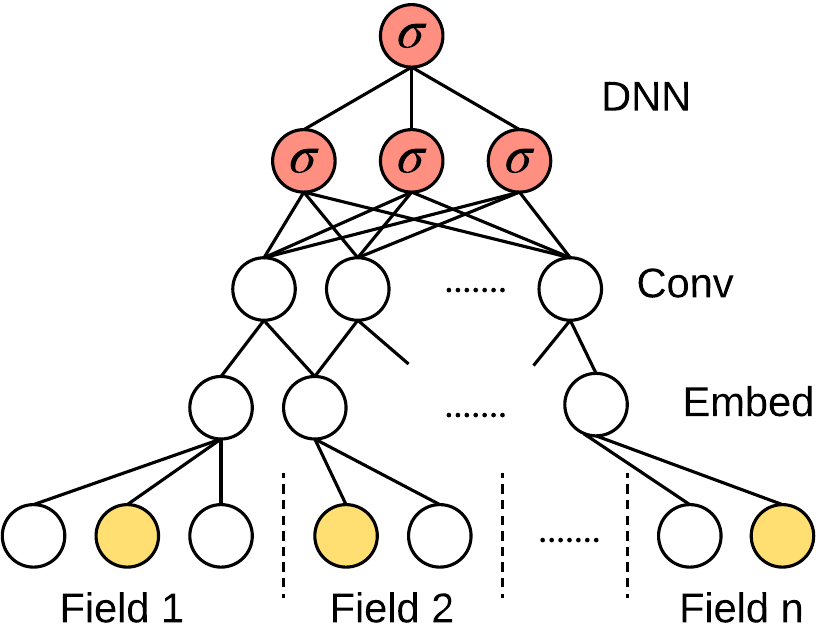}
}
\subfigure[DeepFM]{
	\label{fig:deepfm}
	\includegraphics[width=.314\textwidth]{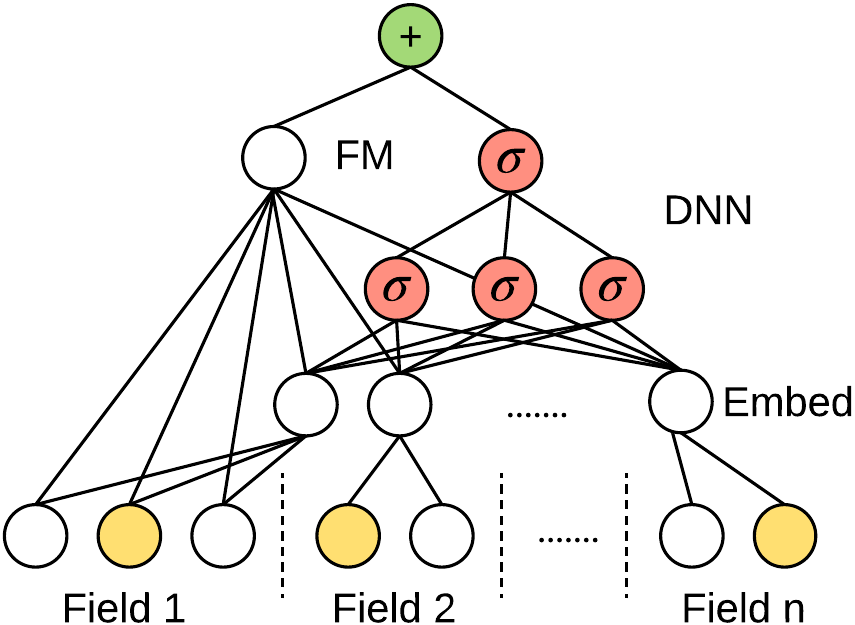}
}
\caption{Illustration of FNN, DeepFM, and CCPM. Yellow node means one-hot input of a field; green `$+$' node means add operation; red `$\sigma$' node means nonlinear operation.}
\label{fig:dnns}
\end{figure}

\subsection{\revise{DNN-based Models}}\label{sec:related-dl}
With the great success of deep learning, it is not surprising there emerge some deep learning techniques for recommender systems \change{\cite{zhang2017deep}}. 
Primary work includes: 
(i) Pretraining autoencoders to extract feature representations, e.g., Collaborative Deep Learning \cite{wang2015collaborative}.
(ii) \revise{Using DNN to model general interactions} \cite{he2017neuralcf, he2017neuralfm, qu2016product}.
\change{(iii) Using DNN \revise{to process images/texts in content-based recommendation \cite{wang2017dynamic}}.}
\change{We mainly focus on (ii) in this paper.}

The input to DNN is usually dense and numerical, while the case of multi-field categorical data has not been well studied. 
FM supported Neural Network (FNN) \cite{zhang2016deep} (Fig.~\ref{fig:dnns}(a)) \revise{has an embedding layer and a DNN classifier. 
Besides, FNN uses FM to pre-train the embedding layer.}
\revise{Other models use DNN to improve FM}, e.g., Neural Collaborative Filtering (NCF) \cite{he2017neuralcf}, Neural FM (NFM) \cite{he2017neuralfm}\change{, Attentional FM (AFM) \cite{xiao2017attentional}. 
\revise{NCF uses DNN to solve collaborative filtering problem.} 
NFM extends NCF to more general \revise{recommendation} scenarios.
Based on NFM, AFM uses attentive mechanism to improve feature interactions, and becomes a state-of-the-art model.}

Convolutional Click Prediction Model (CCPM) \cite{liu2015convolutional} (Fig.~\ref{fig:dnns}(b)) uses \revise{convolutional layers} to explore local-global dependencies. 
CCPM performs convolutions on the neighbor fields in a certain alignment, but fails to model convolutions among non-neighbor fields.
RNNs are leveraged to model \revise{historical user behaviors \cite{zhang2014sequential}.
In this paper, we do not consider sequential patterns.}

Wide \& Deep Learning (WDL) \cite{cheng2016wide} trains a wide model and a deep model jointly.
\change{The wide part uses LR to ``memorize'', meanwhile, the deep part uses DNN to ``generalize''.} 
Compared with single models, WDL achieves \revise{better AUC in offline evaluations and higher} CTR in online A/B test. 
\revise{WDL requires human efforts for feature engineering on the input to the wide part, thus is not end-to-end.}
DeepFM \cite{guo2017deepfm}, as illustrated in Fig.~\ref{fig:dnns}(c), can both utilize the strengths of WDL and avoid expertise in feature engineering. 
It replaces the wide part of the WDL with FM. 
Besides, 
\revise{the FM component and the deep component share same embedding parameters.}
DeepFM is regarded as one state-of-the-art model in user response prediction.

\revise{To complete our discussion of DNN-based models, we list some less relevant work, such as the following.}
Product Unit Neural Network \cite{engelbrecht1999training} defines the output of each neuron as the product of all its inputs. 
Multilinear FM \cite{lu2017multilinear} studies FM in a multi-task setting. 
DeepMood \cite{cao2017deepmood} presents a neural network view for FM and Multi-view Machine. 


\subsection{Net-in-Net Architecture}\label{sec:related-nin}
Network In Network (NIN) \cite{lin2013network} is originally proposed in CNN.
NIN builds micro neural networks between convolutional layers to abstract the data within the receptive field. 
Multilayer perceptron as a potent function approximator is used in micro neural networks of NIN.
GoogLeNet \cite{szegedy2015going} makes use of the micro neural networks suggested in \cite{lin2013network} and achieves great success. 
NIN is powerful in modeling local dependencies.
In this paper, we borrow the idea of NIN, and propose to explore inter-field feature interactions with flexible micro networks.

\section{Methodology}\label{sec:method}
\revise{As in Fig.~\ref{fig:decom}, the difficulties in learning multi-field categorical data are decomposed into 2 phases: representation and classification.}
\revise{Following this idea, we first explain field-aware feature interactions, then we study the deficiency of DNN classifiers}, finally, we present our Product-based Neural Networks (PNNs) in detail.

A commonly used objective for user response prediction is to minimize cross entropy, or namely log loss, defined as
\begin{align}
\bL(y, \sigma(\hat{y})) = -y \log(\sigma(\hat{y})) - (1-y) \log(1-\sigma(\hat{y}))~,
\end{align}
where $y \in \{0, 1\}$ is the label and $\sigma(\hat{y}) \in (0, 1)$ is the predicted probability of $y = 1$, more specifically, the probability of a user giving a positive response on the item. 
We adopt this training objective in all experiments. 

\subsection{Field-aware Feature Interactions}\label{sec:method-fafi}
In user response prediction, the input data contains multiple fields, e.g., \ttt{WEEKDAY}, \ttt{GENDER}, \ttt{CITY}. 
A field contains multiple categories 
and takes one category in each data instance. 
Table~\ref{tab:example} shows 4 data examples, each of which contains 3 fields, 
and each field takes a single value.
For example, a \ttt{Male} customer located in \ttt{London} buys some beer on \ttt{Tuesday}.
\revise{From this record we can extract a useful feature combination}: ``\ttt{Male} and \ttt{London} and \ttt{Tuesday} implies True''.
The efficacy of feature combinations (a.k.a., cross features) has already been proved \cite{menon2011response, rendle2010factorization}.
In FM, the $2$nd order combinations are called feature interactions.

Assume the input data has $n$ categorical fields, \change{ $\bx \in \bN^n$, where $x_i$ 
is an ID indicating a category of the $i$-th field.}
\revise{The feature representations learned by parametric models}
could be weight coefficients (e.g., in LR) or latent vectors (e.g., in FM).
For an input instance $\bx$, each field is converted into corresponding weight $x_i \rightarrow w_i$ or latent vector $x_i \rightarrow \bv_i$. 
For an output $\hat{y}$, the probability is obtained from sigmoid function $\hat{y} \rightarrow \sigma(\hat{y}) = 1 / (1+\exp(-\hat{y}))$.
For simplicity, we use $w_i$ and $\bv_i$ to represent the input, and we use $\hat{y}$ to represent the output.

\subsubsection{A Coupled Gradient Issue of Latent Vector-based Models}\label{sec:method-fafi-couple}
The prediction of FM \cite{rendle2010factorization} can be formulated as
\begin{align}
\hat{y}_{FM} & = \sum_{i=1}^{n} w_i + \sum_{i=1}^{n-1} \sum_{j=i+1}^{n} \langle \bv_i, \bv_j \rangle + b ~,
\end{align}
where $w_i \in \bR$ is the weight of category $x_i$, $\bv_i \in \bR^k$ is the latent vector of $x_i$, and $b \in \bR$ is the global bias. 
Take the first example in Table~\ref{tab:example}, 
\begin{align}
\hat{y}_{FM} =w_{\bfm} + w_{\bfl} +  w_{\bft} + \langle \bv_\bfm, \bv_\bfl \rangle + \langle \bv_\bfl, \bv_\bft \rangle + \langle \bv_\bfm, \bv_\bft \rangle + b \nonumber ~.
\end{align} 
The gradient of the latent vector $\bv_\bfm$ is $\nabla_{\bv_\bfm} \hat{y}_{FM}= \bv_\bfl + \bv_\bft$. 
FM makes an implicit assumption that a field interacts with different fields in the same manner, which may not be realistic. 
\revise{Suppose \ttt{GENDER} is independent of \ttt{WEEKDAY}, it is desirable to learn $\bv_\bfm \perp \bv_\bft$. 
However, the gradient $\nabla_{\bv_\bfm} \hat{y}_{FM}$ continuously updates $\bv_\bfm$ in the direction of $\bv_\bft$.
Conversely, the latent vector $\bv_\bft$ is updated in the direction of $\bv_\bfm$.
To summarize, FM uses the same latent vectors in different types of inter-field interactions, which is an over-simplification and degrades the model capacity. 
We call this problem a coupled gradient issue.}

\revise{The gradients of latent vectors could be decoupled by assigning different weights to different interactions, such as the attentive mechanism in Attentional FM (AFM) \cite{xiao2017attentional}:}
\begin{align}
\hat{y}_{AFM} & = \sum_{i=1}^{n} w_i + \sum_{i=1}^{n-1} \sum_{j=i+1}^{n} f(\bv_i, \bv_j) \langle \bv_i, \bv_j \rangle + b ~,
\end{align}
where $f$ denotes the attention network which takes embedding vectors as input and assigns weights to interactions, $f(\bv_i, \bv_j) \in \bR$. 
In AFM, the gradient of $\bv_\bfm$ becomes $\nabla_{\bv_\bfm} \hat{y}_{AFM} = f(\bv_\bfm, \bv_\bfl)\bv_\bfl + f(\bv_\bfm, \bv_\bft)\bv_\bft + others$, where the gradients of $\bv_\bfm$ and $\bv_\bft$ \revise{are decoupled when $f(\bv_\bfm, \bv_\bft)$ approaches 0.
However, when the attention score approaches 0, the attention network becomes hard to train.}

This problem is solved by Field-aware FM (FFM) \cite{juan2016field}
\begin{align}
\hat{y}_{FFM} = \sum_{i=1}^{n} w_i + \sum_{i=1}^{n-1} \sum_{j=i+1}^{n} \langle \bv_{i,j}, \bv_{j,i} \rangle + b ~,
\end{align}
where $\bv_{i} \in \bR^{n \times k}$, meaning the $i$-th category has $n$ independent $k$-dimensional latent vectors when interacting with $n$ fields.
\revise{Excluding intra-field interactions}, we usually use $\bv_i \in \bR^{(n-1) \times k}$ in practice.
Using field-aware latent vectors, the gradients of different interactions are decoupled, e.g., $\nabla_{\bv_{\bfm,\ttt{CITY}}} \hat{y}_{FFM} = \bv_{\bfl,\ttt{GENDER}}$, $\nabla_{\bv_{\bfm,\ttt{WEEKDAY}}} \hat{y}_{FFM} = \bv_{\bft,\ttt{GENDER}}$.
This leads to the main advantage of FFM over FM and brings a higher capacity.

FFM makes great success in various data science contests. 
However, it has a memory bottleneck, because its latent vectors have $O(N n k)$ parameters (FM is $O(Nk)$), where $N$, the total number of categories, is usually in million to billion scales in practice. 
When $Nn$ is large, $k$ must be small enough to fit FFM in memory, which severely restricts the expressive ability of latent vectors. 
\revise{This problem is also discussed in Section~\ref{sec:exp-fafi} through visualization.}
To tackle the problems of FM and FFM, we propose kernel product.

\subsubsection{Kernel Product}\label{sec:method-fafi-kernel}

Matrix factorization (MF) learns low-rank representations of a matrix.
A matrix $A$ can be represented by the product of two low-rank matrices $P$ and $Q$, i.e., $A = PQ^\top$.
MF estimates each observed element $A_{i,j}$ with the inner product of two latent vectors $\bp_{i}$ and $\bq_{j}$.
The target of MF is to find optimal latent vectors which can minimize the empirical error
\begin{align}
\hat{A}_{i,j} & = \langle \bp_{i}, \bq_{j} \rangle \\
P^*, Q^* & = \arg\min_{P, Q} \sum_{i,j} \bL(A_{i,j}, \hat{A}_{i,j}) ~,
\end{align}
where $\langle \bp, \bq \rangle = \sum_{s} p_s q_s$ is the inner product of two vectors, $\bL$ represents the loss function (e.g., root mean square error, log loss).

MF has \revise{another} form, $A = UD V^\top$, where $D$ can be regarded as a projection matrix. 
$U$ and $V$ factorize $A$ in the projected space like $P$ and $Q$ do in the original space.
We define the inner product in a projected space, namely kernel product,  $\langle \bp, \bq \rangle_\phi = \bp^\top \phi \bq$, then we can extend MF
\begin{align}
\hat{A}_{i,j}^{\phi} & = \langle \bp_{i}, \bq_{j} \rangle_\phi \\
P^*, Q^*, \phi^* & = \arg\min_{P,Q,\phi} \sum_{i,j} \bL_{\mathcal{F}}(A_{i,j}, \hat{A}_{i,j}^{\phi}) ~, \label{eq:kernel}
\end{align}
where $\phi \in \mathcal{F}$ is a projection matrix, namely kernel, and $\mathcal{F}$ is the parameter space.
\revise{Vector inner product can be regarded as a special case of kernel product when $\phi = I$.
MF can be regarded as a special case of kernel MF when $\mathcal{F} = \{I\}$.}
\revise{Kernel product also generalizes vector outer product.
The convolution sum of an outer product is equivalent to a kernel product}
\begin{align}
\bp \bq^\top \odot \phi  & = \sum_{s=1}^{k}\sum_{t=1}^{k} p_s \phi_{s,t} q_t  = \bp^\top \phi \bq ~,
\end{align}
where $\odot$ denotes convolution sum.
It is worth noting that, the outer product form $\bp \bq^\top \odot \phi$ has $2k^2$ multiplication and $k^2$ addition operations, while the kernel product form $\bp^\top \phi \bq$ has $k^2+k$ multiplication and $k^2+k$ addition operations.
\revise{Therefore, kernel product generalizes both vector inner product and outer product.}
In Eq.~\eqref{eq:kernel}, the kernel matrix is optimized in a parameter space, \revise{and a kernel matrix maps two vectors to a real value.}
\change{From this point, a kernel is equivalent to a function.
We can define kernel product in parameter or function spaces to adjust to different problems. 
In this paper, we study (i) linear kernel, and (ii) micro network kernel.}

In practice, field size (number of categories in a field) varies dramatically (e.g., \ttt{GENDER}=2, \ttt{CITY}=7).
Field size reflects the amount of information contained in one field. 
It is natural to represent a large (small, respectively) field in a large (small, respectively) latent space, and we call it \revise{adaptive embedding.}
In \cite{covington2016deep}, \revise{an adaptive embedding }size is decided by the logarithm of the field size.
The challenge is how to combine \revise{adaptive embeddings with MF}, since inner product requires the latent vectors to have the same length.
Kernel product can solve this problem
\begin{align}
\langle \bp, \bq \rangle_\phi & = \sum_{s=1}^{k_1}\sum_{t=1}^{k_2} p_s \phi_{s,t} q_t ~,
\end{align}
where $\bp \in \bR^{k_1}$, $\bq \in \bR^{k_2}$, and $\phi \in \bR^{k_1 \times k_2}$.

\subsubsection{Field-aware Feature Interactions}\label{sec:method-fafi-fafi}
\revise{The coupled gradient issue of latent vector-based models can be solved by field-aware feature interactions.}
FFM learns field-aware feature interactions with field-aware latent vectors.
However, the space complexity of FFM is $O(Nnk)$, which restricts FFM from using large latent vectors.
\revise{A relaxation of FFM is projecting latent vectors into different kernel spaces.
Corresponding to $O(n^2)$ inter-field interactions, the $O(n^2)$ kernels require $O(n^2k^2)$ extra space.
Since $n^2k^2 \ll Nk$, the total space complexity is still $O(Nk)$.
In this paper, we extend FM with (i) linear kernel, and (ii) micro network kernel.}

Kernel FM (KFM):
\begin{align}
\hat{y}_{KFM} & = \sum_{i=1}^{n} w_i + \sum_{i=1}^{n-1} \sum_{j=i+1}^{n} \langle \bv_i, \bv_j \rangle_{\phi_{i,j}} + b ~,
\end{align}
where $\phi_{i,j} \in \bR^{k\times k}$ is the kernel matrix of field $i$ and $j$.

Network in FM (NIFM):
\begin{align}
\hat{y}_{NIFM} & = \sum_{i=1}^{n} w_i + \sum_{i=1}^{n-1} \sum_{j=i+1}^{n} f_{i, j}(\bv_i, \bv_j) + b \\
f_{i,j}(\bv_i, \bv_j) & = \sigma([\bv_i, \bv_j]^\top \bw_{i,j}^1 + \bb_{i,j}^1)^\top \bw_{i,j}^2 ~, \label{eq:sub-net}
\end{align}
where $f_{i, j}$ denotes a micro network taking latent vectors as input and producing feature interactions with nonlinearity.
In Eq.~\eqref{eq:sub-net}, the micro network output has no bias term because it is redundant with respect to the global bias $b$.
This model is inspired by net-in-net architecture \cite{lin2013network, szegedy2015going}.
\change{With flexible micro networks, we can control the complexity and take the advantages of enormous deep learning techniques.}

\change{
Recall the first example in Table~\ref{tab:example}.
Suppose \ttt{GENDER} is independent of \ttt{WEEKDAY}, we can have $\langle \bv_\bfm, \bv_\bft \rangle_{\phi_{\ttt{GENDER}, \ttt{WEEKDAY}}} = 0$ if (i) $\phi_{\ttt{GENDER}, \ttt{WEEKDAY}} \bv_\bft \perp \bv_\bfm$, or (ii) $\phi_{\ttt{GENDER}, \ttt{WEEKDAY}} = 0$.
Thus, the gradients of latent vectors are decoupled.
\revise{Comparing KFM with FM/FFM: 
(i) KFM bridges FM with FFM. 
(ii) The capacity of KFM is between FM and FFM, because KFM shares kernels among certain types of inter-field interactions.
(iii) KFM re-parametrizes FFM, which is ``time for space''.
Comparing KFM/NIFM with AFM, AFM uses a universal attention network which is field-sharing, while KFM/NIFM use multiple kernels which are field-aware.
If we share one kernel among all inter-field interactions, it will become an attention network. 
Thus, KFM/NIFM generalize AFM.
Comparing kernel product with CNN, their kernels are both used to extract feature representations. 
Besides, kernel product shares projection matrices/functions among interactions, and CNN shares kernels in space.}
}

\subsection{Training Feature Interactions with Trees or DNNs is Difficult}\label{sec:method-diff}

In the previous section, we propose kernel product to solve the coupled gradient issue in latent vector-based models. 
On the other hand, trees and DNNs can approximate very complex functions.
In this section, we analyze the difficulties of trees and DNNs in learning feature interactions.

\subsubsection{\change{A Sparsity Issue of Trees}}\label{sec:method-diff-tree}
Growing a decision tree performs greedy search among categories and splitting points \cite{quinlan1996improved}. 
Tree models encounter a sparsity issue in multi-field categorical data. Here gives an example.
\change{A tree with depth 10 has at most $10^3$ leaf nodes ($2^{10} \approx 10^{3}$), and a tree with depth 20 has at most $10^6$ leaf nodes ($2^{20}\approx 10^{6}$). 
Suppose we have a dataset 
with 10 categorical fields, each field contains $100$ categories, and the input dimension is $10^3$ after one-hot encoding.
This dataset has $C_{10}^1 100$ categories, $C_{10}^2 100^2$ $2$nd order feature combinations, and $C_{10}^{10} 100^{10}$ full order feature combinations. 
From this example, we can see that even a very deep tree model can only explore a small fraction of all possible feature combinations on such a small dataset. 
Therefore, modeling capability of tree models, e.g., Decision Tree, Random Forest and Gradient Boosting Decision Tree (GBDT) \cite{chen2016xgboost}, is restricted in multi-field categorical settings.}

\subsubsection{\change{An Insensitive Gradient Issue of DNNs}}\label{sec:method-diff-nn}
\begin{figure}[tbp]
	\centering
	\includegraphics[width=0.6\textwidth]{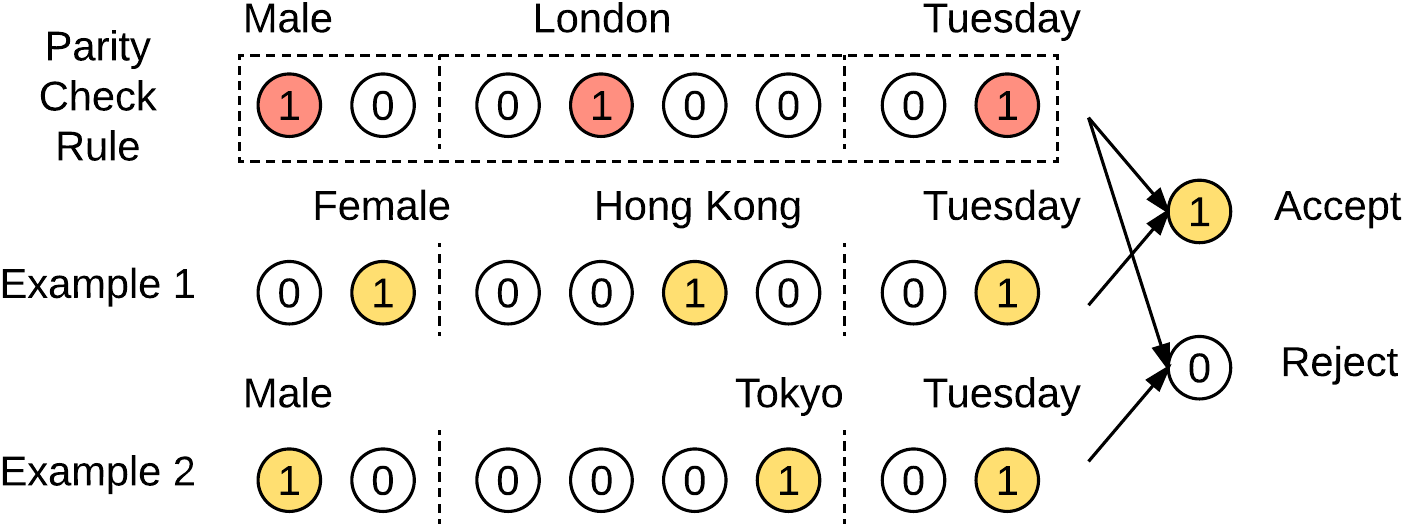}
	\caption{Two examples of parity check from Table~\ref{tab:example}.}
	\label{fig:parity}
\end{figure}

Gradient-based DNNs refer to the DNNs based on gradient descent and backpropagation.
Despite the universal approximation property \cite{hornik1989multilayer}, there is no guarantee that a DNN naturally converges to any expected functions using gradient-based optimization.
\revise{In user response prediction, the target function can be represented by a set of rules, e.g., ``\ttt{Male} and \ttt{London} and \ttt{Tuesday} implies True''.}
Here we show the periodic property of the \revise{target function} via parity check.
Recall the examples in Table~\ref{tab:example}, a feasible classifier is $parity(\bx,\bh), \bh \in \mathcal{H}$, where $\bx$ is the input, $\bh = \{\ttt{Male}, \ttt{London}, \ttt{Tuesday}\}$ is the checking rule, and $\mathcal{H}$ is the hypothesis space.
$\bx_1 = \{\ttt{Male}, \ttt{London}, \ttt{Tuesday}\}$ is accepted by the predictor because 3 (which is odd) conditions are matched and $\bx_3 = \{\ttt{Female}, \ttt{Hong Kong}, \ttt{Tuesday}\}$ is also accepted because 1 (which is also odd) condition is matched. 
In contrast, $\bx_2 = \{\ttt{Male}, \ttt{Tokyo}, \ttt{Tuesday}\}$ and $\bx_4 = \{\ttt{Female}, \ttt{New York}, \ttt{Monday}\}$ are rejected since 2 and 0 (which are even) conditions are matched.
Two examples are shown in Fig.~\ref{fig:parity}.

From this example, we observe that, in multi-field categorical data: 
(i) Basic rules can be represented by feature combinations, and several basic rules induce a parity check.
(ii) The periodic property \revise{reveals that }a feature set giving positive results does not conclude its superset nor its subset being positive.
(iii) \revise{Any target functions should also reflect the periodic property.}


A recent work \cite{shalev2017failures} proves an insensitive gradient issue of DNN: 
(i) If the target is a large collection of uncorrelated functions, the variance (sensitivity) of DNN gradient to the target function decreases linearly with $|\mathcal{H}|$.
(ii) When variance decreases, the gradient at any point will be extremely concentrated around a fixed point independent of $\bh$.
(iii) When the gradient is independent of the target $\bh$, it contains little useful information to optimize DNN, thus gradient-based optimization fails to make progress.
The authors in \cite{shalev2017failures} use the variance of gradient w.r.t. hypothesis space to measure the useful information in gradient, and explain the failure of gradient-based deep learning with an example of parity check.

\change{Considering the large hypothesis space of DNNs, we conclude that learning feature interactions with gradient-based DNNs is difficult.
In another word, DNNs can hardly learn feature interactions implicitly or automatically.
We conduct a synthetic experiment to support this idea in Section~\ref{sec:exp-diff}.
And we propose product layers to help DNNs tackle this problem.}

\subsection{Product-based Neural Networks}\label{sec:method-pnn}


\subsubsection{DNN-based Models}\label{sec:method-pnn-dl}
\revise{For consistency, we introduce DNN-based models according to the 3 components:}
the embedding module, \change{the interaction module}, and the DNN classifier
\begin{align}
\hat{y}_{DNN} = \net(\inter(\embed(\bx))) ~.
\end{align}

\revise{AFM \cite{xiao2017attentional} has already been discussed in Section~\ref{sec:method-fafi}.
AFM uses an attention network to improve FM, yet its prediction is simply based on the sum of interactions,}
\begin{align}
\hat{y}_{AFM} = \text{sum}(\text{attend}(\embed(\bx)))~.
\end{align}

FM supported Neural Network (FNN) \cite{zhang2016deep} (Fig.~\ref{fig:dnns}(a)) is formulated as
\begin{align}
\hat{y}_{FNN} = \net(\embed(\bx)) ~,
\end{align}
where $\embed(\cdot)$ is initialized from a pre-trained FM model.
Compared with shallow models, FNN has a powerful DNN classifier, which gains it significant improvement. 
However, without the interaction module, FNN may fail to learn expected feature interactions automatically.

Similarly, Convolutional Click Prediction Model (CCPM) \cite{liu2015convolutional} (Fig.~\ref{fig:dnns}(b)) is formulated as
\begin{align}
\hat{y}_{CCPM} = \net(\text{conv}(\embed(\bx))) ~,
\end{align}
where $\text{conv}(\cdot)$ denotes convolutional layers which are expected to explore local-global dependencies. 
CCPM only performs convolutions on the neighbor fields in a certain alignment but fails to model the full convolutions among non-neighbor fields.
 
DeepFM \cite{guo2017deepfm} (Fig.~\ref{fig:dnns}(c)) improves Wide \& Deep Learning (WDL) \cite{cheng2016wide}. 
It replaces the wide model with FM and gets rid of feature engineering expertise.
\begin{align}
\hat{y}_{DeepFM} =\hat{y}_{FM} + \hat{y}_{FNN} ~.
\end{align}
Note that the embedding vectors of the FM part and the FNN part are shared in DeepFM.
WDL and DeepFM follows a joint training scheme.
In another word, \revise{other single models can also be integrated into a mixture model}, yet the joint training scheme is beyond our discussion.

\begin{figure}[tbp]
	\subfigure[Product-based Neural Network]{
		\includegraphics[height=0.2\textheight]{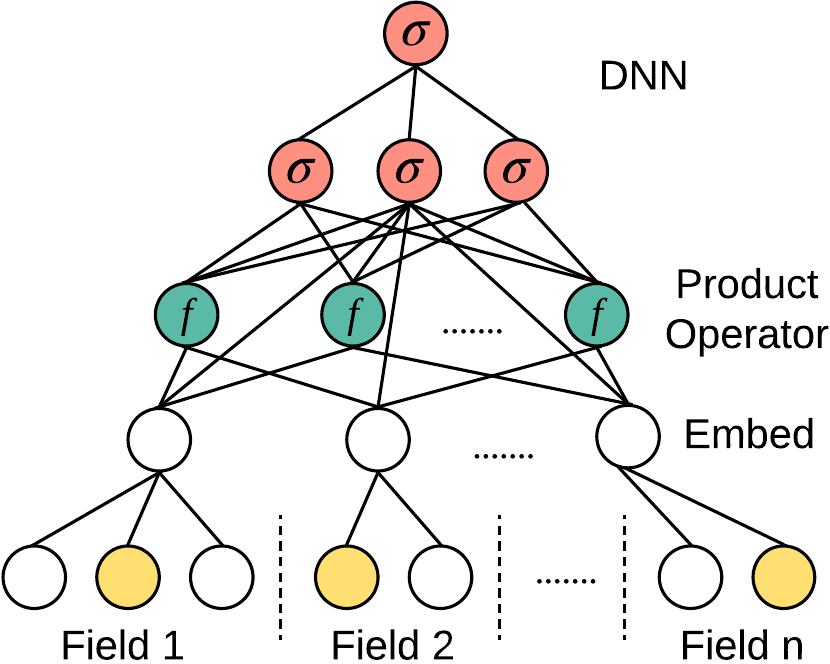}
	}\\
	\subfigure[Inner Product]{
		\includegraphics[height=0.15\textheight]{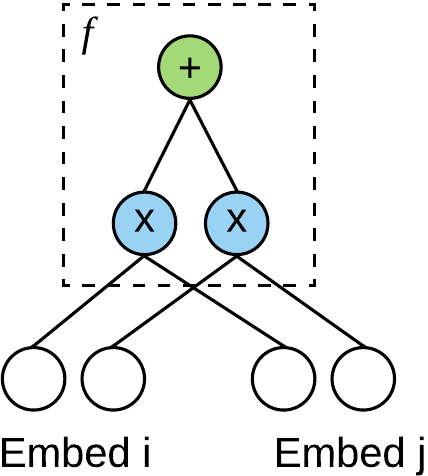}
	}
	\subfigure[Kernel Product]{
		\includegraphics[height=0.15\textheight]{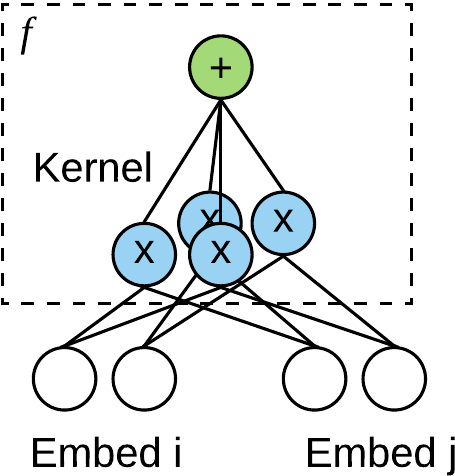}
	}
	\subfigure[Micro Network]{
		\includegraphics[height=0.15\textheight]{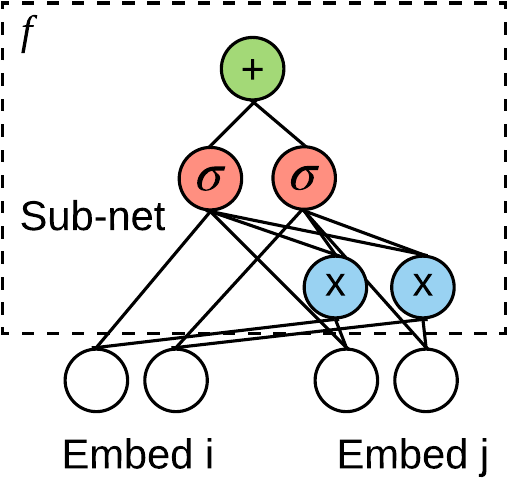}
	}	
	\caption{Product-based Neural Networks. \emph{Note:} Yellow node means one-hot input of a field; blue `$\times$' node means multiply operation; green `$+$' node means add operation; red `$\sigma$' node means nonlinear operation; \change{cyan `$f$' node means some product operator. \revise{PNN uses multiple product operators in different inter-field interactions.} There are 3 types of product operators: (b) inner product, (c) kernel product, and (d) micro network. In (b)-(d), we use 2 white nodes to represent an embedding vector of length 2.}}
	\label{fig:pnn_all}
\end{figure}

\change{FNN has a linear feature extractor (without pre-training) and a DNN classifier. 
CCPM additionally explores local/global dependencies with convolutional layers, but the exploration is limited in neighbor fields.
\revise{DeepFM has a bi-linear feature extractor, yet the bi-linear feature representations are not fed to its DNN classifier.
}
\revise{The insensitive gradient issue of DNN-based models has already been discussed in Section~\ref{sec:method-diff}.
To solve this issue, we propose Product-based Neural Network (PNN). 
The general architecture of PNN is illustrated in Fig.~\ref{fig:pnn_all}(a). 
}}
\begin{align}
\bv & = \text{embed}(\bx) \label{eq:embed} \\
\bp & = \text{product}(\bv) \label{eq:product}\\
\hat{y}_{PNN} & = \net(\bv, \bp) \label{eq:net}
\end{align}
The embedding layer \eqref{eq:embed} is field-wisely connected.
This layer looks up the corresponding embedding vector for each field $x_i \rightarrow \bv_i$, and produces dense representations of the sparse input, $\bv^\top = [\bv_1, \dots, \bv_n]$. 
The product layer \eqref{eq:product} uses multiple product operators to explore feature interactions $\bp^\top = [\bp_{1,2}, \dots, \bp_{n-1,n}]$. 
The DNN classifier \eqref{eq:net}, takes both the embeddings $\bv$ and the interactions $\bp$ as input, and conduct the final prediction $\hat{y}_{PNN}$.

Using FM, KFM, and NIFM as feature extractors, we develop 3 PNN models: Inner Product-based Neural Network (IPNN), Kernel Product-based Neural Network (KPNN), and Product-network In Network (PIN), respectively.
\revise{We decompose all discussed parametric models in Table~\ref{tab:relation}. 
A component level comparison is in Table~\ref{tab:compose}, e.g., FM $+$ kernel product $\rightarrow$ KFM.
}

\revise{One may concern the complexity, initialization, or training of PNNs.
As for the complexity, there are $O(n^2)$ interactions, yet the complexity may not be a bottleneck:
(i) In practice, $n$ is usually a small number. In our experiments, the datasets involved contain at most 39 fields.
(ii) The computation of interactions is independent and can speed up via parallelization.
PNN concatenates embeddings and interactions as the DNN input.
The embeddings and the interactions follow different distributions, which may cause problems in initialization and training.
One solution to this potential risk is careful initialization and normalization.
These practical issues are discussed in Section~\ref{sec:issue}, and corresponding experiments are in Section~\ref{sec:exp-prac}.}

\begin{table}[tbp]
\centering
\caption{Model decomposition.}
\label{tab:relation}
\begin{tabular}{c|c|c|c}
& \tabincell{c}{linear \\ features} & 
\tabincell{c}{feature \\ interactions} & \tabincell{c}{field-aware \\ feature  interactions} \\ \hline
\tabincell{c}{linear \\ classifier} & LR & 
FM & \tabincell{c}{FFM, KFM, NIFM} \\ \hline
\tabincell{c}{DNN \\ classifier} & \tabincell{c}{\revise{FNN (w/o pre-}\\train), CCPM} & 
DeepFM, IPNN & KPNN, PIN
\end{tabular}
\end{table}

\begin{table}[tbp]
\centering
\caption{Component-level comparison.}
\label{tab:compose}
\begin{tabular}{p{2cm}|p{1cm}p{1cm}p{1cm}p{1cm}p{1cm}p{1cm}}
	& FM & KFM & NIFM & FNN & KPNN & PIN \\ \hline
	embedding & yes & yes & yes & yes & yes & yes \\ \hline
	kernel & & yes & & & yes & \\ \hline
	net-in-net & & & yes & & & yes \\ \hline
	DNN & & & & yes & yes & yes \\
\end{tabular}
\end{table}

\subsubsection{Inner Product-based Neural Network}\label{sec:method-pnn-ipnn}

IPNN uses FM as the feature extractor, where the feature interactions are defined as inner products of the embeddings, as illustrated in Fig.~\ref{fig:pnn_all}(b). 
The $n$ embeddings of $\bv$, and the $n(n-1)/2$ interactions of $\bp$ are flattened and fully connected with the successive hidden layer
\begin{align}
\bp^\top & = [\langle \bv_1, \bv_2 \rangle, \dots, \langle \bv_{n-1}, \bv_n \rangle] \\
\hat{y}_{IPNN} & = \net({[\bv_1, \dots, \bv_n, p_{1,2}, \dots, p_{n-1,n}]})~.
\end{align}

\change{
Comparing IPNN with Neural FM (Section~\ref{sec:related-dl}), 
their inputs to DNN classifiers are quite different: (i) In Neural FM, the interactions are summed up and passed to DNN. (ii) In IPNN, the interactions are concatenated and passed to DNN.
\revise{Since AFM has no DNN classifier, it is compared with KFM/NIFM in Section~\ref{sec:method-fafi}.
Comparing IPNN with FNN: (i) FNN explores feature interactions through pre-training). (ii) IPNN explores feature interactions through the product layer.
Comparing IPNN with DeepFM: (i) DeepFM adds up the feature interactions to the model output. (ii) IPNN feeds the feature interactions to the DNN classifier.}
}

\subsubsection{Kernel Product-based Neural Network}\label{sec:method-pnn-kpnn}
KPNN utilizes KFM as the feature extractor, where the feature interactions are defined as kernel products of the embeddings, as illustrated in Fig.~\ref{fig:pnn_all}(c).
Since kernel product generalizes outer product, we use kernel product as a general form.
The formulation of KPNN is similar to IPNN, except that
\begin{align}
\bp^\top & = [\langle \bv_1, \bv_2 \rangle_{\phi_{1,2}}, \dots, \langle \bv_{n-1}, \bv_n \rangle_{\phi_{n-1,n}}] \\
\hat{y}_{KPNN} & = \net({[\bv_1, \dots, \bv_n, p_{1,2}, \dots, p_{n-1,n}]}) ~.
\end{align}

\subsubsection{Product-network In Network}\label{sec:method-pnn-pin}
A micro network is illustrated in Fig.~\ref{fig:pnn_all}(d)\footnote{We test several sub-net architectures and the structure in Fig.~\ref{fig:pnn_all}(d) is a relatively good choice.}. 
\revise{In PIN, the computation of several sub-network\footnote{In this paper, we use micro network and sub-network interchangeably.} forward passes are merged into a single tensor multiplication}
\begin{align}
\bH_1^\top & = [\bh_1, \bh_2, \dots, \bh_m] \\
\bH_2^\top & = \sigma([\bw_1, \bw_2, \dots, \bw_m]^\top \bH_1) \nonumber \\
& = \sigma([\bw_1^\top \bh_1, \bw_2^\top \bh_2, \dots, \bw_m^\top \bh_m]) ~,
\end{align}
where $\bh_i \in \bR^{d_1}$ is the input to sub-network $i$, 
and $d_1$ is the input size.
$\bH_1$ concatenates all $\bh_i$ together, $\bH_1 \in \bR^{m \times d_1}$, where $m$ is the number of sub-networks.
The weights $\bw_i$ of sub-network $i$ are also concatenated to a weight tensor $[\bw_1, \dots, \bw_m] \in \bR^{d_2 \times m \times d_1}$, where $d_2$ is the output size of a sub-network.
Applying tensor multiplication on dimension $d_1$ and element-wise nonlinearity $\sigma$, we get $\bH_2 \in \bR^{m \times d_2}$.

\revise{Layer normalization (LN) \cite{ba2016layer} is proposed to stabilize the activation distributions of DNN.} 
In PIN, we use fused LN on sub-networks \revise{to stabilize training.}
\revise{For each data instance, LN collects statistics from different neurons and normalizes the output of these neurons.
However, the sub-networks are too small to provide stable statistics.  
Fused LN instead collects statistics from all sub-networks, thus is more stable than LN.
More detailed discussions are in Section~\ref{sec:issue-reg}, and corresponding experiments are in Section~\ref{sec:exp-prac}.}
\begin{align}
\text{LN}(\bw_i^\top \bh_i) & = \frac{\bw_i^\top \bh_i - \text{mean}_{k=1}^{d_2}(\bw_i^\top \bh_i)}{\text{std}_{k=1}^{d_2}(\bw_i^\top \bh_i)} \mathbf{g}_i + \bb_i\\
\text{fused-LN}(\bw_i^\top \bh_i) & = \frac{\bw_i^\top \bh_i - \text{mean}_{i=1,k=1}^{m,d_2}(\bw_i^\top \bh_i)}{\text{std}_{i=1,k=1}^{m,d_2}(\bw_i^\top \bh_i)} \mathbf{g} + \bb~.
\end{align}

Replacing $\{\bh_1, \dots, \bh_m\}$ with $\{\bh_{1,2}, \dots, \bh_{n-1,n}\}$, the PIN model is presented as follows,
\begin{align}
\bh_{i,j} & = [\bv_i, \bv_j, \bv_i \odot \bv_j] \\
f_{i,j}(\bv_i, \bv_j) & = \sigma(\bh_{i,j}^\top \bw_{i,j}^1 + \bb_{i,j}^1)^\top \bw_{i,j}^2 + \bb_{i,j}^2 \\
\bp_{i,j} & = f_{i,j}(\bv_i, \bv_j), ~j \neq i \\
\hat{y}_{PIN} & = \net([\bp_{1,2}, \dots, \bp_{n-1,n}]) ~,
\end{align}
where $\odot$ denotes element-wise product instead of convolution sum. 
\revise{To stabilize micro network outputs, LN can be inserted into the hidden layers of the micro networks.}

Compared with NIFM, the sub-networks of PIN are slightly different.
(i) Each sub-network takes an additional product term as the input. 
(ii) The sub-network bias $\bb_2$ is necessary because there is no global bias like NIFM.
(iii) The sub-network output is a scaler in NIFM, however, it \revise{could be} a vector in PIN.
Compared with other PNNs, the embedding vectors are no longer fed to the DNN classifier because the embedding-DNN connections are redundant.
The embedding-DNN connections:
\begin{align}
[\bv_1, \dots, \bv_n]^\top [\bw_1, \dots, \bw_n] = \sum_{i=1}^{n} \bv_i^\top \bw_i ~.
\end{align}

\change{In terms of the embedding-subnet connections, the input contains several concatenated embedding pairs $[[\bv_1, \bv_2], [\bv_1, \bv_3], \dots, [\bv_i, \bv_j], \dots, [\bv_{n-1}, \bv_n]]$, and each pair is passed through some micro network. For simplicity, we regard the micro networks as linear transformations, thus the weight matrix can be represented by $[\bw_{i,j}, \bw_{i,j}']$, where the input dimension is twice the embedding size, and the output dimension is determined by the following classifier. 
}
\begin{align}
&\sum_{i=1}^{n-1}\sum_{j=i+1}^{n} [\bv_i, \bv_j]^\top [\bw_{i,j}, \bw_{i,j}'] \nonumber \\
= &\sum_{i=1}^{n-1}\sum_{j=i+1}^{n} \bv_i^\top \bw_{i,j} + \bv_j^\top \bw_{i,j}' \nonumber \\
= &\sum_{i=1}^{n-1} \bv_i^\top \sum_{j=i+1}^{n} \bw_{i,j} + \sum_{j=2}^{n} \bv_j^\top \sum_{i=1}^{j-1} \bw_{i,j}' \nonumber \\
= &\sum_{i=1}^{n} \bv_i^\top (\sum_{j<i} \bw_{j,i}' + \sum_{j>i} \bw_{i,j}) ~.
\end{align}

From these two formulas, we find the embedding-DNN connections can be yielded from the embedding-subnet connections: \change{$\bv_i^\top (\sum_{j<i} \bw_{j,i}' + \sum_{j>i} \bw_{i,j}) \rightarrow \bv_i^\top \bw_i''$.}

\section{Practical Issues}\label{sec:issue}
This section discusses several practical issues, some of which are mentioned in Section~\ref{sec:method-pnn} (e.g., initialization), the others are related to applications (e.g., data processing).
Corresponding experiments are located at Section~\ref{sec:exp-prac}.

\subsection{Data Processing}\label{sec:issue-num}

\revise{The data in user response prediction is usually categorical, 
and sometimes are numerical or multi-valued.
When the model input contains both one-hot vectors and real values, this model is hard to train:}
(i) One-hot vectors are usually sparse and high-dimensional while real values are dense and low-dimensional, \change{therefore, they have quite different distributions.}
(ii) Different from real values, one-hot vectors are not comparable in numbers. 
\change{For these reasons, categorical and numerical data should be processed differently.}

\revise{Numerical data is usually processed by bucketing/clustering:}
First, a numerical field is partitioned by a series of thresholds according to its histogram distribution. 
Then a real value is assigned to a bucket.
Finally, the bucket identifier is used to replace the original value.
For instance, $\text{age} < 18 \rightarrow \text{Child}$, $18 \le \text{age} < 28 \rightarrow \text{Youth}$, $\text{age} \ge 28 \rightarrow \text{Adult}$.
\revise{Another solution is trees~\cite{he2014practical}. 
Since decision trees split data examples into several groups, each group can be regarded as a category.}

\revise{Different from numerical and categorical data, set data takes multiple values in one data instance, e.g., an item has multiple tags.
The permutation invariant property of set data has been studied in \cite{zaheer2017deep}.
In this paper, the set data embeddings are averaged before feeding to DNN.}

\subsection{Initialization}\label{sec:issue-init}

Initializing parameters with small random numbers is widely adopted, e.g., uniform or normal distribution with 0 mean and small variance. 
For a hidden layer in DNN, an empirical choice for standard deviation is $\sqrt{1 / n_{in}}$, where $n_{in}$ is the input size of that layer.
Xavier initialization \cite{glorot2010understanding} takes both forward and backward passes into consideration:
taking uniform distribution $[-v_{max}, v_{max}]$ as an example, the upper bound $v_{max}$ should be 
$\sqrt{{6}/{(n_{in} + n_{out})}}$, 
where $n_{in}$ and $n_{out}$ are the input and output sizes of a hidden layer.
This setting stabilizes activations and gradients of a hidden layer at the early stage of training.

\begin{figure}[tbp]
\centering
\includegraphics[width=0.5\textwidth]{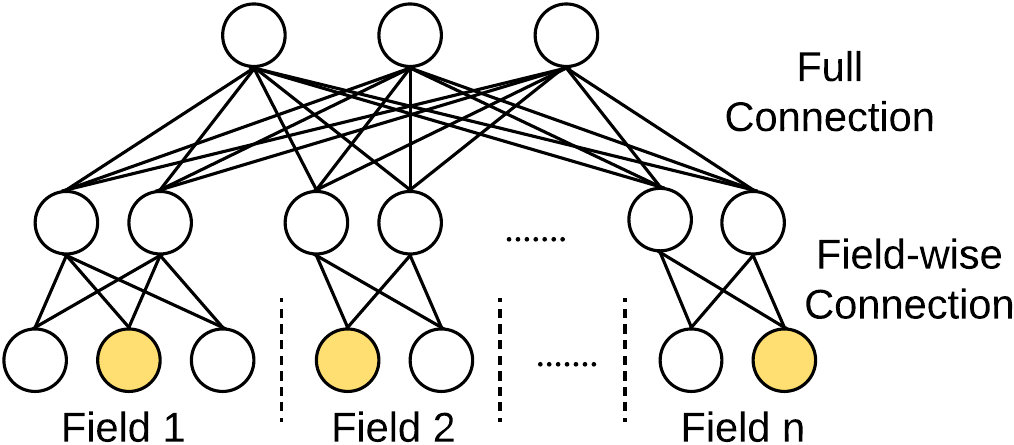}
\caption{An example of field-wise connection.}
\label{fig:nin}
\end{figure}

The above discussion is limited to (i) dense input, and (ii) fully connected layers. 
\change{Fig.~\ref{fig:nin} shows an embedding layer followed by a fully connected layer.
An embedding layer has sparse input and is field-wisely connected, i.e.,} \revise{each field is connected with only a part of the neurons.
Since an embedding layer usually has extremely large input dimension, typical standard deviations include: $\sqrt{c/Nk}$, $\sqrt{c/nk}$, $\sqrt{c/k}$, and pre-training, where $c$ is a constant, $N$ is the input dimension, $n$ is the number of fields, and $k$ is the embedding size.
Pre-training is used in \cite{zhang2016deep, xiao2017attentional}.
For convenience, we use random initialization in most experiments for end-to-end training.
And we compare random initialization with pre-training in Section~\ref{sec:exp-prac}.}

\subsection{Optimization}\label{sec:issue-opt}
In this section, we discuss potential risks of adaptive optimization algorithms in the scenario of sparse input. 
Compared with SGD, adaptive optimization algorithms converge much faster, e.g., AdaGrad \cite{duchi2011adaptive}, Adam \cite{kingma2014adam}, FTRL \cite{mcmahan2013ad, ta2015factorization}, among which Adam is an empirically good choice \cite{xu2015show, goodfellow2016deep}. 

Even though adaptive algorithms speed up training and sometimes escape from local minima in practice, there is no theoretical guarantee of better performance. 
Take Adam as an example, 
\begin{align}
m_t & = \beta_1 m_{t-1} + (1 - \beta_1) g_t \\
v_t & = \beta_2 v_{t-1} + (1 - \beta_2) g_t^2 \\
g_t' & = \frac{m_t / (1-\beta_1^t)}{\sqrt{v_t / (1-\beta_2^t)} + \epsilon} ~,
\end{align}
where $m_t$ and $v_t$ are estimations of the first and the second moments, $g_t$ is the real gradient at training step $t$, $g'_t$ is the estimated gradient at training step $t$, $\beta_1$ and $\beta_2$ are decay coefficients, and $\epsilon$ is a small constant for numerical stability.
Empirical values for $\beta_1$, $\beta_2$ and $\epsilon$ are 0.9, 0.999, $10^{-8}$, respectively.

\begin{figure}[tbp]
	\subfigure[SGD]{
		\centering
		\includegraphics[width=0.48\textwidth]{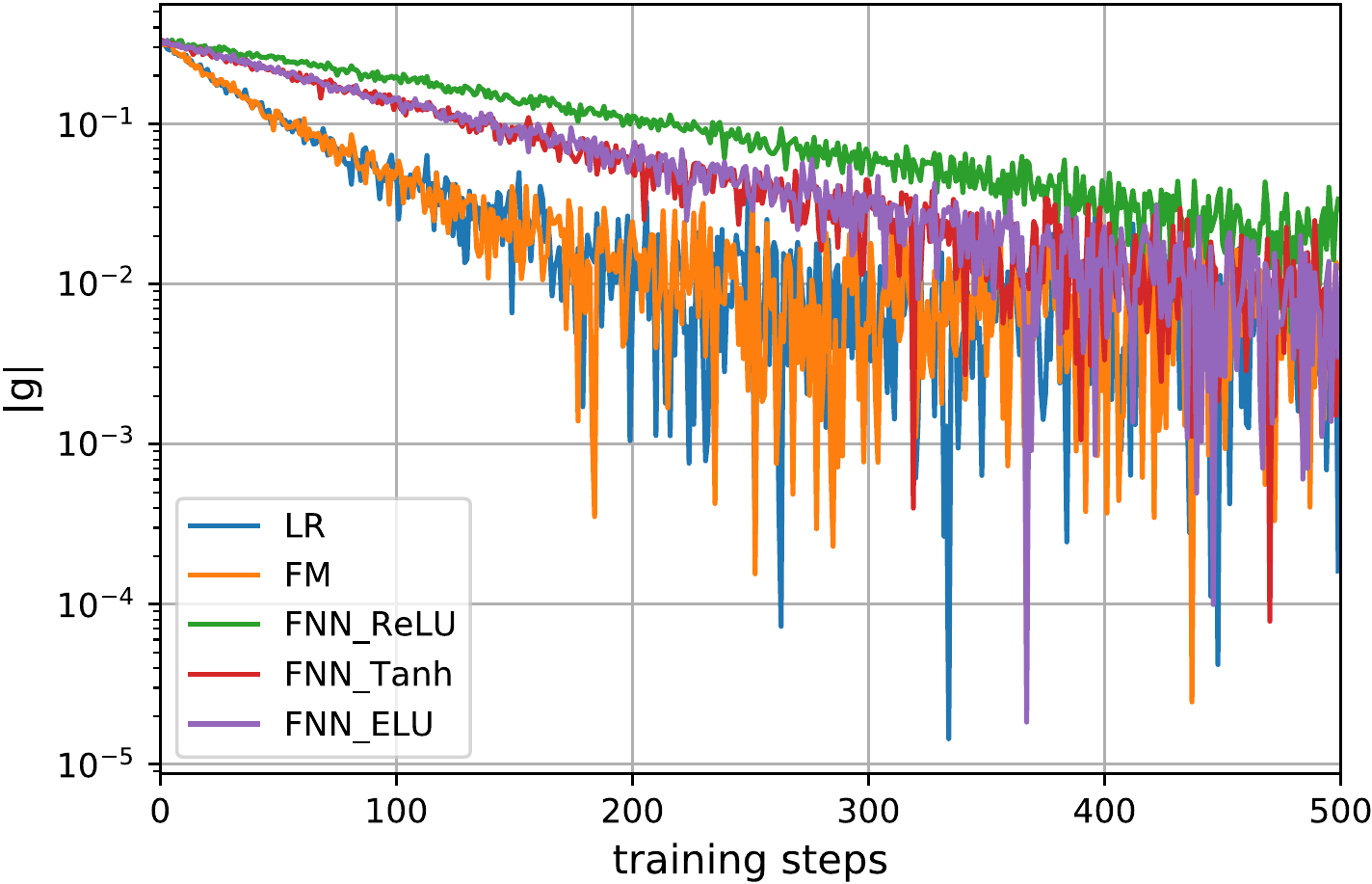}
	}
	\subfigure[Adam]{
		\centering
		\includegraphics[width=0.48\textwidth]{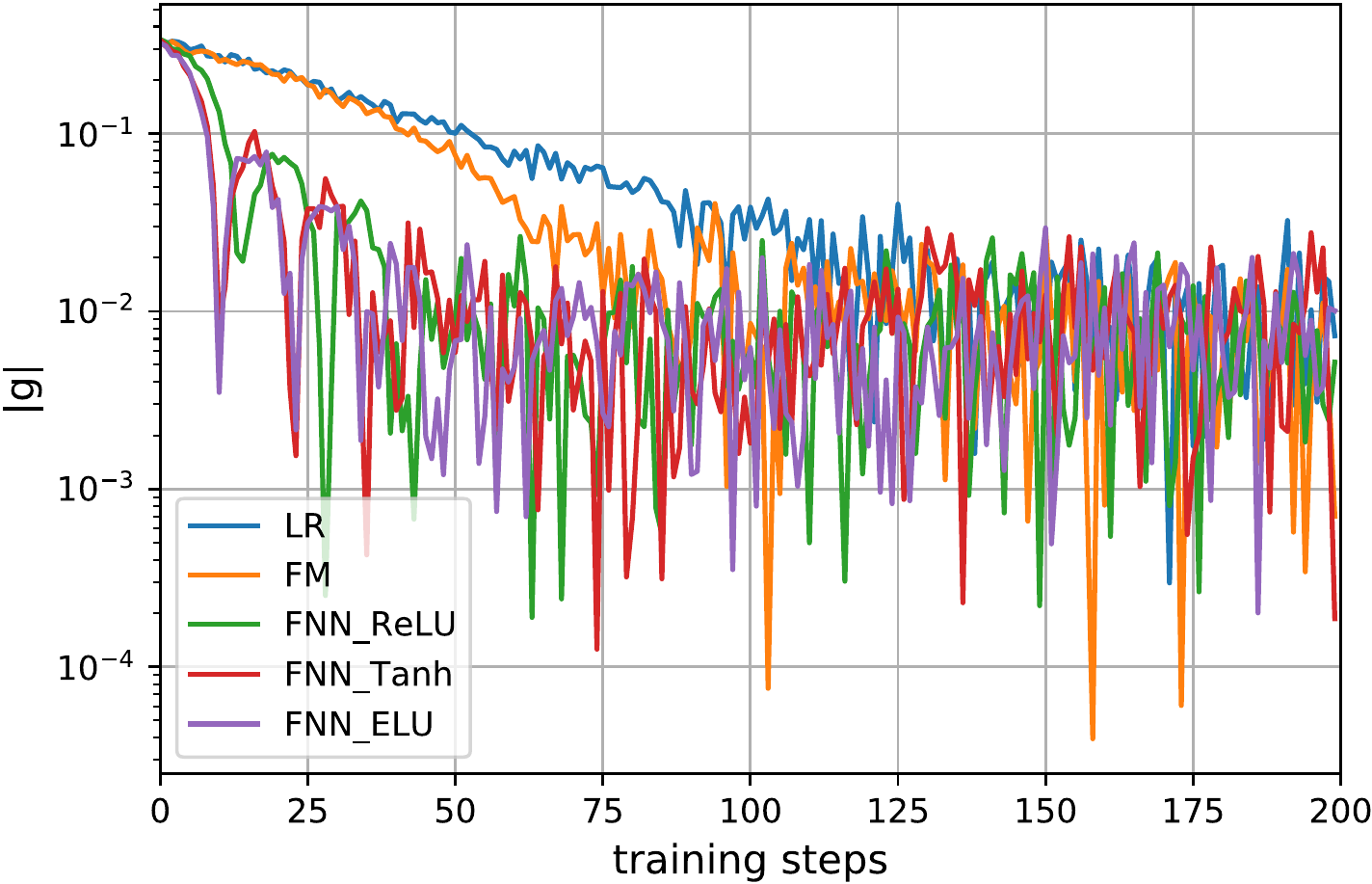}
	}
	\caption{The gradient magnitude of logit decays exponentially at the early stage of training. \emph{Note:} FNN\_ReLU, FNN\_Tanh, FNN\_ELU are FNNs with different activation functions. The x-axis means the number of mini-batches fed in a model, and the mini-batch size is 2000. The y-axis means the absolute gradient of the logit.}
	\label{fig:log_grad}
\end{figure}

Before our discussion, we should notice that the gradient of the logit $\hat{y}$
\begin{align}
\nabla_{\hat{y}}\bL & = \sigma(\hat{y}) - y
\end{align}
is bounded in $(-1, 1)$.  
Fig.~\ref{fig:log_grad} shows $\nabla_{\hat{y}}\bL$ of typical models with SGD/Adam. 
From this figure, we find the logit gradient decays exponentially at the early stage of training.
Because of chain rule and backpropagation, the gradients of any parameters depend on $\nabla_{\hat{y}} \bL$.
\revise{This indicates the gradients of any parameters decrease dramatically at the early stage of training.}

\revise{The following discussion uses Adam to analyze the behaviors of adaptive optimization algorithms on (unbalanced) sparse dataset, and the parameter sensitivity of Adam is studied in Section~\ref{sec:exp-prac}.
Considering an input $\bv_\bfl$ appears for first time in training examples} at time $t$, without loss of generality, we assume $g_t(\bv_\bfl) > 0$.

\subsubsection{Sensitive Gradient}\label{sec:issue-opt-sens}
\begin{figure}[tbp]
	\subfigure[$t=1$]{
		\includegraphics[width=0.31\textwidth]{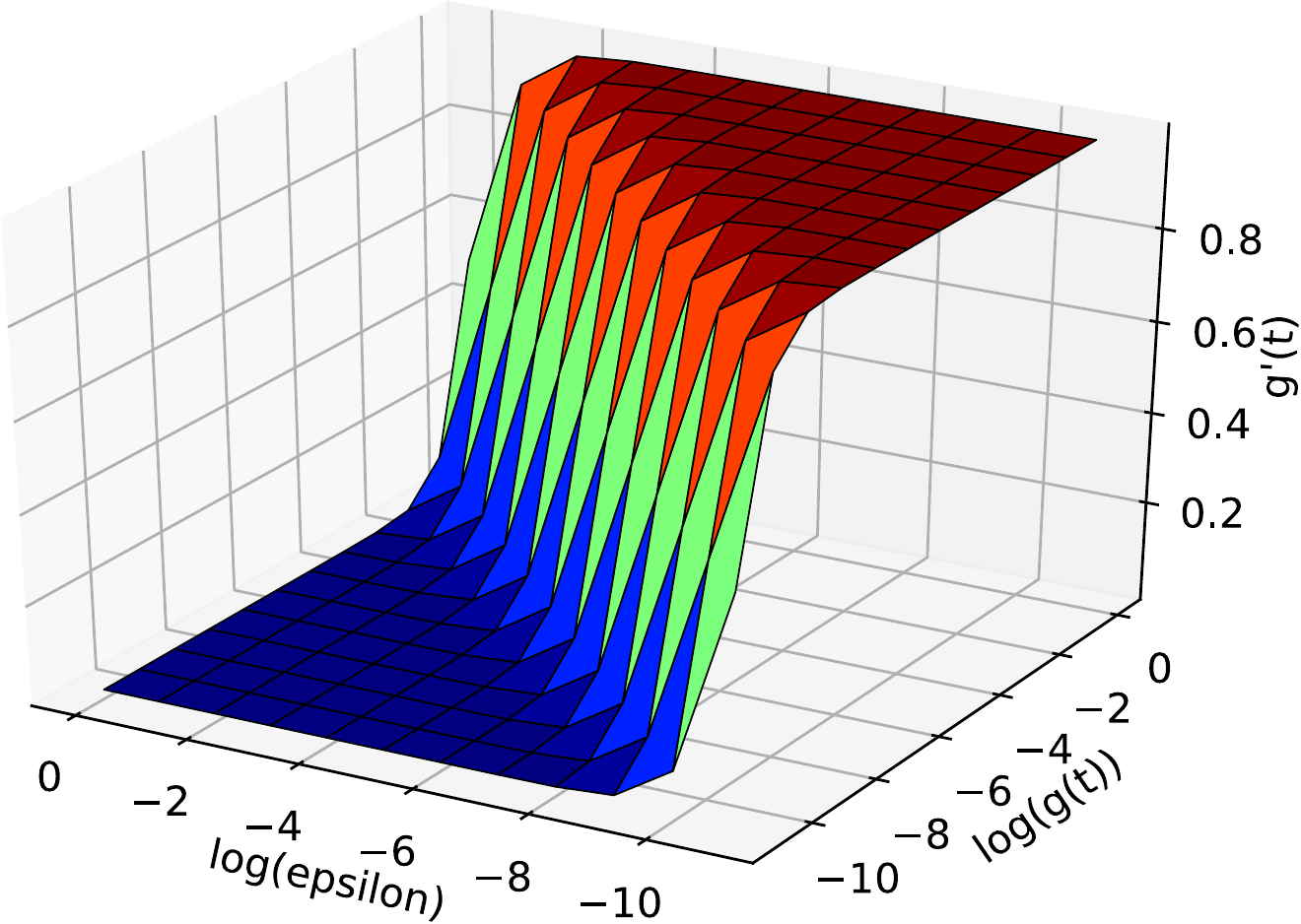}
	}
	\subfigure[$t=100$]{
		\includegraphics[width=0.31\textwidth]{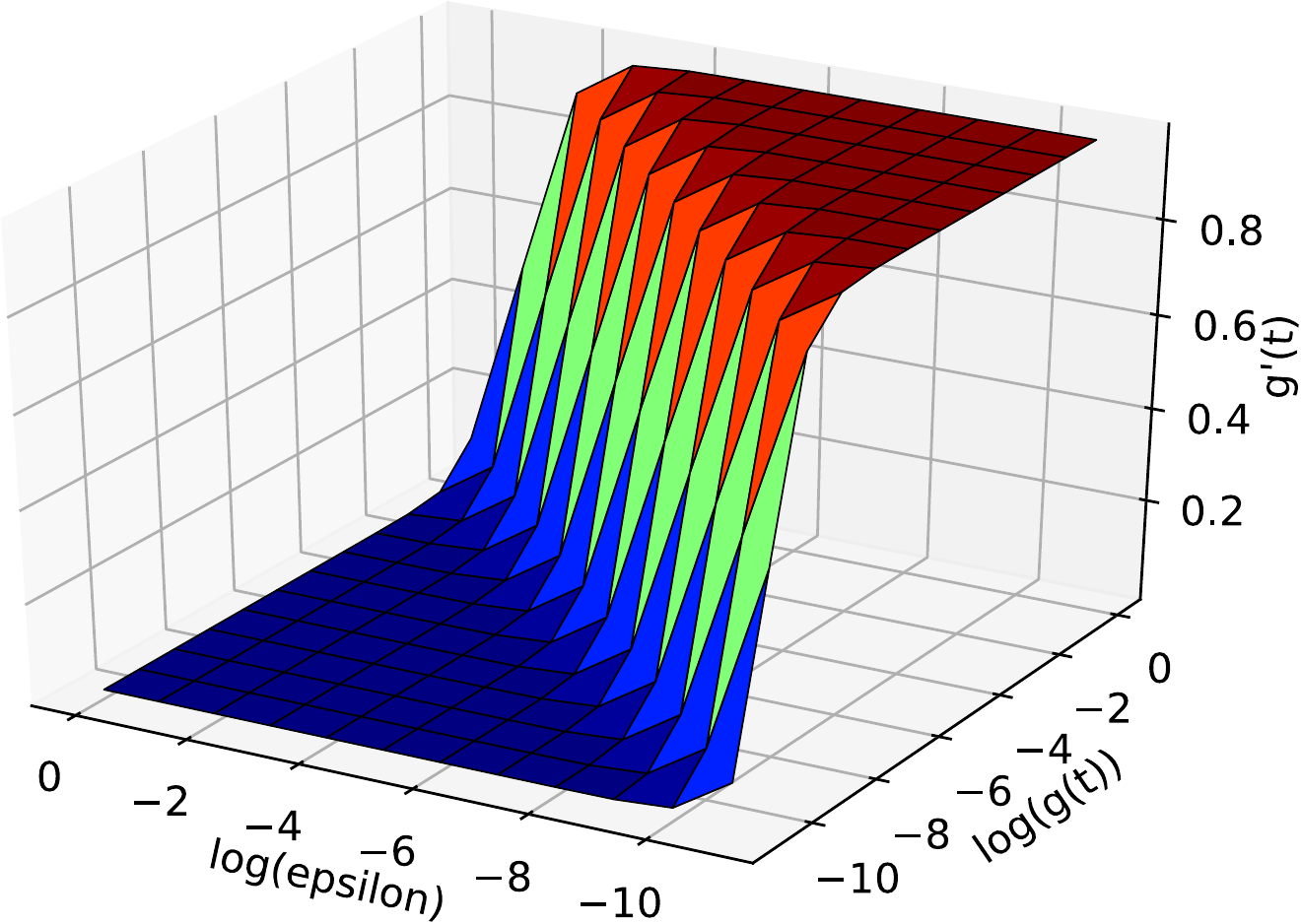}
	}
	\subfigure[$t=10000$]{
		\includegraphics[width=0.31\textwidth]{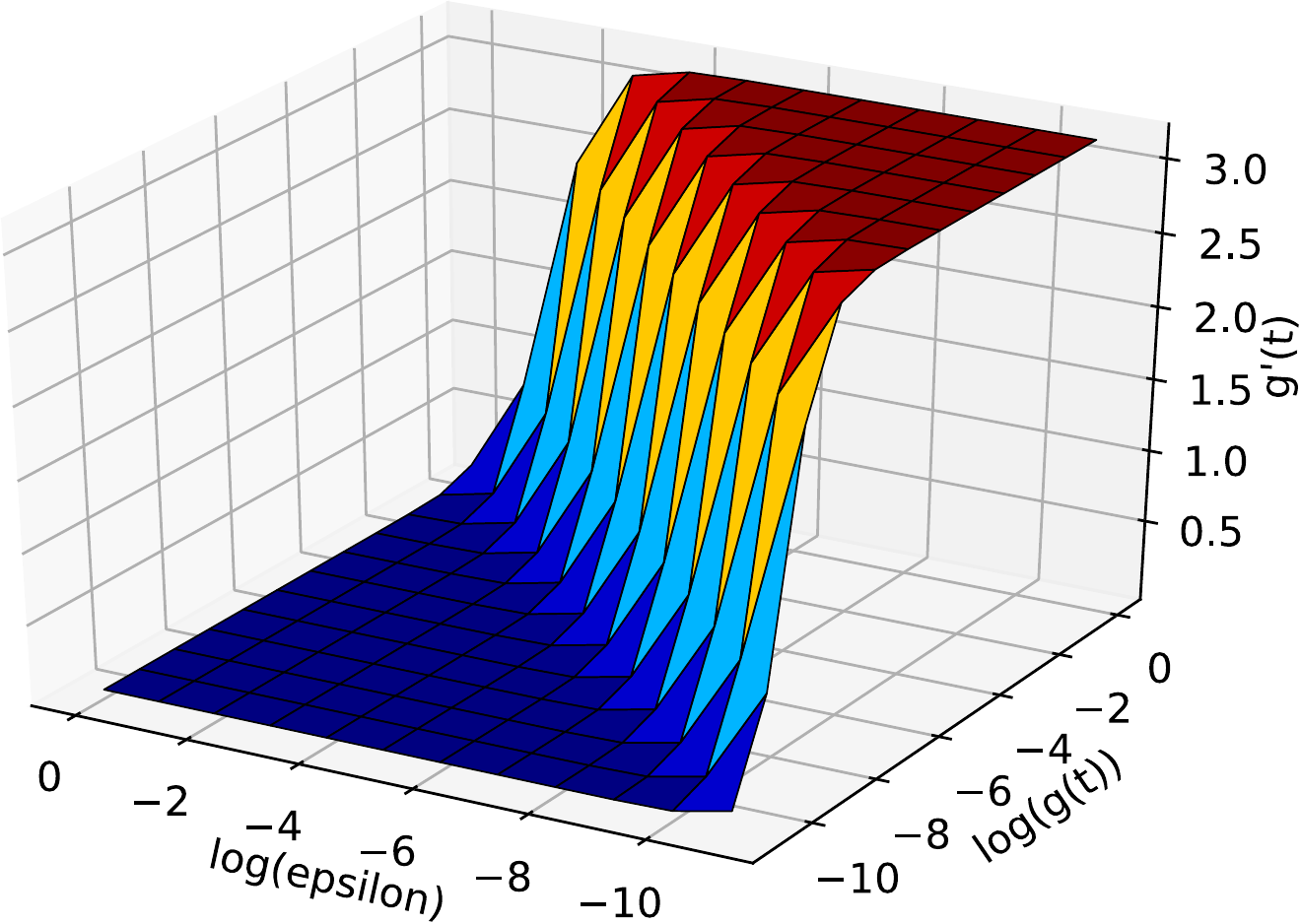}
	}
	\subfigure[$10^{-4} < \epsilon \le 1$]{
		\includegraphics[width=0.31\textwidth]{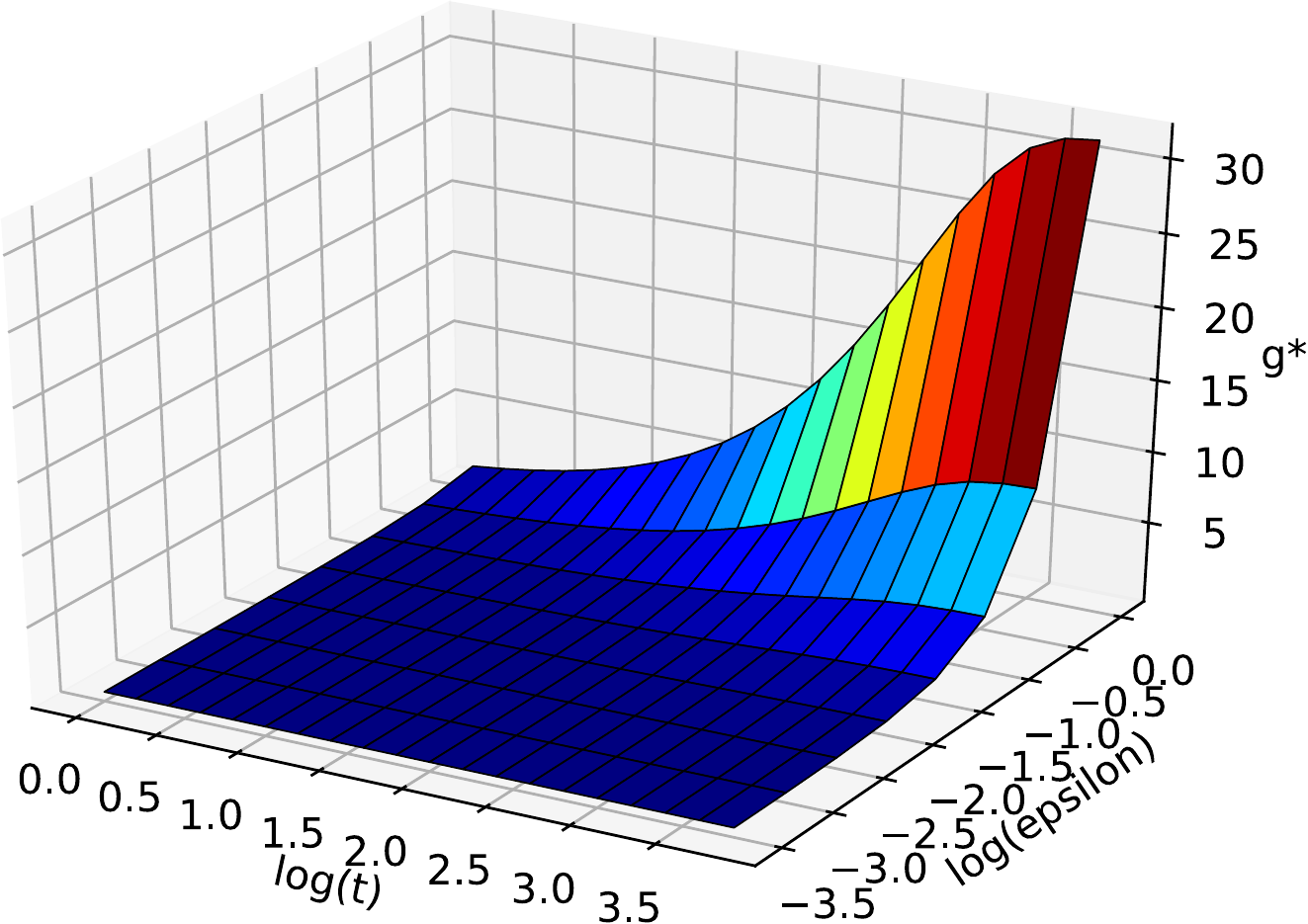}
	}
	\subfigure[$10^{-8} < \epsilon \le 10^{-4}$]{
		\includegraphics[width=0.31\textwidth]{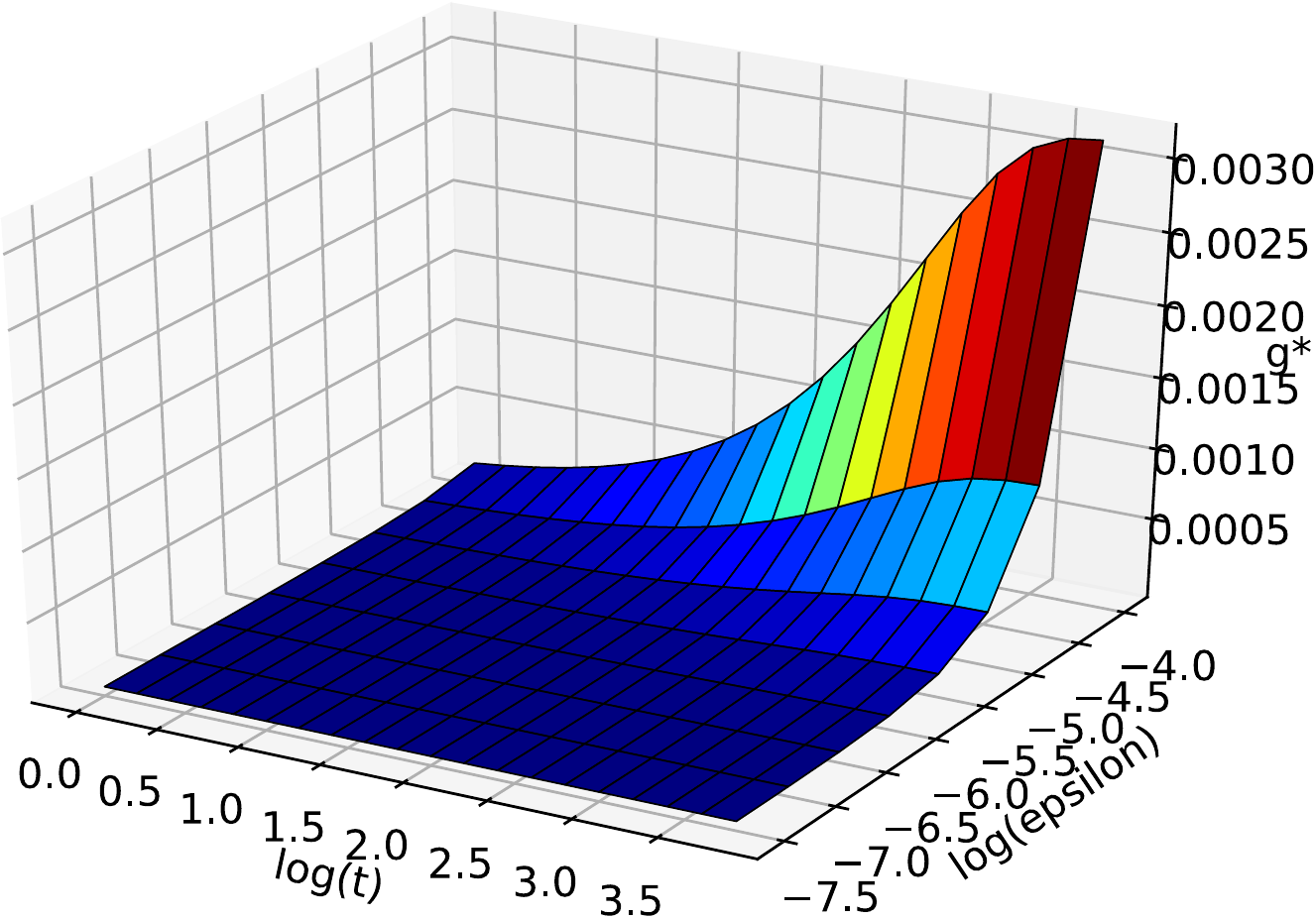}
	}
	\subfigure[$10^{-12} < \epsilon \le 10^{-8}$]{
		\includegraphics[width=0.31\textwidth]{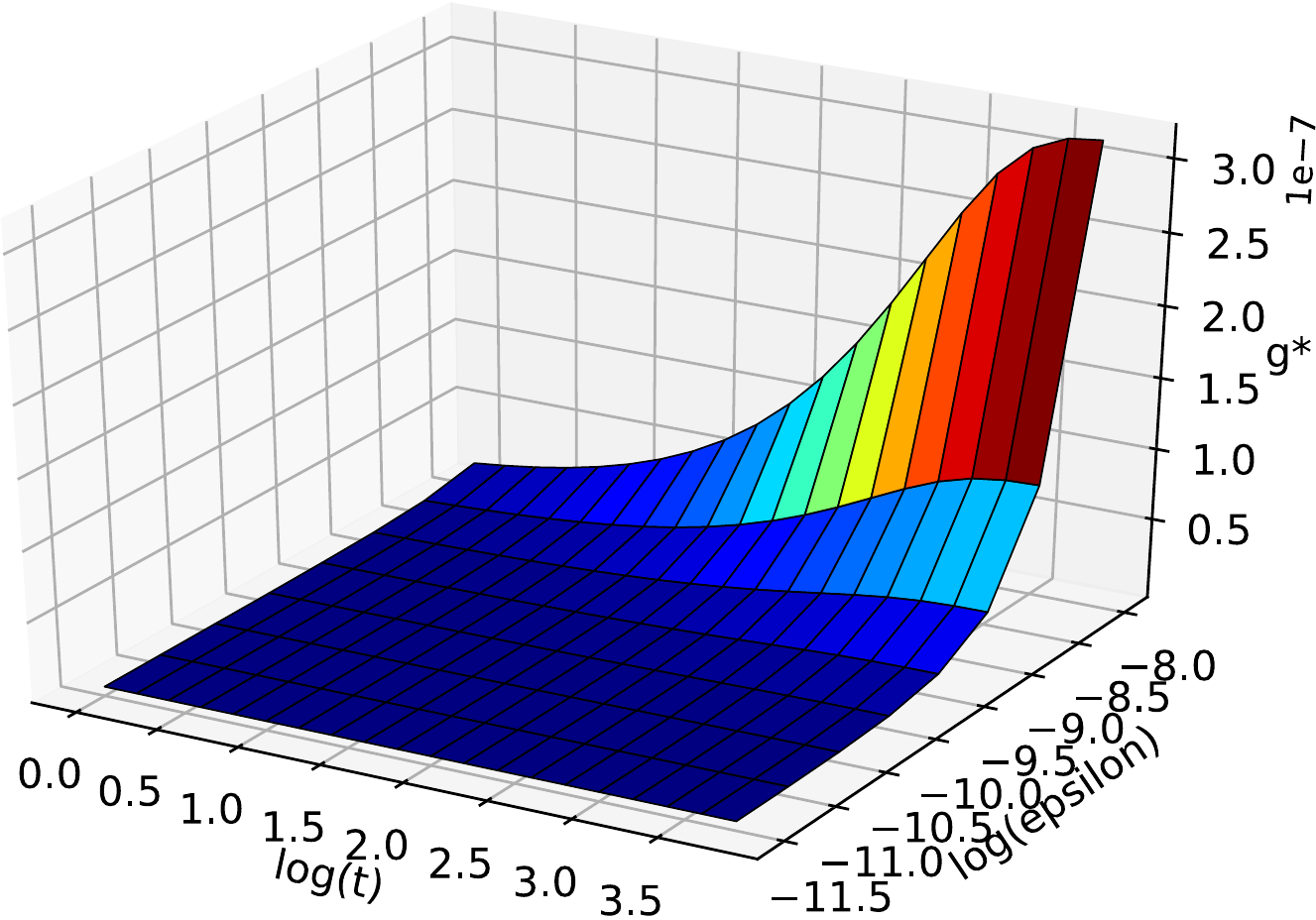}
	}
	\caption{Adam gradient estimation with respect to $\epsilon$ and training steps $t$. \emph{Note:} In (a)-(c), the x- and y-axis represent logarithm of real gradient $g_t$ and constant $\epsilon$, the z-axis represents the estimated gradient of Adam $g'_t$. In (d)-(f), the x- and y-axis represent logarithm of constant $\epsilon$ and training step $t$, the z-axis represents thresholds $g^*$ of real gradient $g_t$.}
	\label{fig:adamgrad}
\end{figure}

Firstly, we discuss the instant behavior of $g_t'$ at time $t$, 
\begin{align}
g_t'(\bv_\bfl) & = \frac{g_t (1-\beta_1)/(1-\beta_1^t)}{g_t \sqrt{(1-\beta_2) /(1-\beta_2^t)} + \epsilon} > 0 ~.
\end{align}
The estimated gradient $g_t'$ mainly depends on $g_t$, $\epsilon$, and $t$, as shown in Fig.~\ref{fig:adamgrad}(a)-(c).
At a certain training step $t$, the estimated gradient $g_t'$ changes dramatically when $g_t$ approaches some threshold.
In another word, $g_t'$ saturates across some of the value domain of $g_t$.
Denoting $x = \log(g_t)$, $a = \frac{1-\beta_1}{1-\beta_1^t}$, $b = \sqrt{\frac{1-\beta_2}{1-\beta_2^t}}$, we have $f(x) = g_t' = \frac{a e^x}{b e^x + \epsilon}$.
Then we can find the threshold $g_t^*$ where $\partial f(x) / \partial x$ is maximal
\begin{align}
\frac{\partial f(x)}{\partial x} & = \frac{a \epsilon}{b^2 e^x + \epsilon^2 e^{-x} + 2b\epsilon} \le \frac{a \epsilon}{4b\epsilon} \\
g^*_t & = \frac{\epsilon}{b} = \frac{\epsilon}{\sqrt{(1-\beta_2)/(1-\beta_2^t)}} ~. \label{eq:gstar}
\end{align}

Fig.~\ref{fig:adamgrad}(d)-(f) show the thresholds $g^*$ within $10^4$ training steps.
From these figures, we find $g^*$ increases with $\epsilon$ and $t$.
As training goes on, more and more gradients will cross the threshold and vanish.
Thus, we conclude $\epsilon$ has a large impact on model convergence and training stability.
And this impact will be amplified if the dataset is unbalanced.
Suppose the positive ratio of a dataset is $\alpha = \text{\# positive}/\text{\# total}$, then
\begin{align}
\bE[\nabla_{\hat{y}}\bL] = \bE[\sigma(\hat{y}) - y] = \alpha (\bE[\sigma(\hat{y})] / \alpha - 1) ~.
\end{align}
Thus $\nabla_{\hat{y}}\bL$ is proportional to $\alpha$ when $\bE[\sigma(\hat{y})] / \alpha \rightarrow 1$.

\subsubsection{Long-tailed Gradient}\label{sec:issue-opt-lt}
Secondly, we discuss the long-tailed behavior of $g_t'$ in a time window $T$ after time $t$.
If a sparse input $\bv_\bfl$ appears only once at training step $t$, then $g_{\neq t}(\bv_\bfl) = 0$ and $g'_{t+T}(\bv_\bfl)$ decays in a time window $T$.
Assume $g'_t(\bv_\bfl) > 0$,
\begin{align}
g_{t+T}'(\bv_\bfl) & = \frac{(1-\beta_1)\beta_1^T/(1-\beta_1^{t+T})}{\sqrt{(1-\beta_2) \beta_2^T/(1-\beta_2^{t+T})} + \epsilon / g_t} > 0 ~.
\end{align}

\begin{figure}[tbp]
	\subfigure[Gradient decay (t=1)]{
		\includegraphics[width=.31\textwidth]{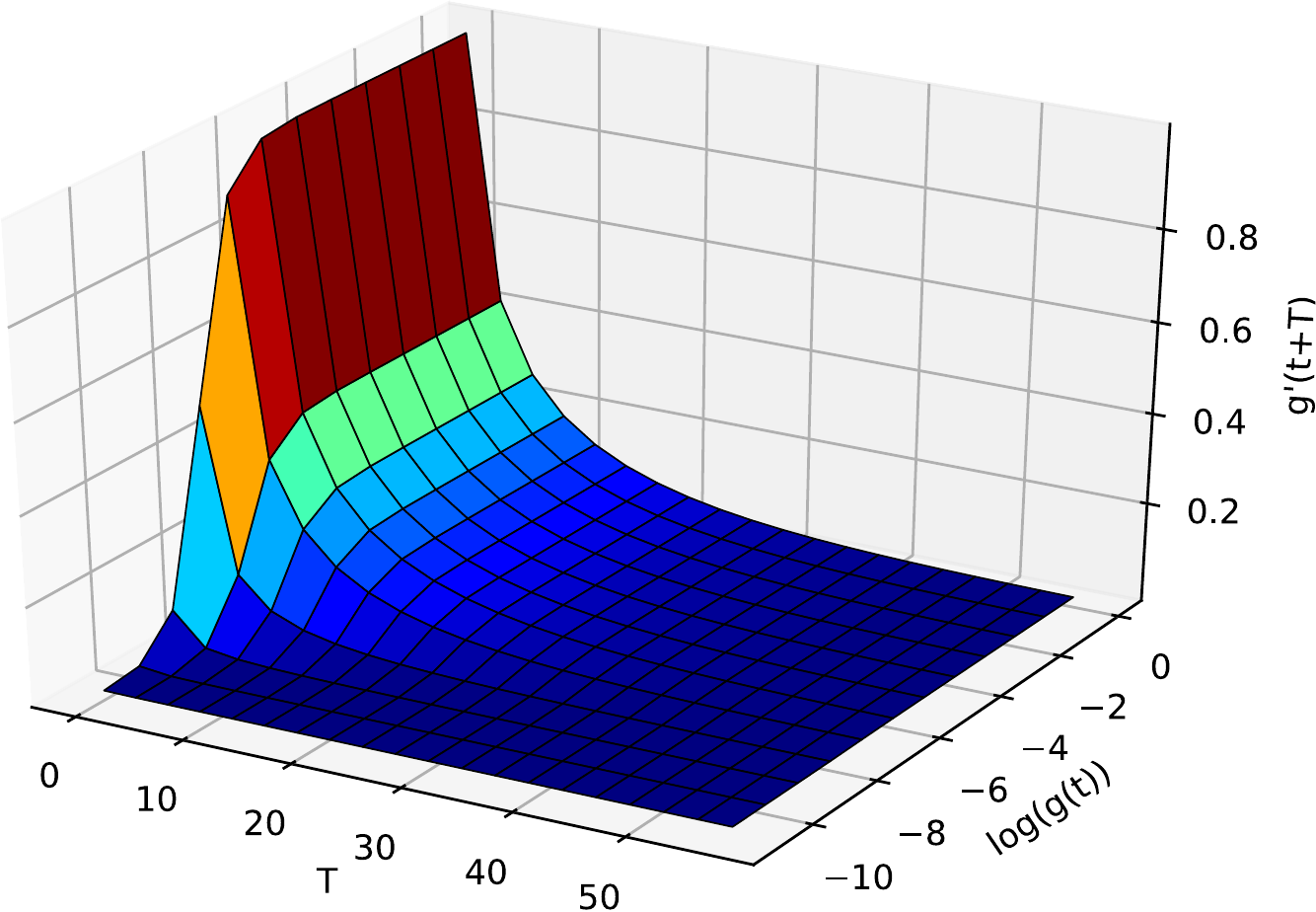}
	}
	\subfigure[Gradient decay (t=100)]{
		\includegraphics[width=.31\textwidth]{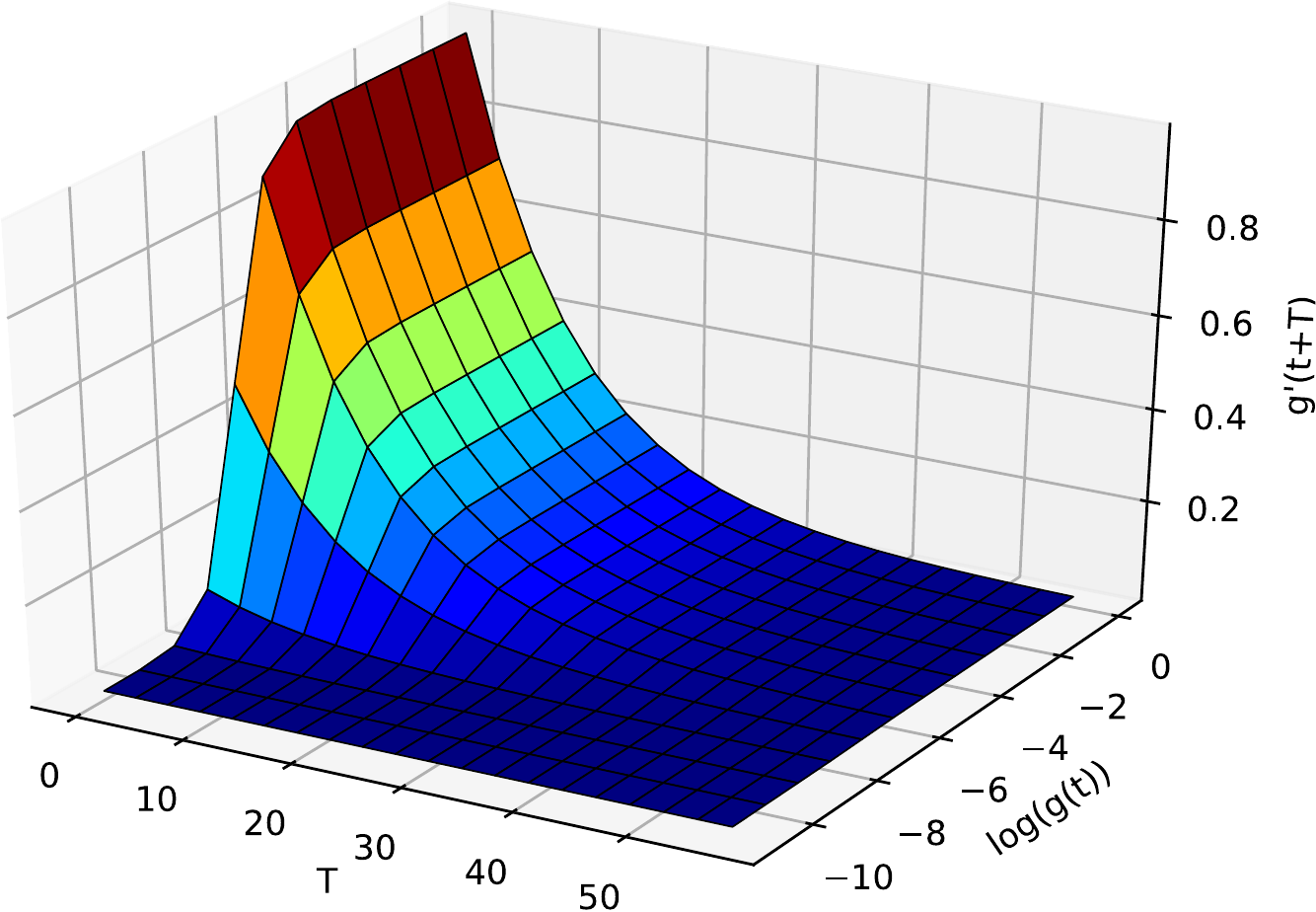}
	}
	\subfigure[Gradient decay (t=10000)]{
		\includegraphics[width=.31\textwidth]{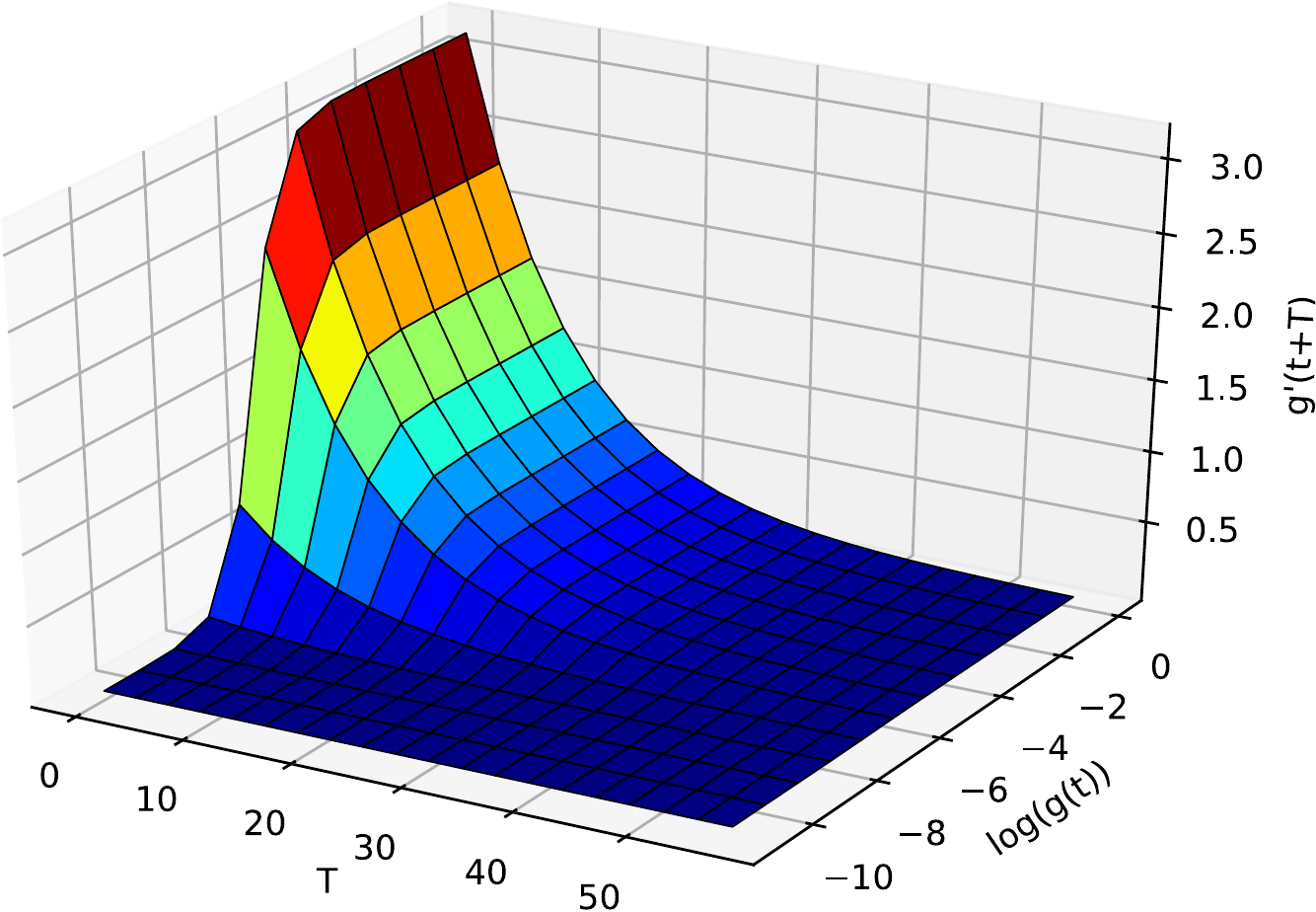}
	}
	\subfigure[Cumulative gradient (t=1)]{
		\includegraphics[width=.31\textwidth]{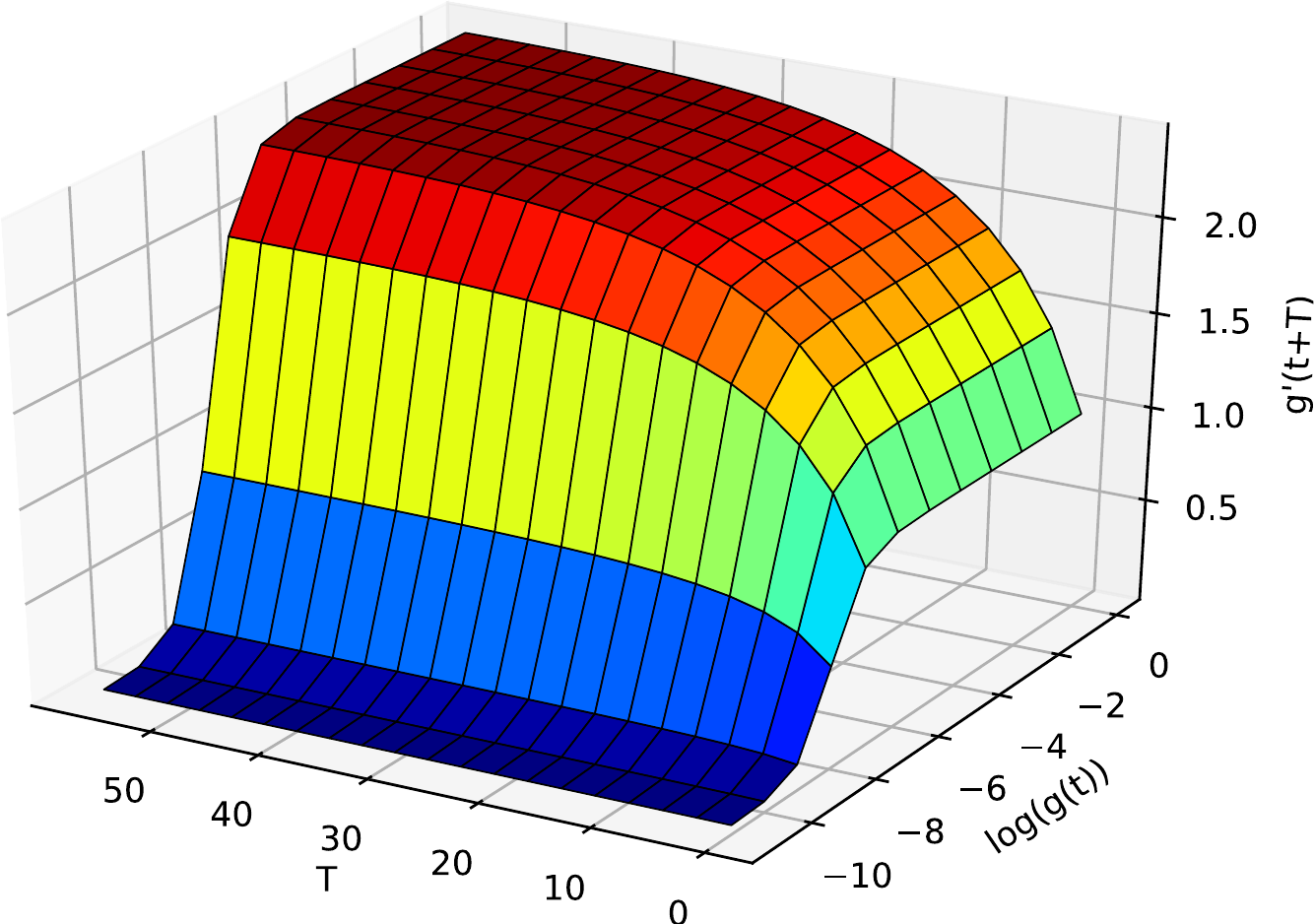}
	}
	\subfigure[Cumulative gradient (t=100)]{
		\includegraphics[width=.31\textwidth]{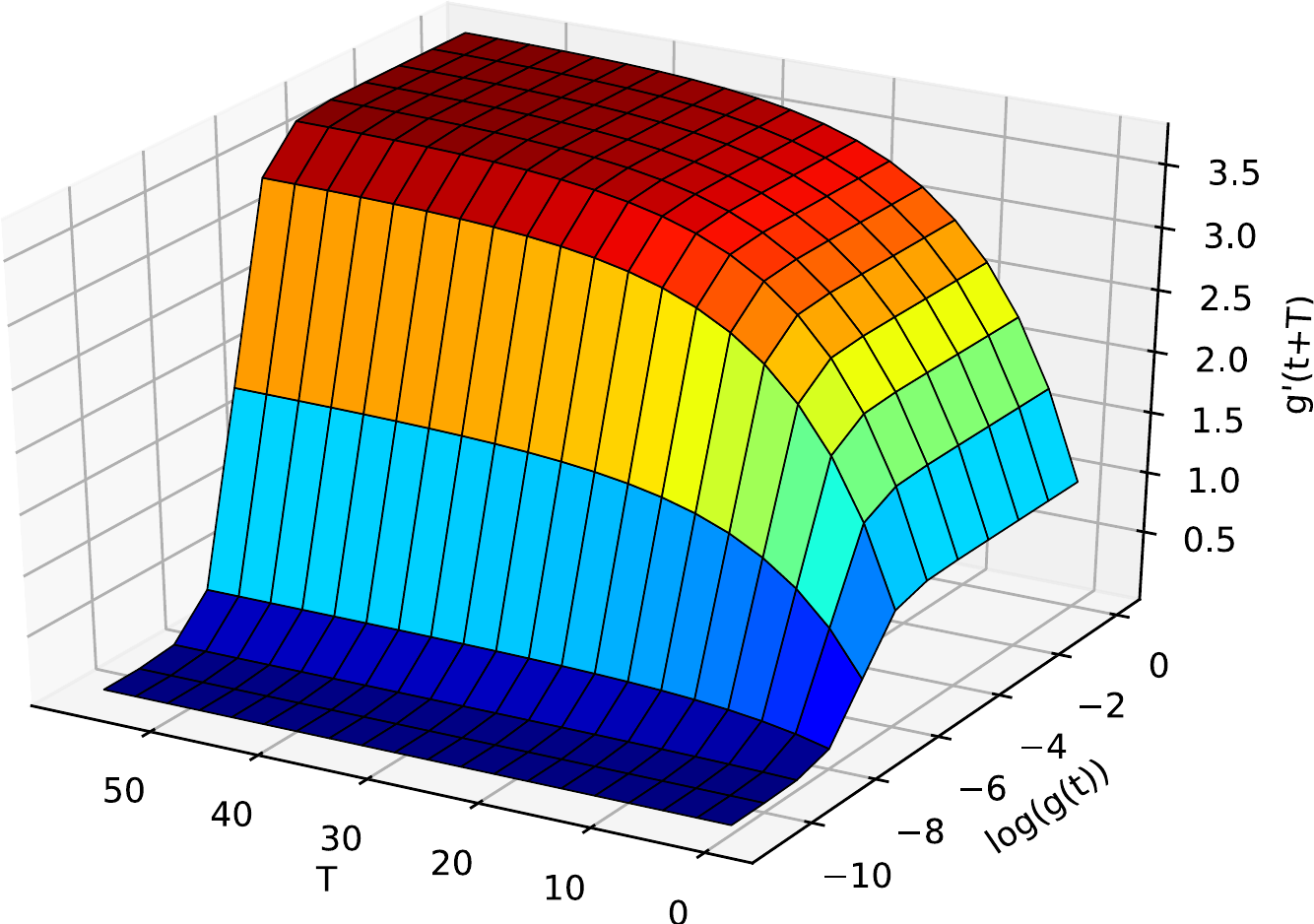}
	}
	\subfigure[Cumulative gradient (t=10000)]{
		\includegraphics[width=.31\textwidth]{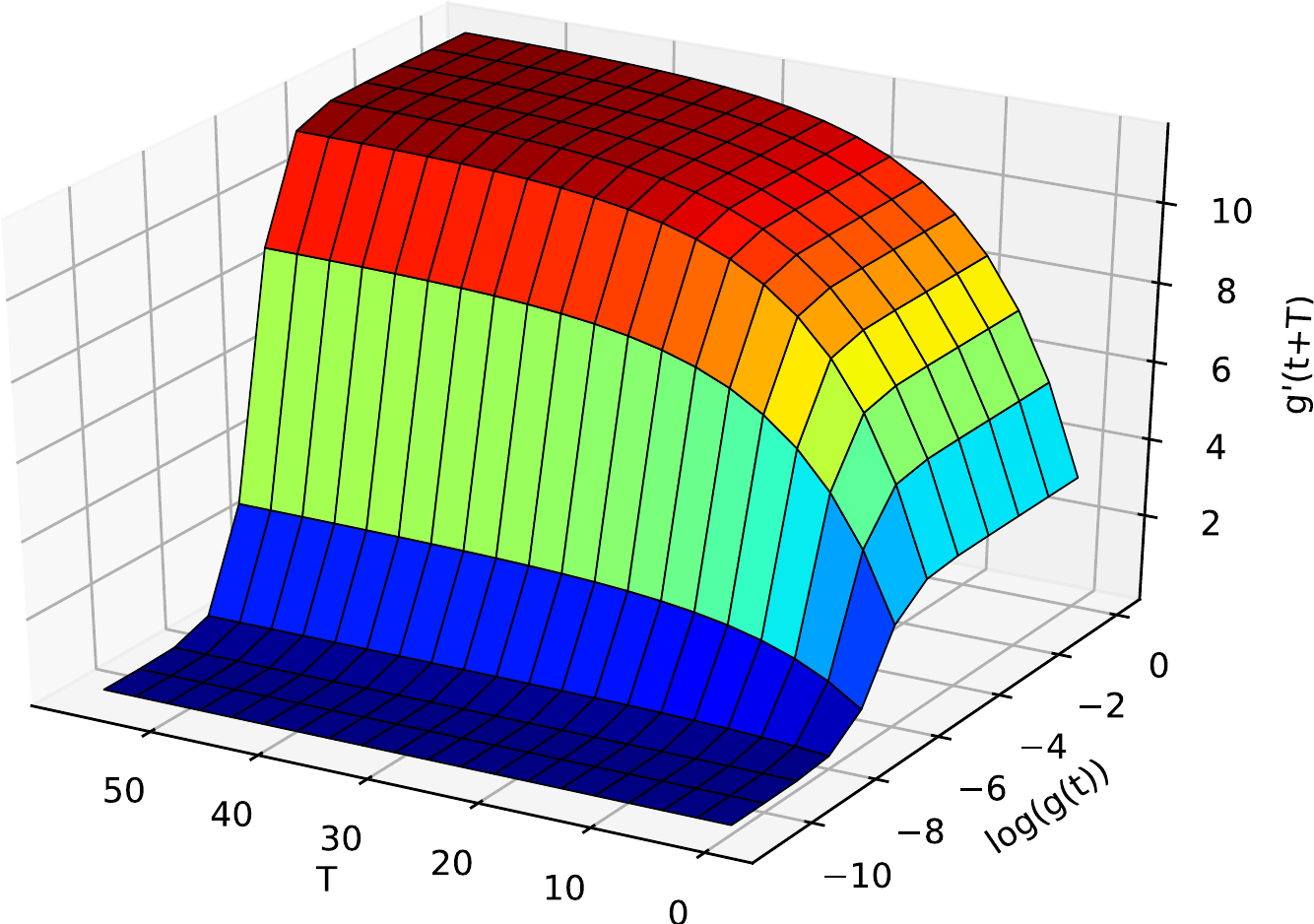}
	}
	\caption{Long-tailed effect of Adam gradient on sparse input. \emph{Note:} In (a)-(c), the x- and y-axis represent logarithm of real gradient $g_t$ and time window $T$, the z-axis represents the estimated gradient of Adam $g'_{t+T}$ within a time window. In (d)-(f), the x- and y-axis are the same as (a)-(c), the z-axis represents the estimated gradient of Adam $g'_{t+T}$ cumulated within time window $T$.}
	\label{fig:decay}
\end{figure}

Fig.~\ref{fig:decay} illustrates gradient decay and cumulative gradient in a window $T \le 60$ at $t=1, 100, 10000$, respectively.
From \ref{fig:decay}(a)-(c) we can see, the gradient larger than a threshold (different from $g^*$) is scaled up to a ``constant'', and the gradient smaller than that threshold shrinks to a ``constant''.
If $g'_{t+T}$ is continuously updated to $\bv_\bfl$, the cumulative effect is shown in \ref{fig:decay}(d)-(f).
The long-tailed effect may result in training instability or parameter divergence on sparse input.
A solution is sparse update, i.e., the estimated moments $m_t$ and $v_t$ are updated normally, but the estimated gradient $g'_t$ is only applied to parameters involved in the forward propagation.


\subsection{Regularization}\label{sec:issue-reg}

\subsubsection{L2 Regularization}\label{sec:issue-reg-l2}
\revise{L2 regularization is usually used to control overfitting of latent vectors, yet it may cause severe computation overhead when the dataset is extremely sparse.}
Denoting $\bV \in \bR^{N \times k}$ as the embedding matrix, L2 regularization adds a term $\frac{1}{2} \Vert \bV \Vert_2^2$ to the loss function.
This term results in an extra gradient term $\nabla_\bV \frac{1}{2} \Vert \bV \Vert_2^2 = \bV$.
For sparse input, L2 regularization is very expensive because it will update all the parameters, usually $10^6 \sim 10^9$ of them, in $\bV$.

An alternative to L2 regularization is sparse L2 regularization \cite{koren2009matrix}, i.e., we only penalize the parameters involved in the forward propagation rather than all of them.
A simple implementation is to penalize $\bv = \embed(\bx)$ instead of $\bV$. Since $\bx$ is a binary input, $\bv$ indicates the parameters involved in the forward propagation.


\subsubsection{Dropout}\label{sec:issue-reg-drop}
\begin{figure}[tbp]
	\subfigure[]{
		\centering
		\includegraphics[width=0.48\textwidth]{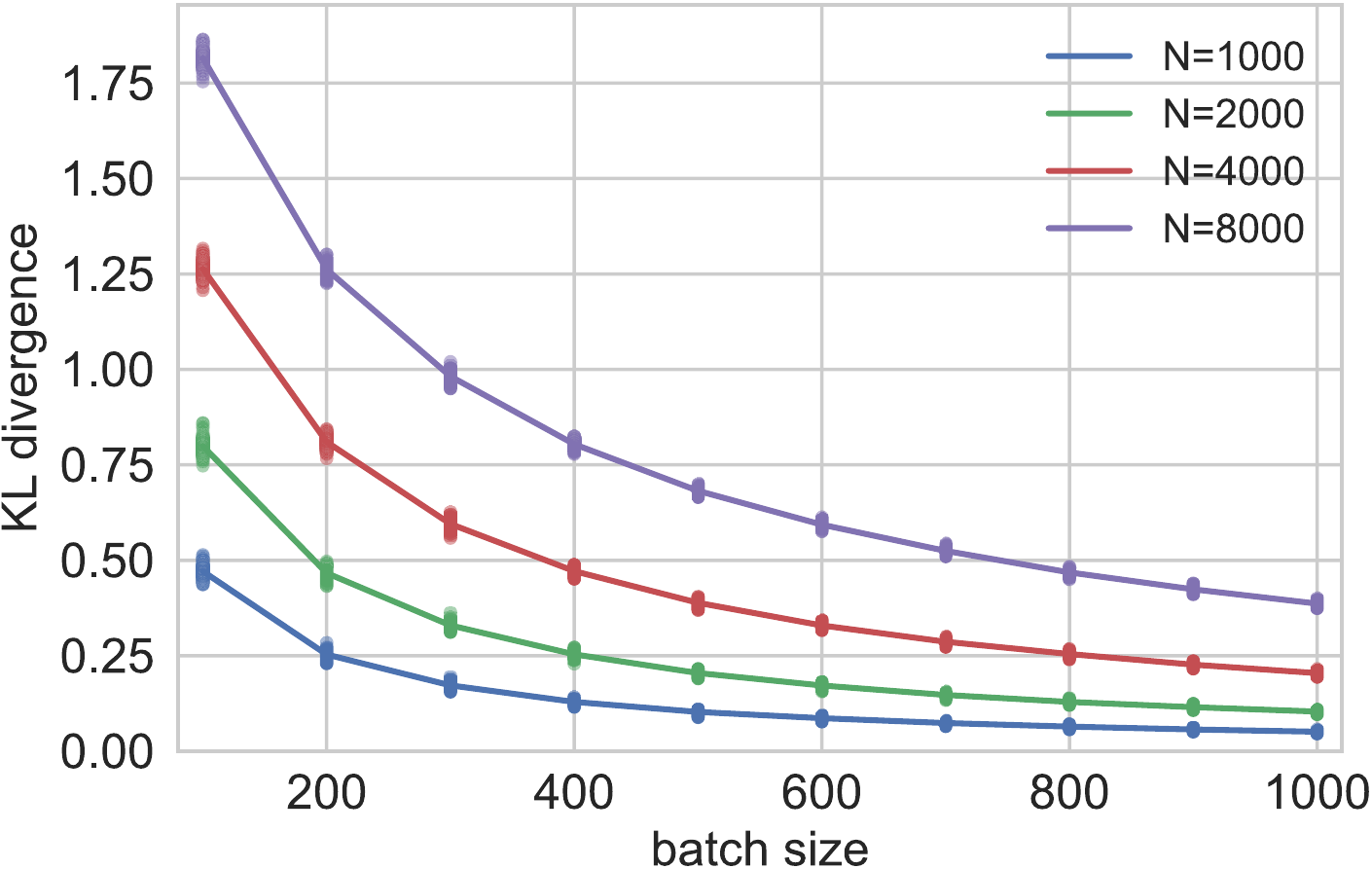}
	}
	\subfigure[]{
		\centering
		\includegraphics[width=0.48\textwidth]{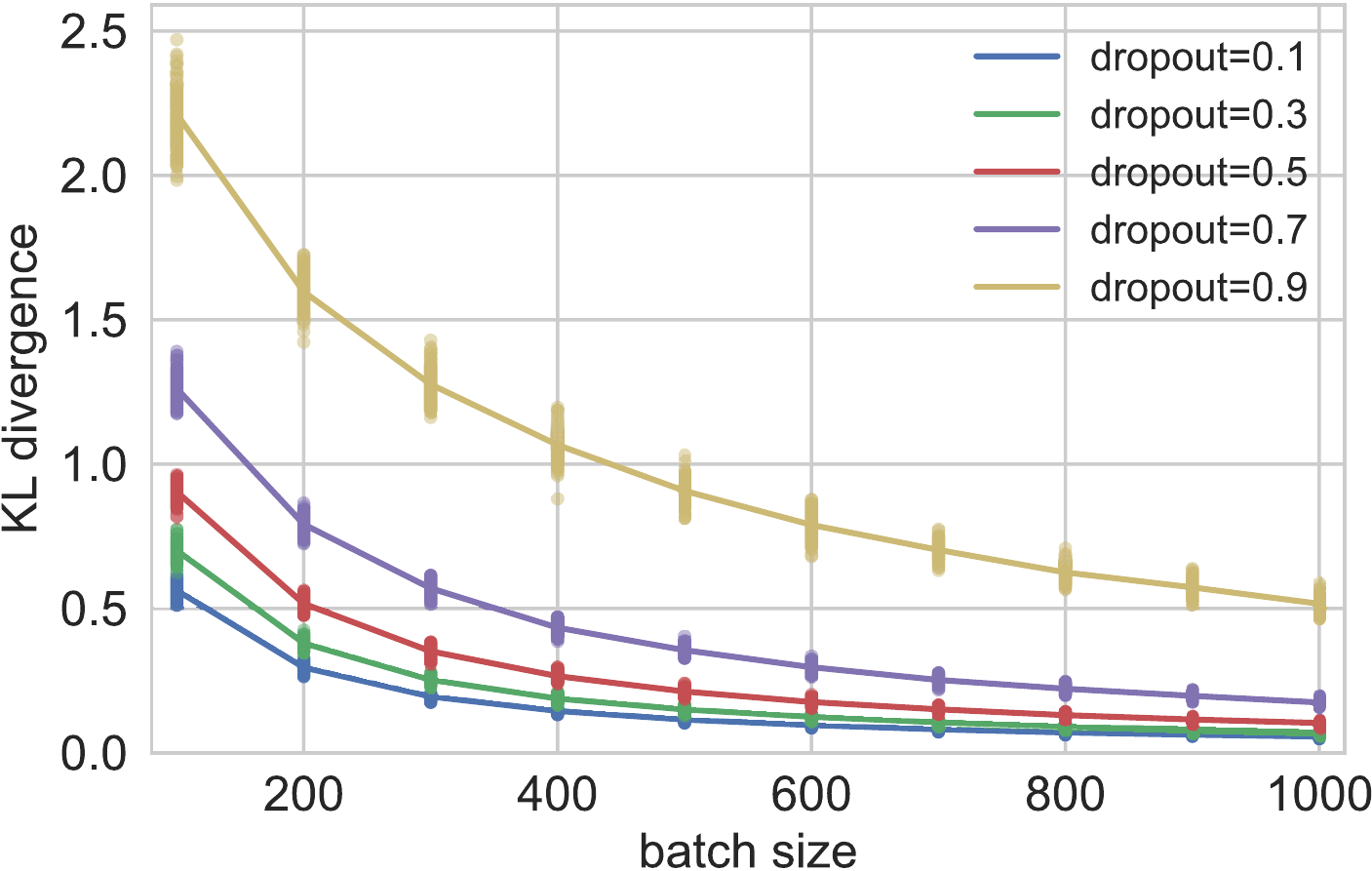}
	}
	\caption{A small mini-batch has a large bias in sparse data, and dropout amplifies this bias. \emph{Note:} The x-axis means the mini-batch size, the y-axis means the KL-divergence of the estimated distribution from a mini-batch with respect to the ground truth distribution. In (a), $N$ denotes the input dimension, and the data is sparser when $N$ is larger. In (b), $dropout$ denotes the probability of an input being dropped, and the input dimension is fixed to 1000.}
	\label{fig:bias}
\end{figure}

Dropout \cite{srivastava2014dropout} is a technique developed for DNN, which introduces noise in training and improves robustness.
However, in sparse data, a small mini-batch tends to have a large bias, and dropout may amplify this bias.
We conduct a simple experiment to show this problem, and the results are shown in Fig.~\ref{fig:bias}.

We generate a categorical dataset from a distribution $Q(X)$, $X \in \{x_1, x_2, \dots, x_N\}$, where every sample has 10 values without replacement.
For a mini-batch size $bs$, we draw $bs$ samples as a batch and use this batch to estimate the distribution, $P(X=x_i) = freq(x_i)$.
The evaluation metric is KL divergence, and we use a constant $10^{-8}$ for numerical stability.
For each $N \in \{1000, 2000, 4000, 8000\}$, and $bs \in \{100, 200, \dots, 1000\}$, we sample 100 batches and the results are shown in Fig.~\ref{fig:bias}(a), where the mean values are calculated and plotted in lines.
From this figure, we conclude: (i) a mini-batch tends to have a large bias on sparse data; (ii) this bias increases with data sparsity and decreases with batch size.

Then we test dropout on $N = 1000$.
Denoting dropout rate as $drop$, we randomly generate a mask of length $N$ with $drop$ elements being 0 while others being $1/drop$ to simulate the dropout process.
Every sample is weighted by a mask before estimating the distribution.
The results are shown in Fig.~\ref{fig:bias}(b).
This simple experiment illustrates the noise sensitivity of a mini-batch in sparse data.
\change{Thus, we turn to normalization techniques when a dataset is extremely sparse.}

\subsubsection{DNN Normalization}\label{sec:issue-reg-norm}
Normalization is carefully studied recently to solve internal covariate shift \cite{ioffe2015batch} in DNNs, and this method stabilizes activation distributions.
Typical methods include batch normalization (BN) \cite{ioffe2015batch}, weight normalization (WN) \cite{salimans2016weight}, layer normalization (LN) \cite{ba2016layer}, self-normalizing network (SNN) \cite{klambauer2017self}, etc.
In general, BN, WN, and LN use some statistics to normalize hidden layer activations, while SNN uses an activation function SELU, making activations naturally converge to the standard normal distribution.

Denote $\bx$ as the input to a hidden layer with weight matrix $\bw$, $\bx_i$ as the $i$-th instance of a mini-batch, and $x_i^j$ as the value of the $j$-th dimension of $\bx_i$. As the following, we discuss BN, WN, LN, and SELU in detail.

\paragraph{Batch Normalization}
BN normalizes activations $\bw^\top\bx$ using statistics within a mini-batch
\begin{align}
\text{BN}(\bw^\top\bx) & = \frac{\bw^\top \bx - \text{avg}_{i}(\bw^\top \bx)}{\text{std}_{i}(\bw^\top \bx)} \mathbf{g} + \mathbf{b} ~,
\end{align}
where $\mathbf{g}$ and $\mathbf{b}$ scale and shift the normalized values.
These parameters are learned along with other parameters and restore the representation power of the network.
Similar parameters are also used in WN and LN.

BN solves internal covariate shift \cite{ioffe2015batch} and accelerates DNN training.
However, BN may fail when the input is sparse, because BN relies on the statistics of a mini-batch. 
As shown in Fig.~\ref{fig:bias}(a), a mini-batch tends to have a large bias when input data is sparse, but large mini-batch is not always practical due to the computation resource limit such as GPU video memory.

\paragraph{Weight Normalization}
WN re-parametrizes the weight matrix $\bw$, and learns the direction and scale of $\bw$ separately
\begin{align}
\text{WN}(\bw^\top\bx) & = \Big(\frac{\bw}{\Vert \bw \Vert} \mathbf{g} \Big)^\top \bx ~.
\end{align}
WN does not depend on mini-batch, thus can be applied to noise-sensitive models \cite{salimans2016weight}.
However, WN is roughly infeasible on high-dimensional data, because WN depends on the L2 norm of parameters, which results in even higher complexity than L2 regularization.
Thus, WN meets similar complexity problem as L2 regularization when input space is extremely sparse.

\paragraph{Layer Normalization}
LN normalizes activations $\bw^\top \bx$ using statistics of different neurons within the same layer
\begin{align}
\text{LN}(\bw^\top \bx) & = \frac{\bw^\top\bx - \text{avg}_{j}(\bw^\top \bx)}{\text{std}_{j}(\bw^\top \bx)} \mathbf{g} + \bb ~.
\end{align}
LN stabilizes the hidden state dynamics in recurrent networks \cite{ba2016layer}.
\revise{In our experiments, we apply LN on fully connected layers and inner/kernel product layers, and we apply fused LN in micro networks.}
Since LN does not work well in CNN \cite{ba2016layer}, we exclude LN in CCPM.

\paragraph{Self-normalizing Network}
SNN uses SELU as the activation function, 
\begin{align}
\text{SELU}(x) & = \lambda \begin{cases}
x & \text{if } x > 0 \\
\alpha e^{x} - \alpha & \text{if } x \le 0
\end{cases} ~.
\end{align}
Based on Banach fixed-point theorem, the activations that are propagated through many SELU layers will converge to zero mean and unit variance.
Besides, SELU declares significant improvement in feed-forward neural networks on a large variety of tasks \cite{klambauer2017self}.

To summarize, we use (sparse) L2 regularization to penalize embedding vectors, and we use dropout, LN, and SELU to regularize DNNs.
BN is not applied because of the mini-batch problem discussed above, and WN is not applied because of its high complexity.
Corresponding experiments are in Section~\ref{sec:exp-prac}.

\section{Experiments}\label{sec:exp}

\revise{In Section~\ref{sec:exp-eval}, we present overall comparison.
In Section~\ref{sec:exp-prac}, we discuss practical issues: complexity, initialization (Section~\ref{sec:issue-init}), optimization (Section~\ref{sec:issue-opt}), and regularization (Section~\ref{sec:issue-reg}).
In Section~\ref{sec:exp-fafi}, we propose a visualization method to analyze feature interactions, corresponding to Section~\ref{sec:method-fafi}.
And finally, in Section~\ref{sec:exp-diff}, we conduct a synthetic experiment to illustrate the deficiency of DNN, corresponding to Section~\ref{sec:method-diff}.}

\subsection{\revise{Offline and Online Evaluations}}\label{sec:exp-eval}

\revise{In this section, we conduct offline and online evaluations to give a thorough comparison:
(i) We compare KFM/NIFM with other latent vector-based models to verify the effectiveness of kernel product methods.
(ii) We compare PNNs with other DNN-based models to verify the effectiveness of product layers.
(iii) We also participate in the Criteo challenge and compete KFM with libFFM directly.
(iv) We deploy PIN in a real recommender system.}

\subsubsection{Datasets}\label{sec:exp-eval-data}


\paragraph{Criteo}
Criteo\footnote{http://labs.criteo.com/downloads/download-terabyte-click-logs/} contains one month of click logs with billions of data examples.
A small subset of Criteo was published in Criteo Display Advertising Challenge, 2013, and FFM was the winning solution \cite{juan2016field}.
We select ``day6-12'' for training, and ``day13'' for evaluation. 
Because of the enormous data volume and serious label unbalance (only 3\% samples are positive), we apply negative down-sampling to keep the positive ratio close to 50\%.
We convert the 13 numerical fields into categorical through bucketing (in Section~\ref{sec:issue-num}). 
And we set the categories appearing less than 20 times as a dummy category ``other''.

\paragraph{Avazu}
Avazu\footnote{http://www.kaggle.com/c/avazu-ctr-prediction} was published in Avazu Click-Through Rate Prediction contest, 2014, and FFM was the winning solution \cite{juan2016field}.
We randomly split the public dataset into training and test sets at 4:1, and remove categories appearing less than 20 times to reduce dimensionality.

\paragraph{iPinYou}
iPinYou\footnote{http://data.computational-advertising.org} was published in iPinYou RTB Bidding Algorithm Competition, 2013. 
We only use the click data from season 2 and 3 because of the same data schema. 
\revise{We follow the data processing of \cite{zhang2014optimal}, and we remove ``user tags'' to prevent leakage.}

\paragraph{Huawei}
Huawei \cite{guo2017deepfm} is collected from the game center of Huawei App Store in 2016, containing app, user, and context features.
We use the same training and test sets as \cite{guo2017deepfm}, and we use the same hyper-parameter settings to reproduce their results.

Table~\ref{tab:data} shows statistics of the 4 datasets\footnote{Datasets: https://github.com/Atomu2014/Ads-RecSys-Datasets and https://github.com/Atomu2014/make-ipinyou-data.}.

\begin{table}[tbp]
	\centering
	\caption{Dataset statistics.}
	\label{tab:data}
	\begin{tabular}{c|cccc}
		Dataset & \# instances & \# categories & \# fields & pos ratio \\ \hline
		Criteo & $1 \times 10^8$ & $1 \times 10^6$ & 39 & 0.5 \\
		Avazu & $4 \times 10^7$ & $6 \times 10^5$ & 24 & 0.17 \\
		iPinYou & $2 \times 10^7$ & $9 \times 10^5$ & 16 & 0.0007 \\
		Huawei & $5.7 \times 10^7$ & $9\times 10^4$ & 9 & 0.008 \\
	\end{tabular}
\end{table}

\subsubsection{Compared Models}\label{sec:exp-eval-base}
We compare 8 baseline models, including LR \cite{lee2012estimating}, GBDT \cite{chen2016xgboost}, FM \cite{rendle2010factorization}, FFM \cite{juan2016field}, FNN \cite{zhang2016deep}, CCPM \cite{liu2015convolutional}, AFM \cite{xiao2017attentional} and DeepFM \cite{guo2017deepfm}, all of which are discussed in Section~\ref{sec:related} and Section~\ref{sec:method}. 
We use XGBoost\footnote{https://xgboost.readthedocs.io/en/latest/} and libFFM\footnote{https://github.com/guestwalk/libffm} as GBDT and FFM in our experiments. 
We implement\footnote{Code: https://github.com/Atomu2014/product-nets-distributed} all the other models with Tensorflow\footnote{https://www.tensorflow.org/} and MXNet\footnote{https://mxnet.apache.org/}. 
We also implement FFM with Tensorflow and MXNet to compare its training speed with other models on GPU. 
In particular, our FFM implementation (Avazu log loss=0.37805) has almost the same performance as libFFM (Avazu log loss=0.37803).

\subsubsection{Evaluation Metrics}\label{sec:exp-eval-metric}
The evaluation metrics are \textbf{AUC}, and \textbf{log loss}. 
AUC is a widely used metric for binary classification because it is insensitive to the classification threshold and the positive ratio. 
If prediction scores of all the positive samples are higher than those of the negative, the model will achieve AUC=1 (separate positive/negative samples perfectly). 
The upper bound of AUC is 1, and the larger the better. 
Log loss is another widely used metric in binary classification, measuring the distance between two distributions. 
The lower bound of log loss is 0, indicating the two distributions perfectly match, and a smaller value indicates better performance.

\subsubsection{\revise{Parameter Setting}}\label{sec:exp-eval-set}

\begin{table}[tbp]
	\centering
	\caption{Parameter settings.}
	\label{tab:param}
	\begin{tabular}{c|c|c|c|c}
		Param & Criteo & Avazu & iPinYou & Huawei \\ \hline
		\tabincell{c}{General} 
		& \tabincell{c}{bs=2000, \\ opt=Adam, \\ lr=$10^{-3}$} 
		& \tabincell{c}{bs=2000, \\ opt=Adam, \\ lr=$10^{-3}$} 
		& \tabincell{c}{bs=2000, \\ opt=Adam, \\ lr=$10^{-3}$, \\ l2=$10^{-6}$} 
		& \tabincell{c}{bs=1000, \\ opt=Adam, \\ lr=$10^{-4}$, \\ l2=$10^{-4}$} \\ \hline
		LR & - & - & - & - \\ \hline
		GBDT 
		& \tabincell{c}{depth=25, \\ \# tree=1300} 
		& \tabincell{c}{depth=18, \\ \# tree=1000}
		& \tabincell{c}{depth=21, \\ \# tree=700}
		& \tabincell{c}{depth=6, \\ \# tree=600} \\ \hline
		FFM & k=4 & k=4 & k=4 & k=4 \\ \hline
		\tabincell{c}{FM, KFM}
		& \tabincell{c}{k=20, }
		& \tabincell{c}{k=40, }
		& \tabincell{c}{k=20, }
		& \tabincell{c}{k=10, }
		\\ AFM & \tabincell{c}{t=0.01, h=32, \\ l2\_a=0.1}& \tabincell{c}{t=1, h=256 \\ l2\_a=0}& \tabincell{c}{t=1, h=256, \\ l2\_a=0.1}& \tabincell{c}{t=0.0001, h=64, \\ l2\_a=0} \\ 
		NIFM & sub-net=[40,1] & sub-net=[80,1]& sub-net=[40,1]&sub-net=[20,1] \\ \hline
		CCPM 
		& \tabincell{c}{k=20, \\ kernel=7$\times$256, \\ net=[256$\times$3,1]}
		& \tabincell{c}{k=40, \\ kernel=7$\times$128, \\ net=[128$\times$3,1]}
		& \tabincell{c}{k=20, \\ kernel=7$\times$128, \\ net=[128$\times$3,1]}
		& \tabincell{c}{k=10, \\ kernel=5$\times$512, \\ net=[512$\times$3,1], \\ drop=0.1 } \\ \hline
		\tabincell{c}{FNN, DeepFM, \\ IPNN, KPNN}
		& \tabincell{c}{k=20, LN=T, \\ net=[700$\times$5,1]}
		& \tabincell{c}{k=40, LN=T, \\ net=[500$\times$5,1]} 
		& \tabincell{c}{k=20, LN=T, \\ net=[300$\times$3,1]}
		& \tabincell{c}{k=10, LN=F, \\ net=[400$\times$3,1], \\ drop=0.1} \\ 
		PIN 
		& \tabincell{c}{sub-net=[40,5]}
		& \tabincell{c}{sub-net=[40,5]} 
		& \tabincell{c}{sub-net=[40,5]}
		& \tabincell{c}{sub-net=[20,1]} \\ 
	\end{tabular}
	
	\bigskip\centering
	\emph{Note:} bs=Batch Size, opt=Optimizer, lr=Learning Rate, l2=L2 Regularization on Embedding Layer, k=Embedding Size, kernel=Convolution Kernel Size, net=DNN Structure, sub-net=Micro Network, t=Softmax Temperature, l2\_a= L2 Regularization on Attention Network, h=Attention Network Hidden Size, drop=Dropout Rate, LN=Layer Normalization (T: True, F: False)
\end{table}

Table~\ref{tab:param} shows key hyper-parameters of the models.

For a fair comparison, on Criteo, Avazu, and iPinYou, we (i) fix the embedding size according to \revise{the best-performed FM (searched among \{10, 20, 40, 80\})}, and (ii) fix the DNN structure according to \revise{the best-performed FNN (width searched in [100, 1000], depth searched in [1, 9])}. 
\revise{In terms of initialization, we initialize DNN hidden layers with xavier~\cite{glorot2010understanding}, and we initialize the embedding vectors from uniform distributions (range selected from $\{\sqrt{c/Nk}, \sqrt{c/nk}, \sqrt{c/k}\}$, $c = \{1, 3, 6\}$, as discussed in Section~\ref{sec:issue-init}.)}.
For Huawei, we follow the parameter settings of \cite{guo2017deepfm}.

\revise{With these constraints, all latent vector-based models have the same embedding size, and all DNN-based models additionally have the same DNN classifier. }
Therefore, all these models have similar amounts of parameters and are evaluated with the same training efforts.  
\revise{We also conduct parameter study on 4 typical models, where grid search is performed.
}


\subsubsection{Overall Performance}\label{sec:exp-eval-overall}

\begin{table*}[tbp]
	\centering
	\caption{Overall performance. (Left-Right: Criteo, Avazu, iPinYou, Huawei)}\label{tab:overall}
	\resizebox{1\columnwidth}{!}{
	\begin{tabular}{c|cc|cc|cc|cc}
			Model & AUC (\%) & Log Loss & AUC (\%) & Log Loss & AUC (\%) & Log Loss & AUC (\%) & Log Loss \\ \hline
			LR & 78.00	& 0.5631	& 76.76	& 0.3868	& 
			76.38	& 0.005691
			& 86.40	& 0.02648 \\
			GBDT & 78.62	& 0.5560	& 77.53	& 0.3824	& 
			76.90 & 0.005578
			& 86.45	& 0.02656 \\
			FM & 79.09	& 0.5500	& 77.93	& 0.3805	& 
			77.17	& 0.005595
			& 86.78	& 0.02633 \\
			FFM & 79.80	& 0.5438	& 78.31	& 0.3781	& 
			76.18	& 0.005695
			& 87.04	& 0.02626 \\
			CCPM & 79.55	& 0.5469	& 78.12	& 0.3800	& 
			77.53	& 0.005640
			& 86.92	& 0.02633 \\
			FNN & 79.87	& 0.5428	& 78.30	& 0.3778	& 
			77.82	& 0.005573
			& 86.83	& 0.02629 \\
			AFM & \change{79.13} &\change{0.5517} & \change{78.06}& \change{0.3794}& \change{77.71}& \change{\underline{0.005562}}& \change{86.89} & \change{0.02649}\\
			DeepFM & \underline{79.91}	& \underline{0.5423}	& \underline{78.36}	& \underline{0.3777}	& 
			\underline{77.92}	& 0.005588	
			& \underline{87.15}	& \underline{0.02618} \\ \hline
			KFM & 79.85	& 0.5427	& 78.40	& 0.3775	& 
			76.90	& 0.005630
			& 87.00	& 0.02624 \\
			NIFM & 79.80	& 0.5437	& 78.13	& 0.3788	& 
			77.07	& 0.005607
			& 87.16	& 0.02620 \\
			IPNN & 80.13	& 0.5399	& 78.68	& 0.3757	& 
			78.17	& 0.005549
			& 87.27	& 0.02617 \\
			KPNN & 80.17	& 0.5394	& 78.71	& 0.3756	& 
			78.21	& 0.005563
			& 87.28	& 0.02617 \\
			\textbf{PIN} & \textbf{80.21}	& \textbf{0.5390}	& \textbf{78.72}	& \textbf{0.3755}	& 
			\textbf{78.22}	& \textbf{0.005547}
			& \textbf{87.30}	& \textbf{0.02614} \\	
	\end{tabular}}
\end{table*}
	
Table~\ref{tab:overall} shows the overall performance. Underlined numbers are best results of baseline models, and bold numbers are best results of all.

\revise{Comparing KFM/NIFM with FM, FFM, and AFM, }FFM is the best baseline model on Criteo, Avazu, and Huawei, and AFM is the best baseline model on iPinYou. 
KFM and NIFM achieve even better performance than FFM and AFM on all datasets.
The scores of KFM and NIFM successfully verify the effectiveness of kernel product methods.
A more detailed discussion about feature interactions is in Section~\ref{sec:exp-fafi}.
\revise{Comparing GBDT with FNN,} we find GBDT performs no better than FNN.
A possible reason is the enormous feature space makes GBDT hard to explore all possible combinations.
\revise{Comparing PNNs with other DNN-based models, }in general, DeepFM is the best baseline model on 4 datasets.
PNNs consistently outperform DeepFM, and PIN achieves the best results on all datasets. \revise{The performance of PNNs verifies the effectiveness of product layers.}

\begin{table}[htbp]
	\centering
	\caption{Adaptive embedding.}
	\label{fig:flexible}
	\begin{tabular}{c|c|cc}
		Model & Embed Size & AUC & Log Loss \\ \hline
		KFM & \tabincell{c}{40 \\ $\min(4 \log(N_i), 40)$} & \tabincell{c}{0.7840 \\ 0.7850 } & \tabincell{c}{0.3775 \\ 0.3769} \\ \hline
		NIFM & \tabincell{c}{40 \\ $\min(4 \log(N_i), 40)$} & \tabincell{c}{0.7813 \\ 0.7819 } & \tabincell{c}{0.3788 \\ 0.3786} \\
	\end{tabular}
\end{table}

Since kernel product has the ability to learn \revise{adaptive embeddings}, we study adaptive embeddings on KFM and NIFM, as suggested in Section~\ref{sec:method-fafi}.
We use $k_i = \min(c\log(N_i), K)$ as the embedding size for field $i$ with $N_i$ categories.
We choose Avazu to compare adaptive embeddings and fixed size embeddings, because Avazu is relatively small and has balanced positive/negative samples.
As shown in Table~\ref{fig:flexible}, adaptive embeddings further improve the performance of KFM and NIFM.
\revise{Note that adaptive embeddings are harder to parallelize, thus are trained much slower.}

\begin{table}[htbp]
	\centering
	\caption{Significance test.}
	\label{tab:sigtest}
	\begin{tabular}{c|c|c}
		$p$-value & KFM, NIFM & IPNN, KPNN, PIN \\ \hline
		FM, FFM & $< 10^{-6}$ & $< 10^{-6}$ \\ \hline
		FNN, CCPM, DeepFM & $< 10^{-6}$ & $< 10^{-6}$ 
	\end{tabular}
\end{table}

Table~\ref{tab:sigtest} presents the significance test.
We use the Wilcoxon signed-rank test to check if the results of our proposed models and baseline models are generated from the same distribution.
The $p$-values validate that the improvements of our models are significant. 
\change{To test model robustness, we train PIN on Criteo for 10 times with different random seeds. The standard deviation of AUC is 0.0085 (average AUC 80.2), and the standard deviation of log loss is 0.0002 (average log loss 0.539).}

\change{Besides, we participate in the Criteo Display Advertising Challenge, libFFM being the winning solution\footnote{\change{https://github.com/guestwalk/kaggle-2014-criteo. This repository has 2 branches, and we use the ``master'' branch. 
}}. 
We download the winners' code and data to repeat their results and generate the training files.
libFFM achieves $\text{log loss} = 0.44506/0.44520$ on the private/public leaderboard, and achieves 0.44493/0.44508 after calibration.
We train KFM with the same training files as libFFM on one 1080Ti.
KFM achieves 0.44484/0.44492 on the private/public leaderboard, and achieves 0.44484/0.44491 after calibration\footnote{Our solution: https://github.com/Atomu2014/product-nets-distributed}.
Besides, KFM uses less memory (64M parameters) than libFFM (272M parameters), and uses similar training time (3.5h)\footnote{\revise{An acceleration trick is used, see Section~\ref{sec:exp-prac-comp}.}} to libFFM (3.5h).}

\subsubsection{Parameter Study}\label{sec:exp-eval-param}
\begin{figure}[tbp]
\subfigure[Criteo]{
\includegraphics[width=0.48\textwidth]{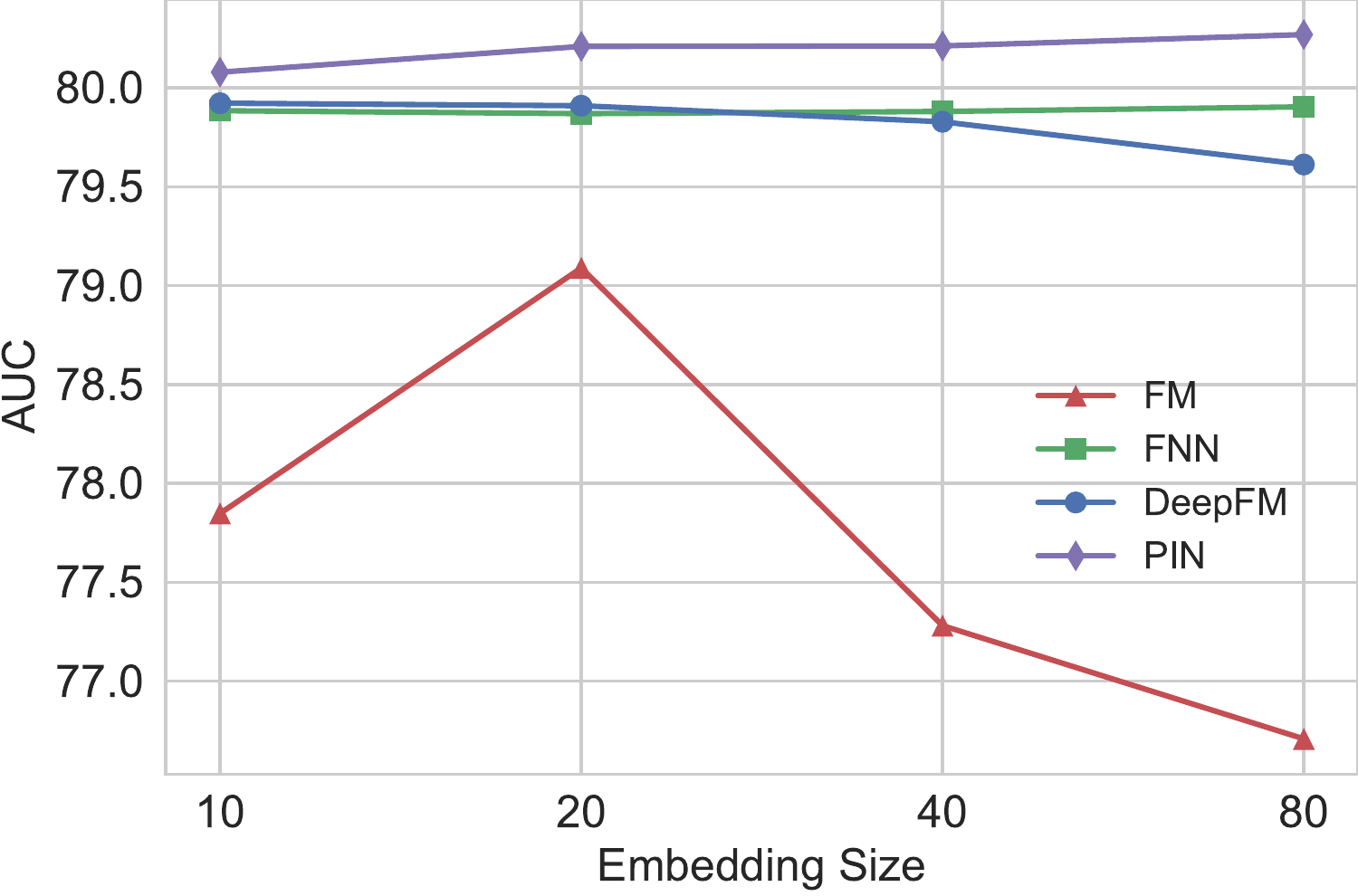}
}
\subfigure[Avazu]{
\includegraphics[width=0.48\textwidth]{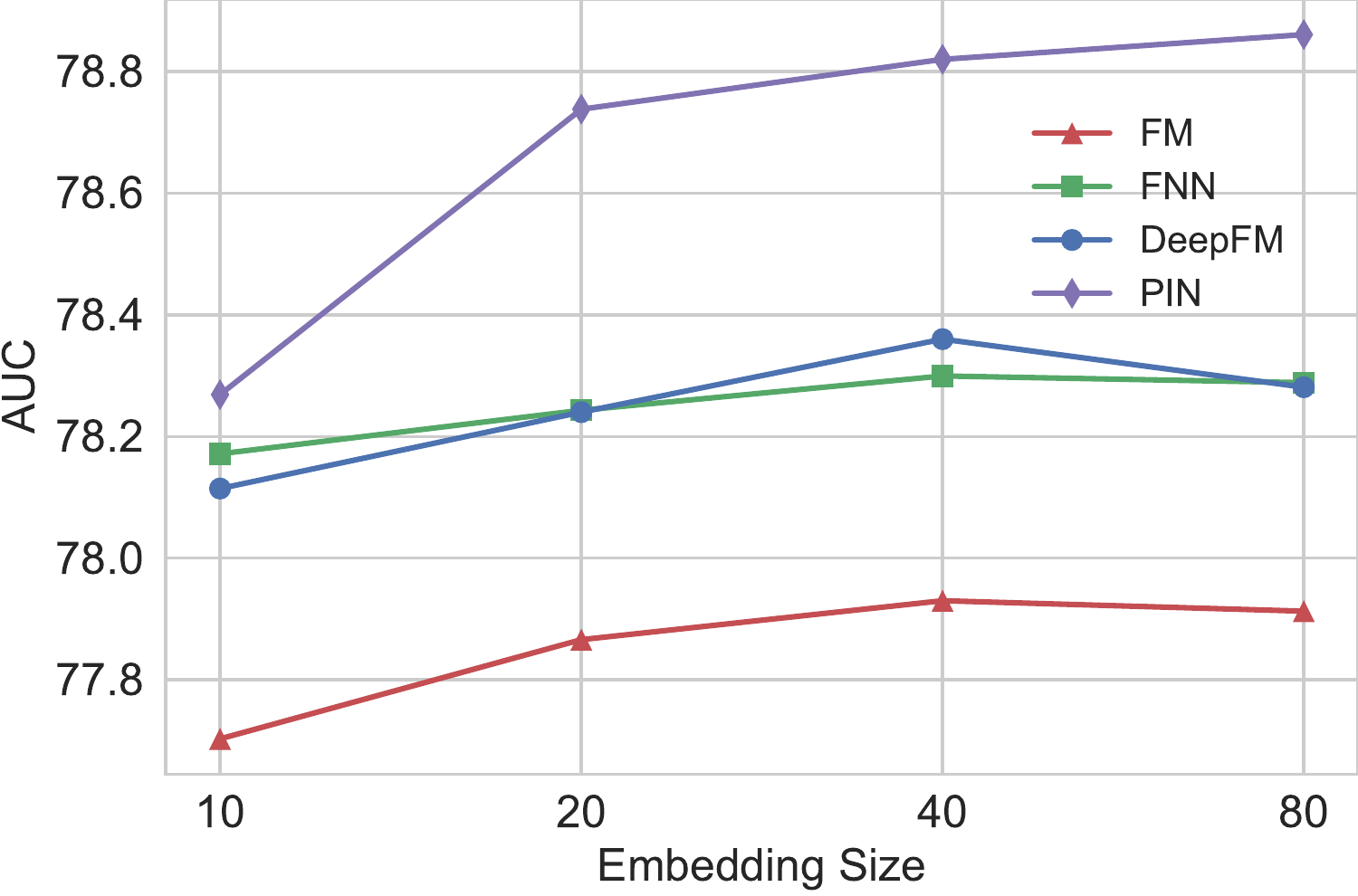}
}
\subfigure[iPinYou]{
\includegraphics[width=0.48\textwidth]{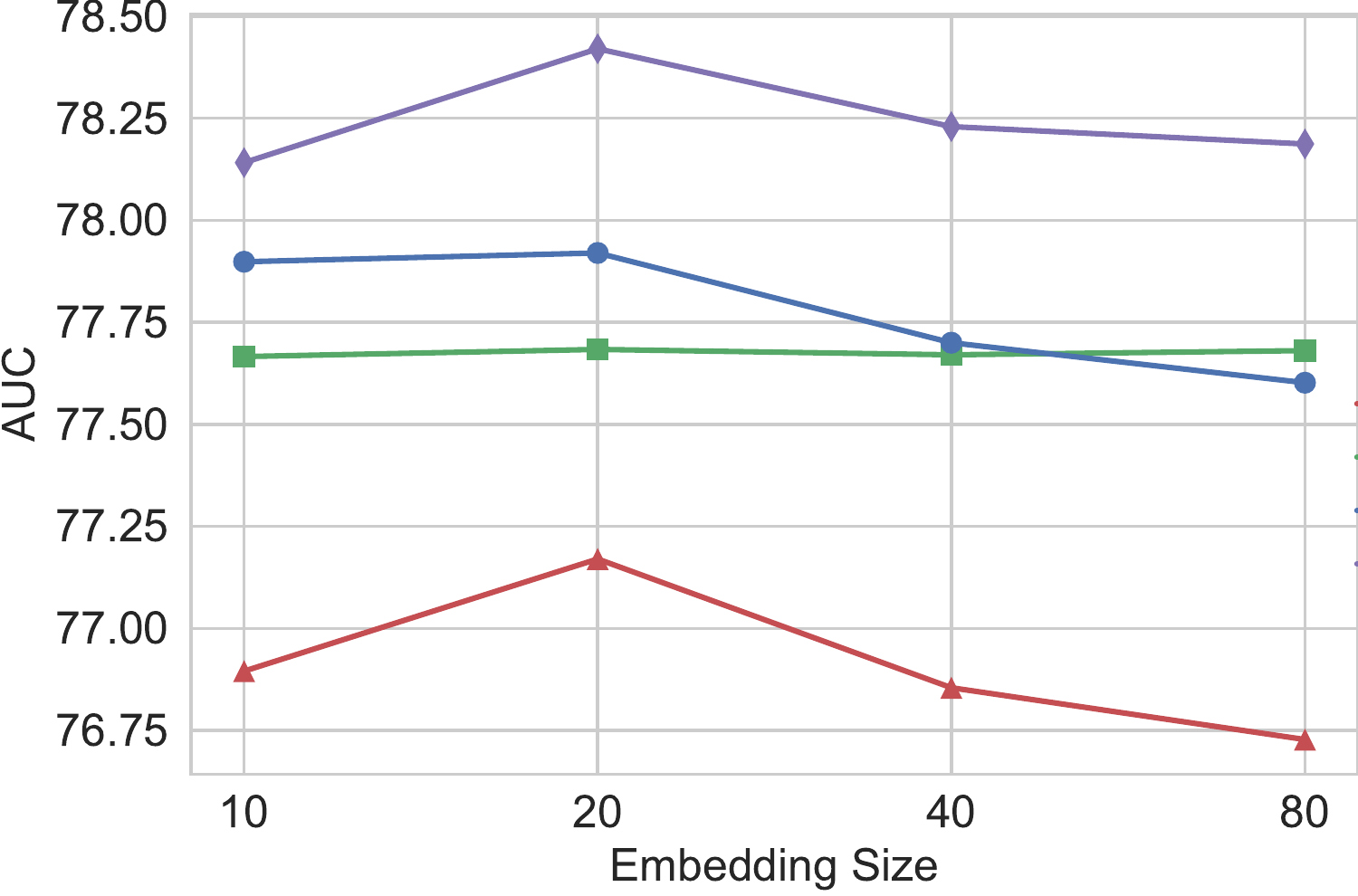}}
\subfigure[Huawei]{
\includegraphics[width=0.48\textwidth]{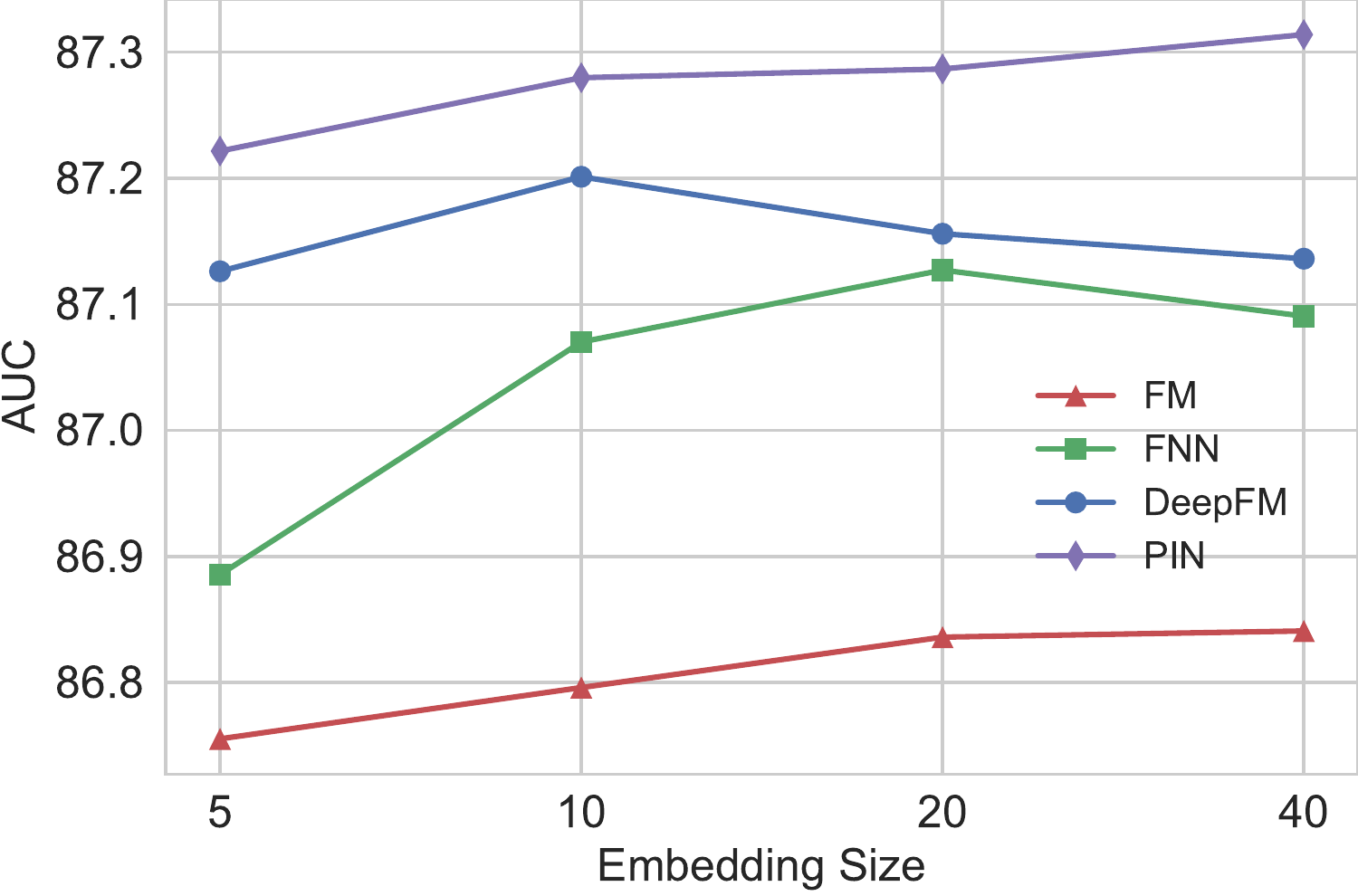}
}
\caption{Embedding size parameter study. \emph{Note:} The network structures of DNN models are fixed according to Table~\ref{tab:param}.}
\label{fig:embed}
\end{figure}

\change{
In this section, we study embedding size, network width, and network depth on FM, FNN, DeepFM, and PIN. 

We test embedding size = \{10, 20, 40, 80\} on Criteo, Avazu, and iPinYou. From Fig.~\ref{fig:embed}, we find 20 (Criteo), 40 (Avazu), 20 (iPinYou) are the best for FM, which is consistent with Table~\ref{tab:param}. 
Since Huawei is relatively low-dimensional, we test embedding size = \{5, 10, 20, 40\}, based on the parameter study of \cite{guo2017deepfm}. }
\revise{An interesting phenomenon is FM and DeepFM are easier to overfit with large embedding sizes.
A possible reason is DNN has higher capacity than FM.}

\begin{figure}[tbp]
	\subfigure[Criteo]{
		\includegraphics[width=0.43\textwidth]{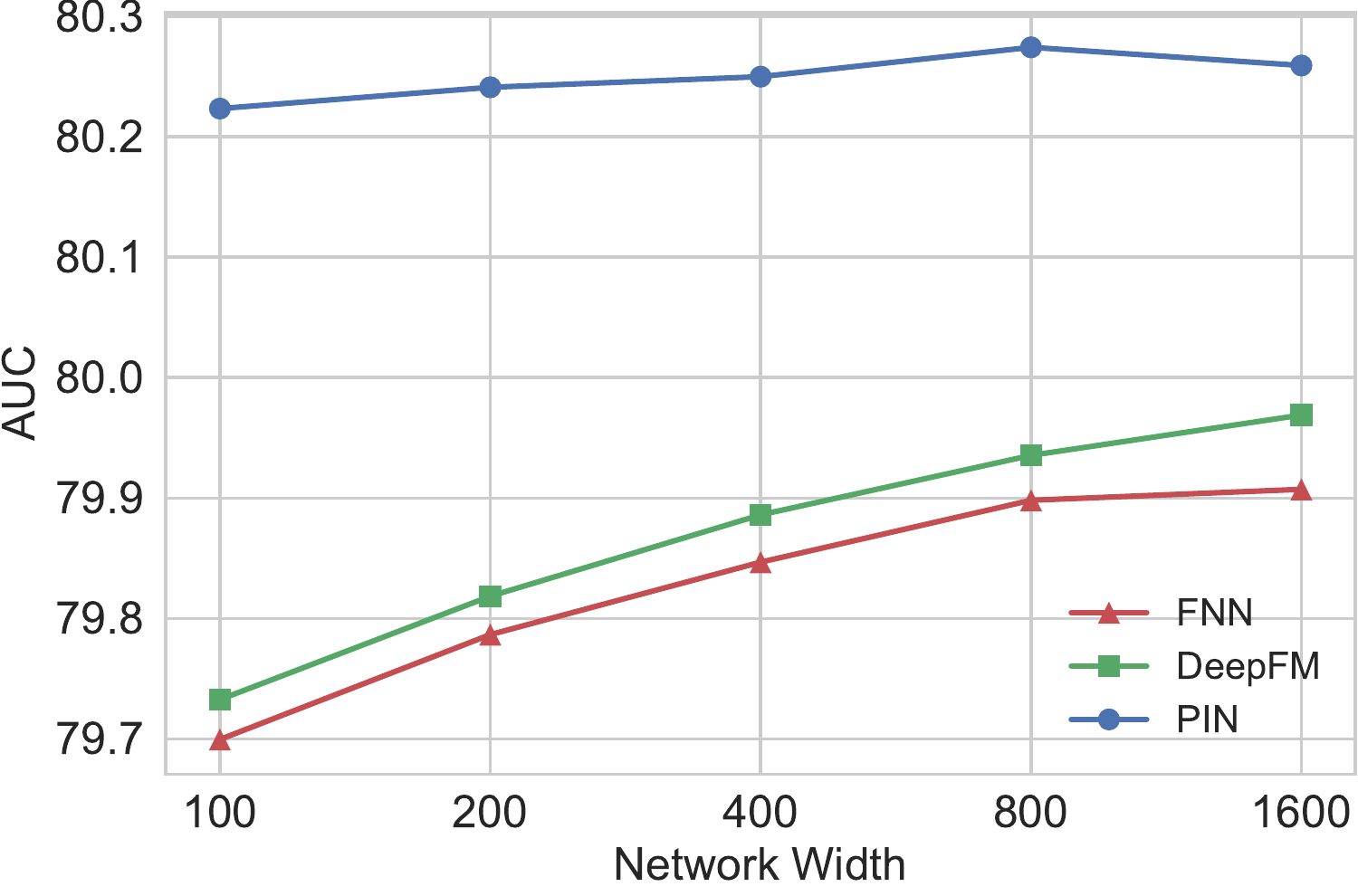}}
	\subfigure[Avazu]{
		\includegraphics[width=0.43\textwidth]{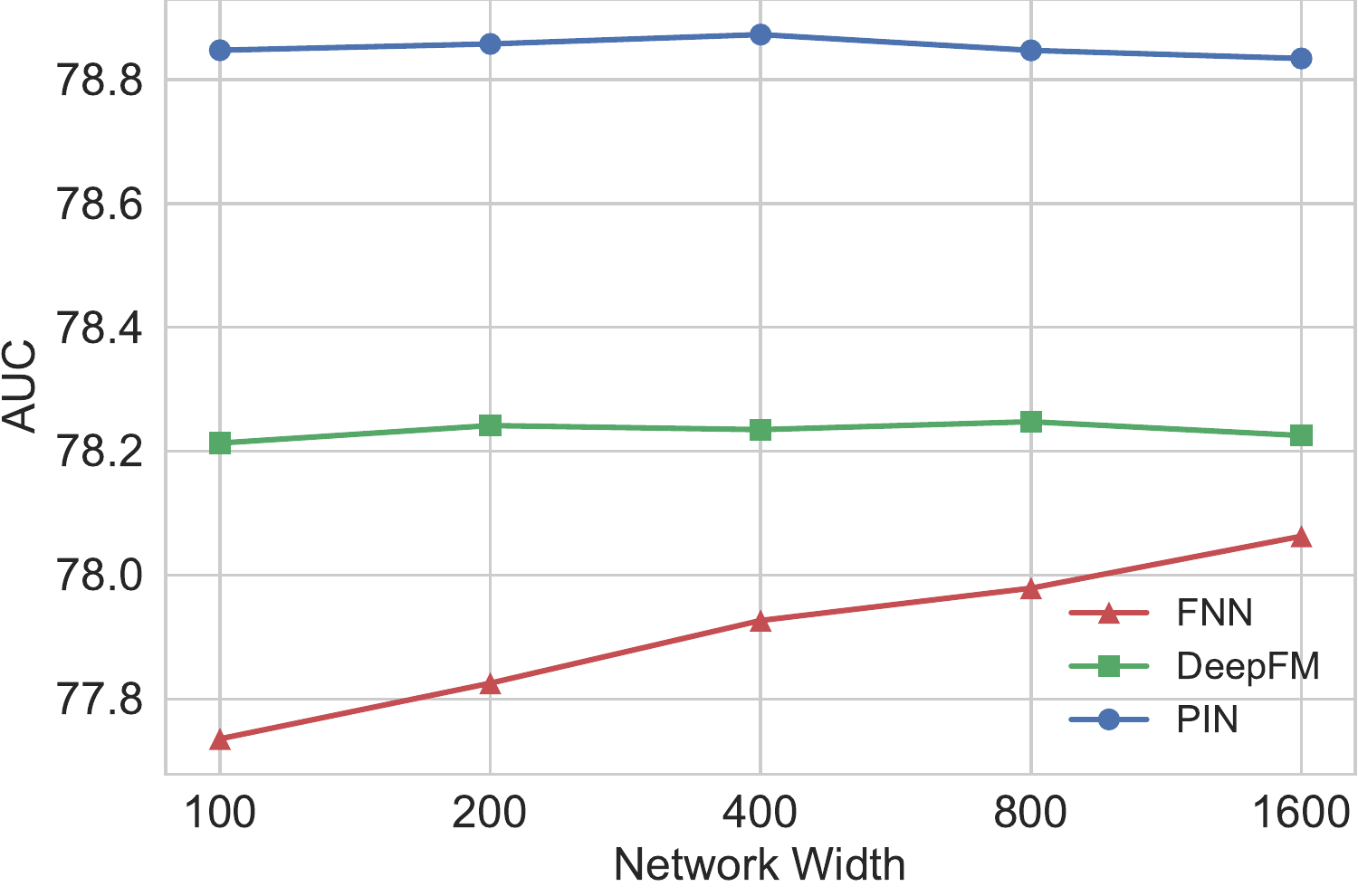}
	}
	\subfigure[iPinYou]{
		\includegraphics[width=0.43\textwidth]{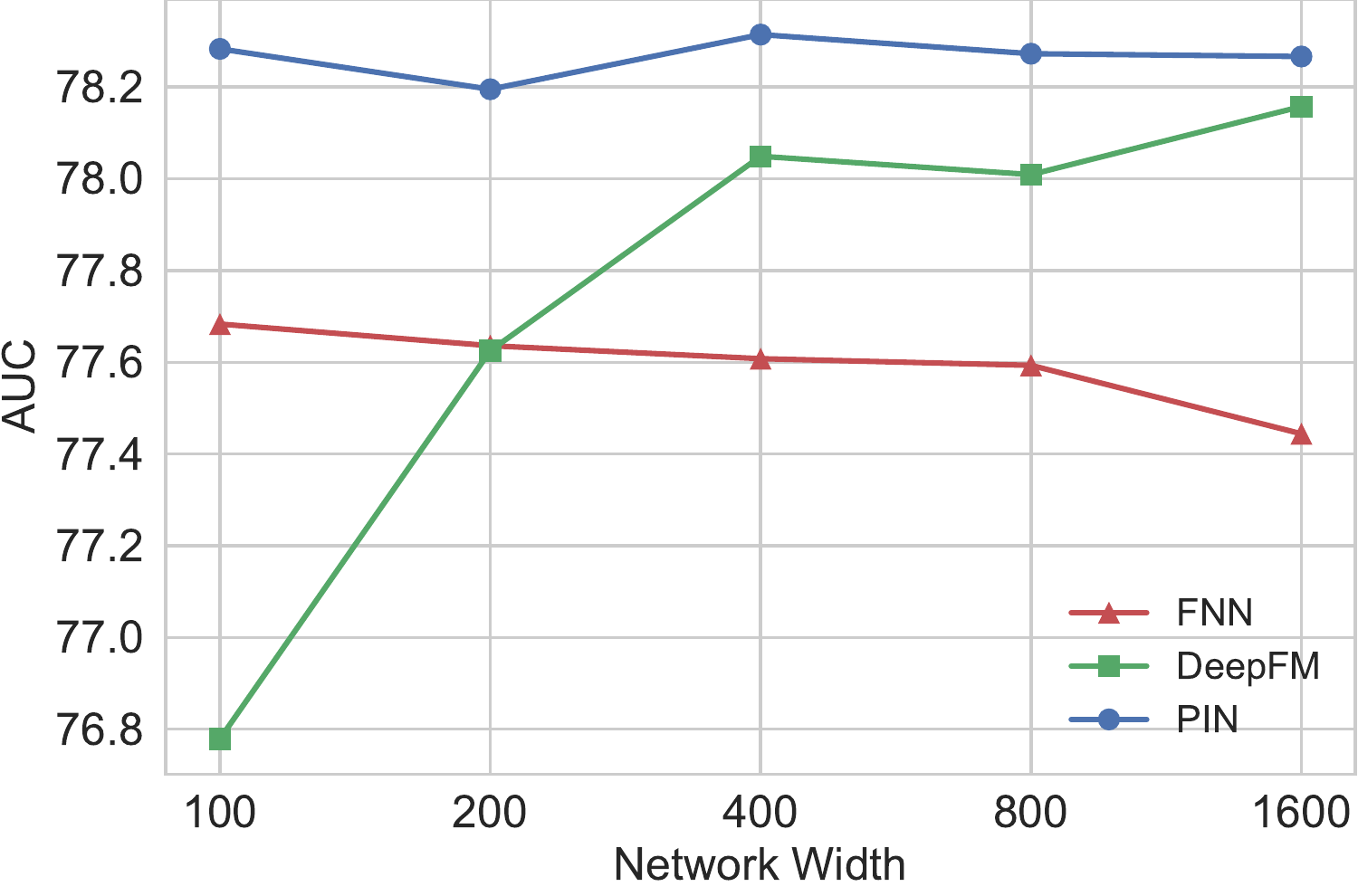}}
	\subfigure[Huawei]{
		\includegraphics[width=0.43\textwidth]{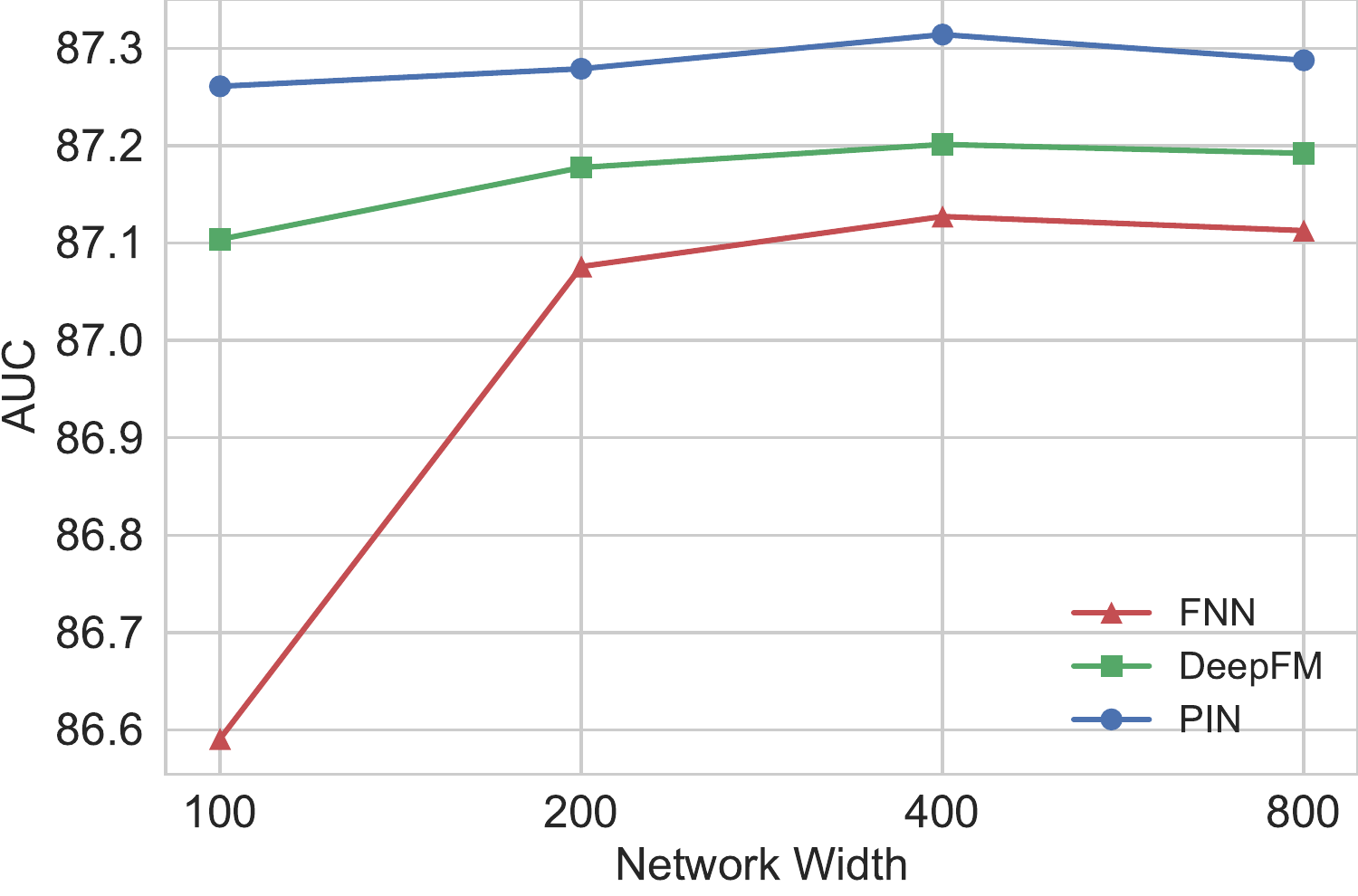}
	}
	\subfigure[Criteo]{
		\includegraphics[width=0.43\textwidth]{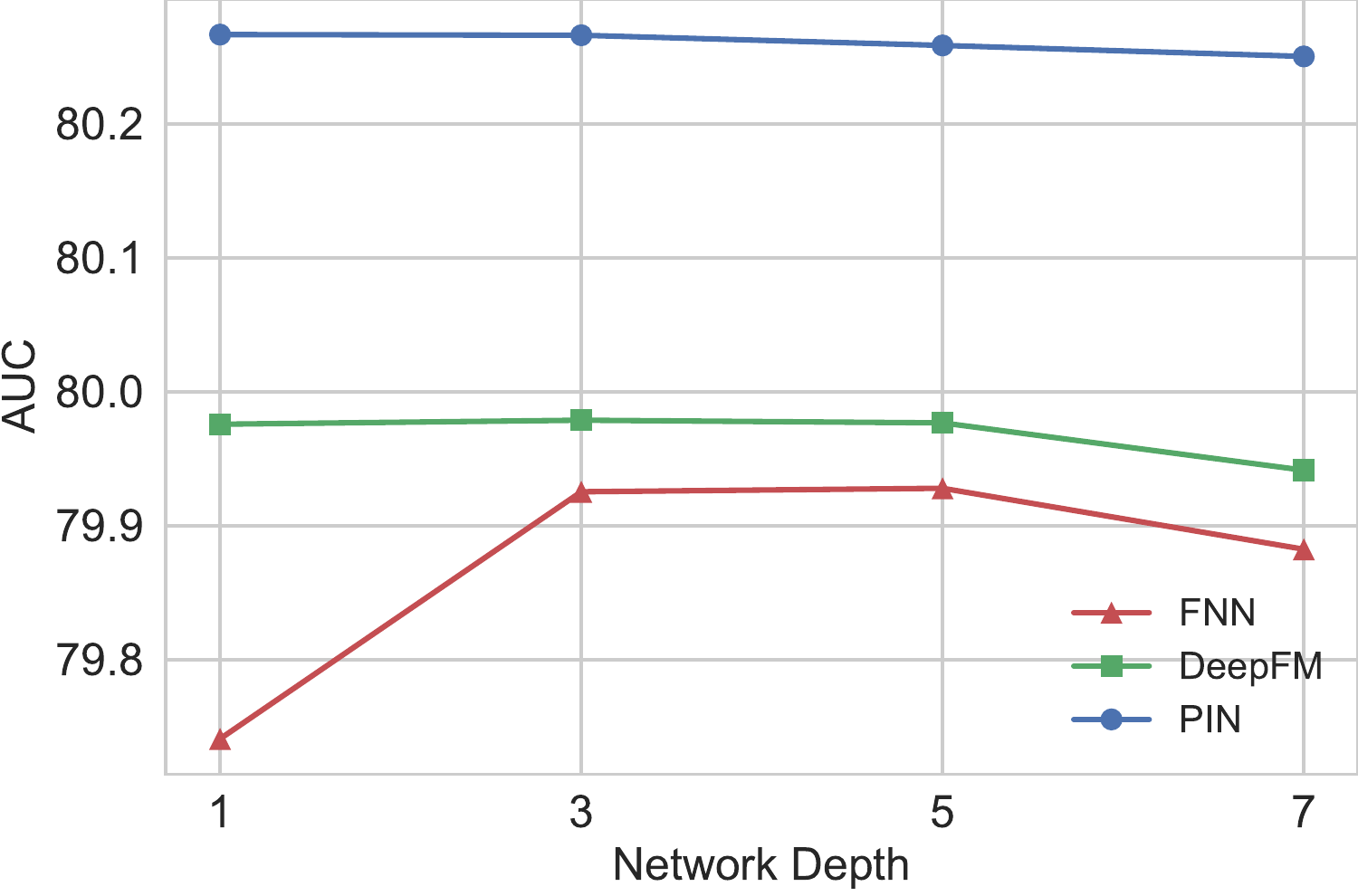}}
	\subfigure[Avazu]{
		\includegraphics[width=0.43\textwidth]{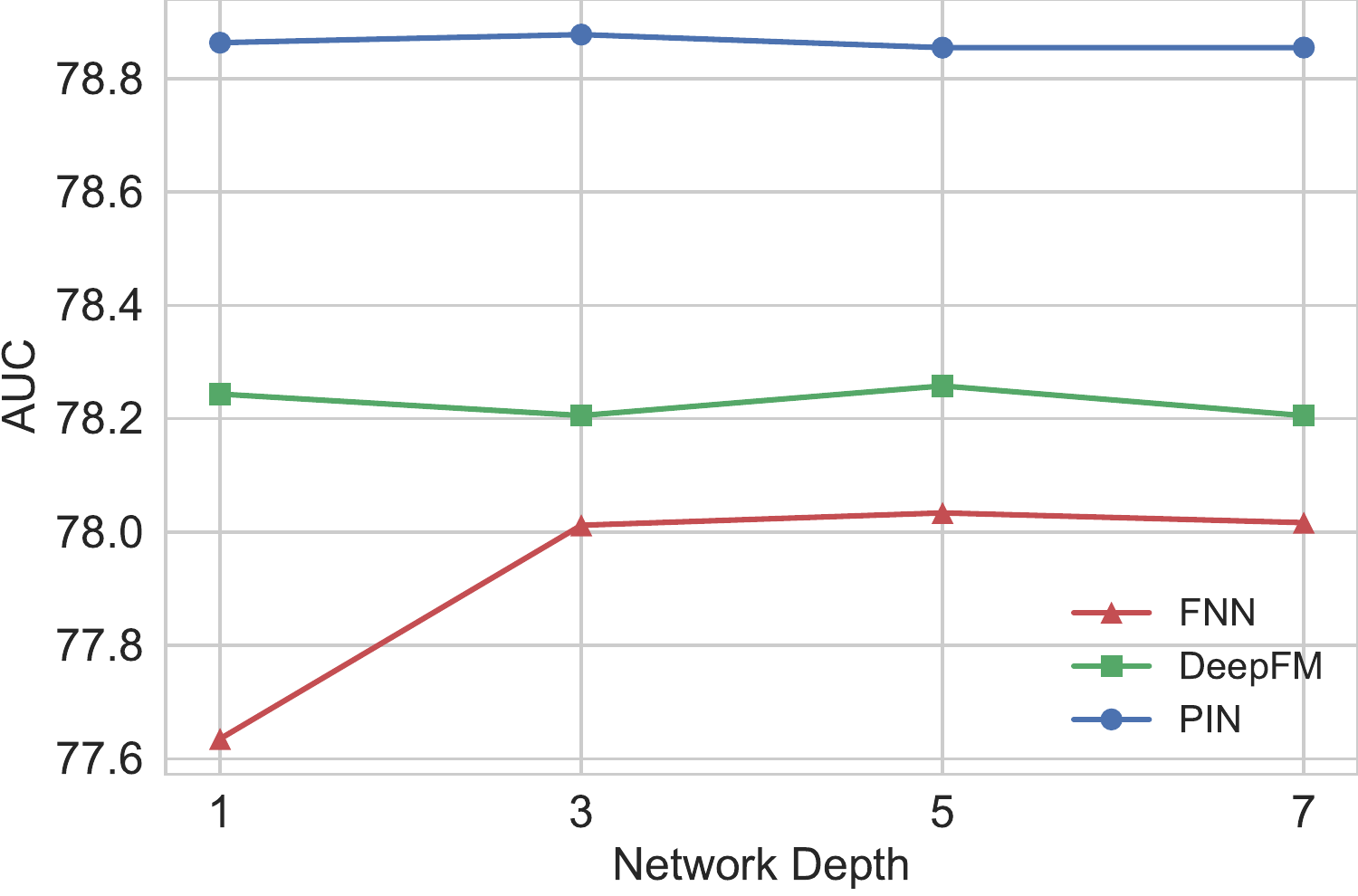}
	}
	\subfigure[iPinYou]{
		\includegraphics[width=0.43\textwidth]{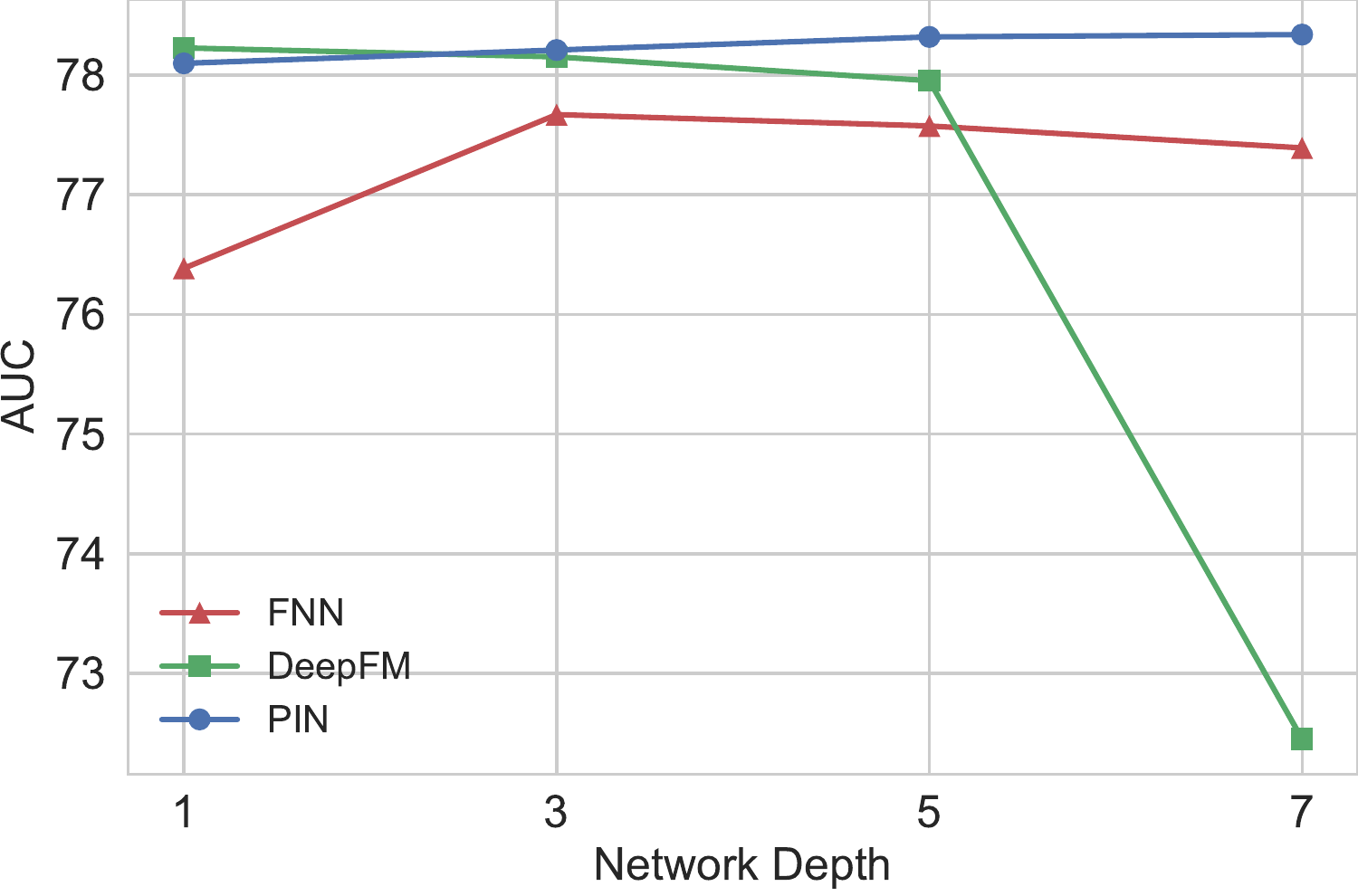}}
	\subfigure[Huawei]{
		\includegraphics[width=0.43\textwidth]{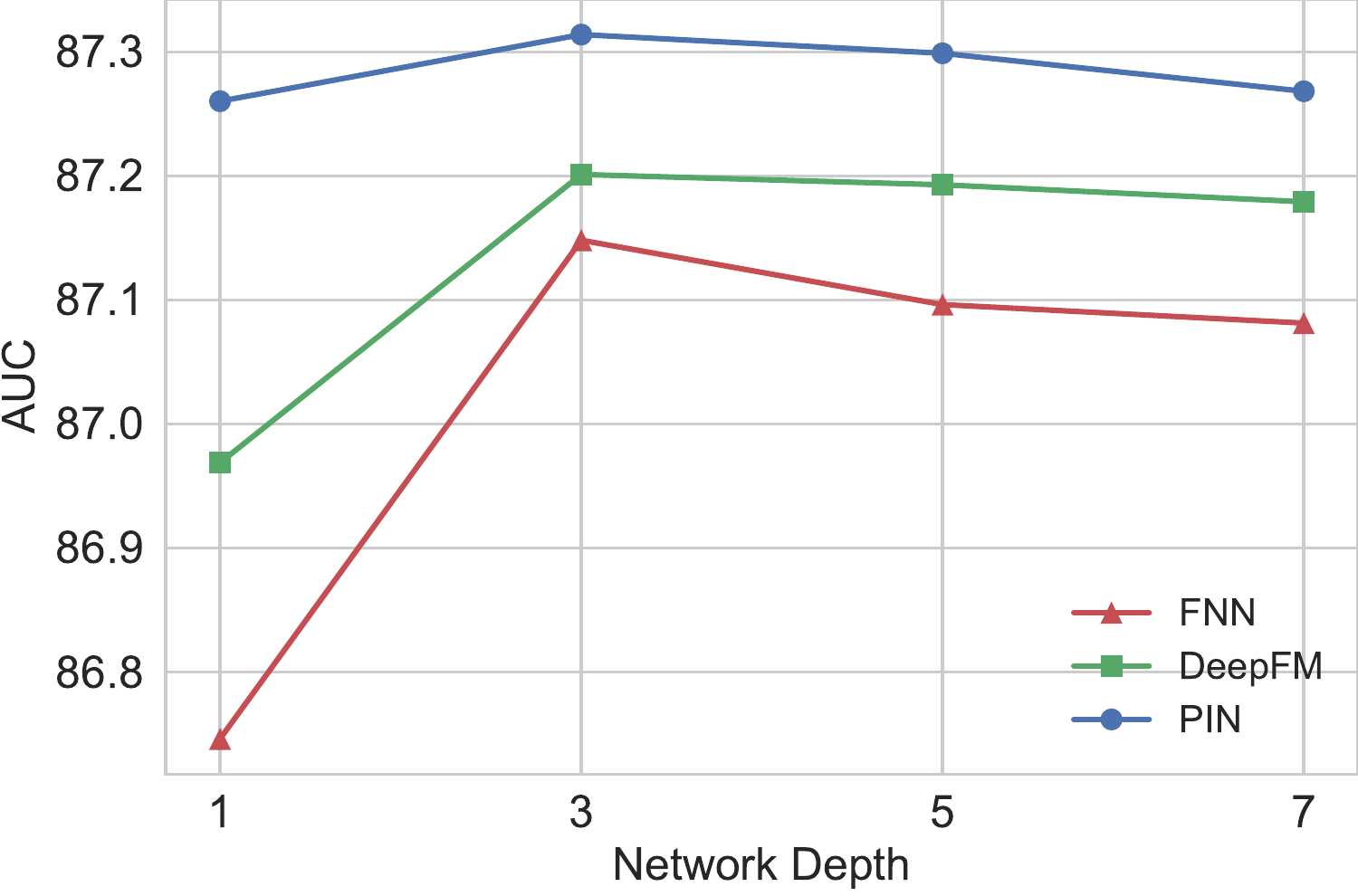}
	}
	\caption{Network structure parameter study. \emph{Note:} In (a)-(d), the network depths are fixed according to Table~\ref{tab:param}, and the embedding sizes are chosen based on the best performance of Fig.~\ref{fig:embed}. In (e)-(h), the embedding sizes and network widths are chosen according to Fig.~\ref{fig:embed} and (a)-(d).}
	\label{fig:structure}
\end{figure}

\change{As for network structure, we first fix the embedding size and the depth according to Fig.~\ref{fig:embed} and Table~\ref{tab:param}, and test network width = \{100, 200, 400, 800, 1600\} on Criteo, Avazu, and iPinYou, \{100, 200, 400, 800\} on Huawei. 
The results are shown in Fig.~\ref{fig:structure}(a)-(d). 
As for network depth, we choose the embedding size and the width according to the above results, and test network depth = \{1, 3, 5, 7\} on all the datasets. The results are shown in Fig.~\ref{fig:structure}(e)-(h).}
\revise{In general, we find: 
(i) PIN consistently outperforms FNN and DeepFM.
(ii) When the network is small, the DNN capacity is restricted, yet PIN performs even better than FNN and DeepFM.
This means PIN learns more expressive feature representations.
(iii) When the network is large, FNN and DeepFM are easier to overfit, which means PIN is more robust.}


\begin{table}[tbp]
\centering
\caption{Best performance in parameter study. \emph{Note:} k=Embedding Size, net=DNN structure.}
\label{tab:best}
\begin{tabular}{c|c|c|c|c}
& Criteo & Avazu & iPinYou & Huawei \\ \hline
FM & \tabincell{c}{k=20 \\ AUC=79.09 \\ log loss=0.5500} & \tabincell{c}{k=40 \\ AUC=77.93 \\ log loss=0.3805} & \tabincell{c}{k=20 \\ AUC=77.17 \\ log loss=0.005595} & \tabincell{c}{k=40 \\ AUC=86.84 \\ log loss=0.02628} \\ \hline
FNN & \tabincell{c}{k=20, net=1600*5 \\ AUC=79.91 \\ log loss=0.5425} & \tabincell{c}{k=40, net=500*5 \\ AUC=78.30 \\ log loss=0.3778} & \tabincell{c}{k=20, net=300*3 \\ AUC=77.68 \\ log loss=0.005585} & \tabincell{c}{k=20 net=400*3 \\ AUC=87.15 \\ log loss=0.02619} \\ \hline
DeepFM & \tabincell{c}{k=10, net=1600*3 \\ AUC=79.98 \\ log loss=	0.5419} & \tabincell{c}{k=40, net=500*5 \\ AUC=78.36 \\ log loss=0.3777} & \tabincell{c}{k=20, net=1600*1 \\ AUC=78.23 \\ log loss=0.005562} & \tabincell{c}{k=10, net=400*3 \\ AUC=87.20 \\ log loss=0.02616} \\ \hline
PIN & \tabincell{c}{k=80, net=800*5 \\ AUC=\textbf{80.27} \\ log loss=\textbf{0.5385}} & \tabincell{c}{k=80, net=400*5 \\ AUC=\textbf{78.88} \\ log loss=\textbf{0.3745}} & \tabincell{c}{k=20, net=300*3 \\ AUC=\textbf{78.42} \\ log loss=\textbf{0.005546}} & \tabincell{c}{k=40, net=400*3 \\ AUC=\textbf{87.31} \\ log loss=\textbf{0.02616}} \\
\end{tabular}
\end{table}

\change{Finally, the best scores and parameters are shown in Table~\ref{tab:best}. 
}

\subsubsection{Online Evaluation}\label{sec:exp-eval-online}
\begin{figure}[tbp]
	\centering
	\includegraphics[width=0.8\textwidth]{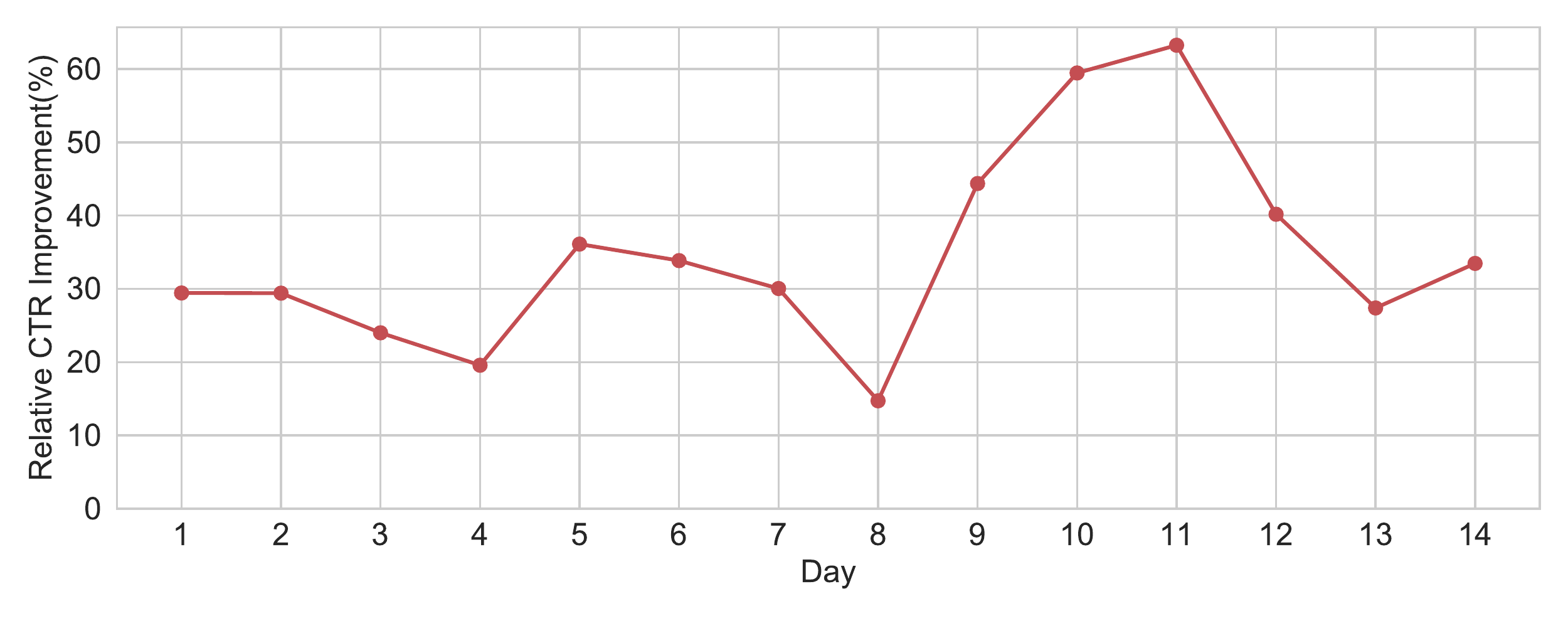}
	\caption{Relative CTR improvements of PIN, calculated by $\frac{CTR of PIN - CTR of FTRL}{CTR of FTRL}$.}\label{fig:ab}
\end{figure}

\revise{Beside offline evaluations, we perform an online A/B test in the game center of Huawei App Market. 
The compared model is FTRL which has been incrementally updated for several years. 
We train PIN on the latest 40-day CTR logs. 
Both models share the same feature map. 
After a 14-day A/B test, we observe an average of 34.67\% relative CTR improvement (maximum 63.26\%, minimum 14.72\%).
The improvements are shown in Fig.~\ref{fig:ab}. 
}

\subsection{Practical Issues}\label{sec:exp-prac}

\subsubsection{\revise{Space and Time Complexity}}\label{sec:exp-prac-comp}
In order to facilitate calculation, we choose Criteo (input dimension $1 \times 10^8$) to compare memory usage and training speed.
Except for LR and GBDT, all the other models have embedding layers. 
\revise{To compare memory usage, we set the embedding size $k=1$ for FFM and $k=20$ for the other models, and we keep other parameters same as Table~\ref{tab:param}.
To compare the training speed, we train these models with 10 million instances (batch size=2000) on an NVidia 1080Ti GPU.
The results are shown in Table~\ref{tab:memory}. 
}

\begin{table}[tbp]
	\centering
	\caption{Parameter number and training speed. \emph{Note:} ``\# params'' includes all trainable weights.}
	\label{tab:memory}
	\begin{tabular}{c|cc|c|cc}
		Model & \# params $(10^6)$ & time (min) & Model & \# params $(10^6)$ & time (min) \\ \hline
		LR & 1 & 1 & FNN & 22.51 & 2 \\
		FM & 21 & 3 & CCPM & 20.23 & 2 \\
		AFM & 21 & 13 & DeepFM & 23.51 & 3 \\
		FFM & $\ge$ 40 & 5 & IPNN & 23 & 3\\
		KFM & 21.3 & 12 & KPNN & 23.3 & 13 \\
		NIFM & 22.22 & 6 & PIN & 26.48 & 6
	\end{tabular}
\end{table}

\change{
	
As for space complexity, LR has the least parameters, since it has no embedding layer.
FFM consumes more memory, because the space complexity of FFM is $O(Nnk)$, where $N = 10^6$ is the input dimension, $n = 39$ is the number of fields, and $k=1$ is the embedding size.
Except for LR and FFM, the space complexities of other models are near $O(Nk)$. 
Among DNN-based models, CCPM uses the least memory due to parameter sharing in convolutional layers.
PNNs use extra feature extractors, thus require more memory than FNN.

In terms of training speed, we find AFM, KFM, and KPNN are relatively slow.
The interactions in FM have a complexity of $O(nk)$, the attention network in AFM has a complexity of $O(n^2kh)$
, and the kernel products in KFM and KPNN have a complexity of $O(n^2k^2)$.
The complexity of kernel products is quadratic to the embedding size, which slows down training if we use large embedding vectors.
In the Criteo Challenge, we use a trick which can reduce the complexity of kernel products to $O(n^2k)$: we replace the kernel matrices of size $k\times k$ with kernel vectors of size $k$.
This trick saves training time dramatically in the contest.
}

\subsubsection{FM Pre-training}\label{sec:exp-prac-init}
\begin{figure}[tbp]
	\subfigure[FNN]{
		\includegraphics[width=0.48\textwidth]{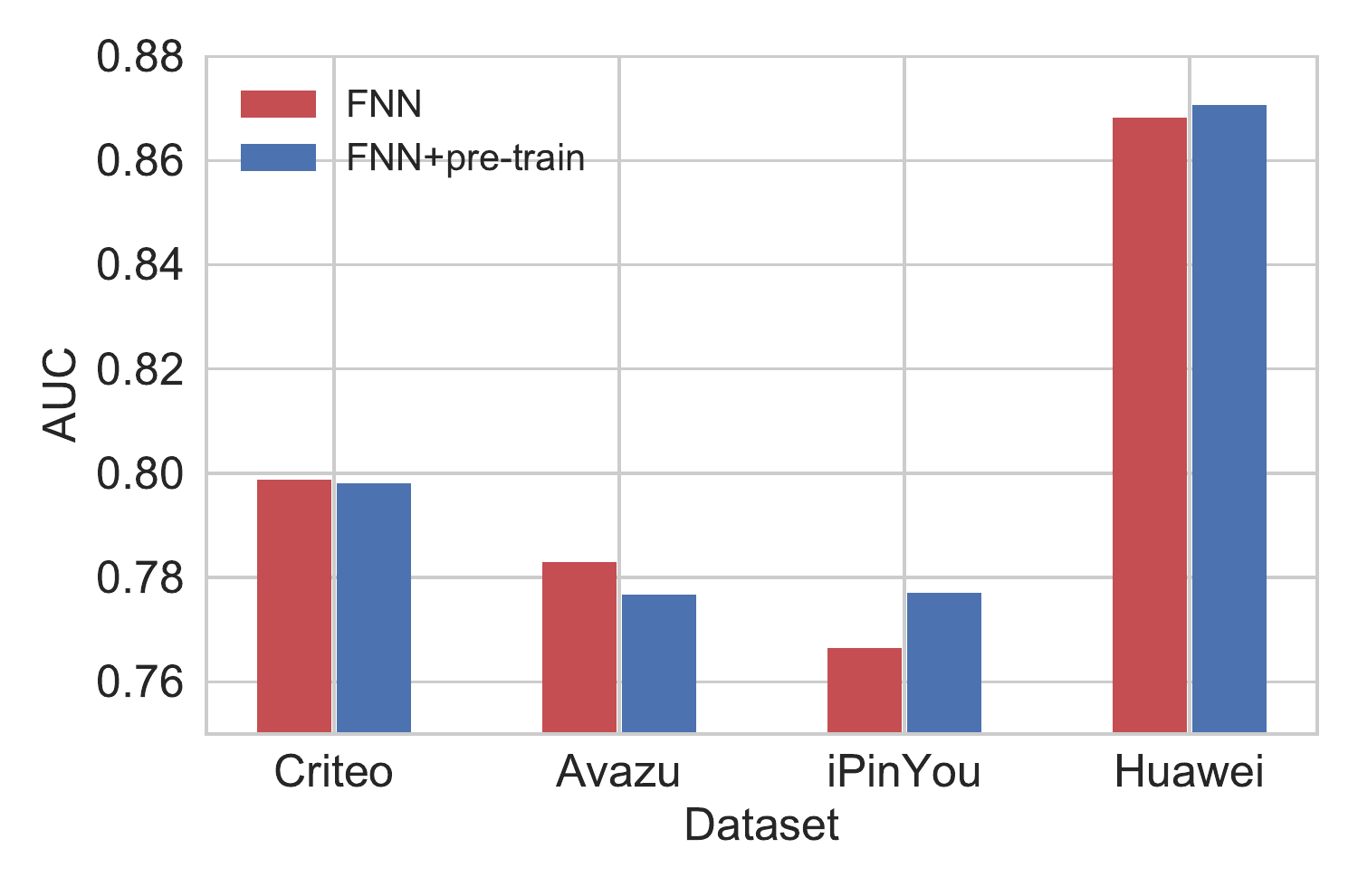}}
	\subfigure[AFM]{
		\includegraphics[width=0.48\textwidth]{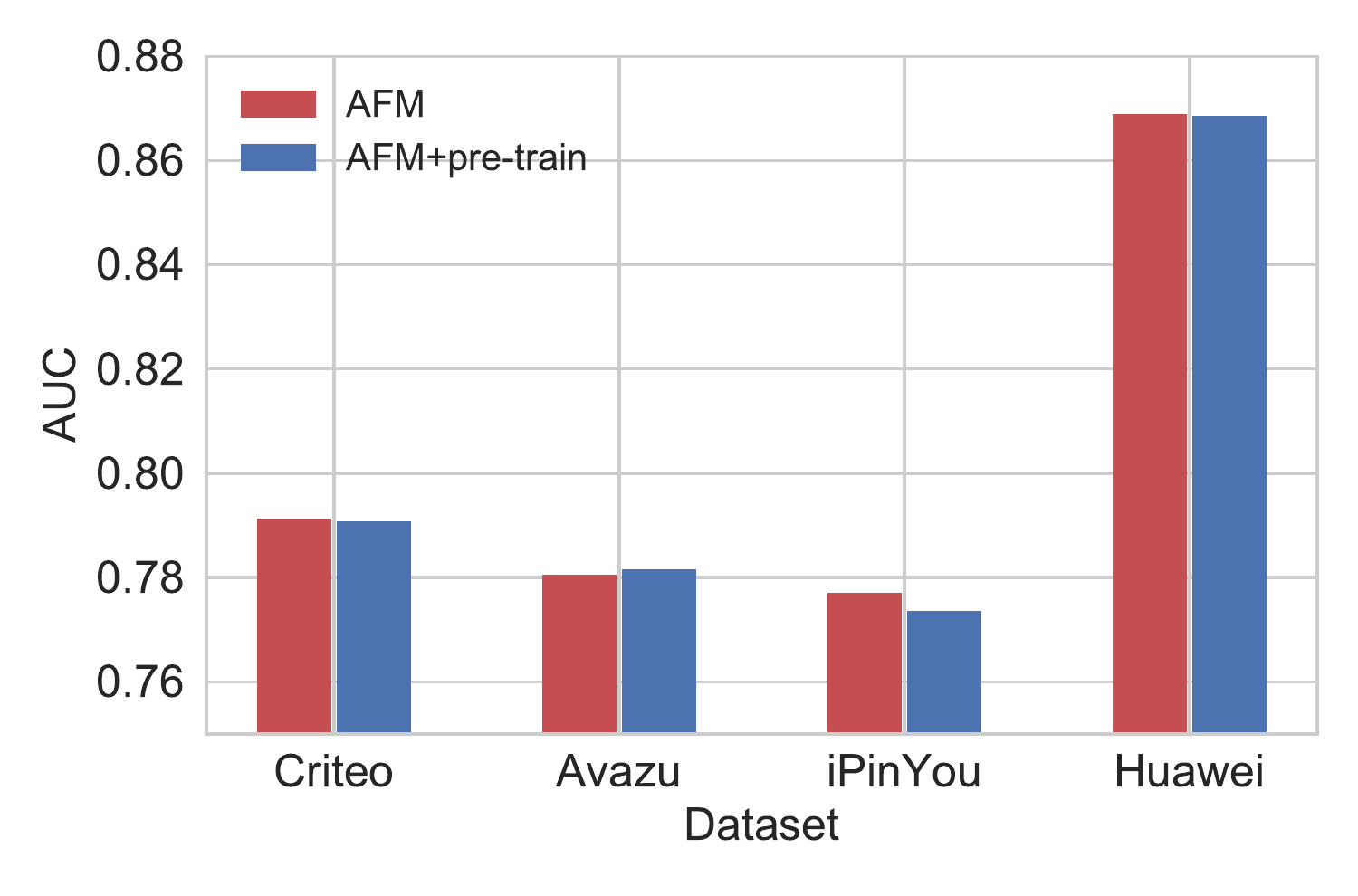}}
	\caption{Comparison of FM pre-training and random initialization. 
	}\label{fig:pre}
\end{figure}

\revise{Section~\ref{sec:issue-init} lists several initialization methods for the embedding vectors. 
In \cite{zhang2016deep, xiao2017attentional}, the authors pre-trained the embedding vectors with FM.
We compare pre-training with random initialization on FNN and AFM, the results are in Fig.~\ref{fig:pre}.
From this figure, we find FM pre-training does not always produce better results compared with random initialization.
Thus we believe the initialization of embedding vectors depends on datasets, which we leave as future work.}

\subsubsection{Optimization}\label{sec:exp-prac-opt}
The data in user response prediction is usually sparse and unbalanced \revise{(positive samples are usually much fewer than negative ones)}.
A sparse dataset is hard to train because the \revise{input categories usually follow long-tailed distributions.}
An unbalanced dataset is also hard to train because it may produce extremely small gradients when updating the model parameters. 
Section~\ref{sec:issue-opt} discusses \revise{the behaviors of Adam on sparse input.} 
In this section, we mainly focus on gradient sensitivity in Section~\ref{sec:issue-opt-sens}.

\begin{figure}[tbp]
	\subfigure[Gradients $> g^*$]{
		\includegraphics[width=0.48\textwidth]{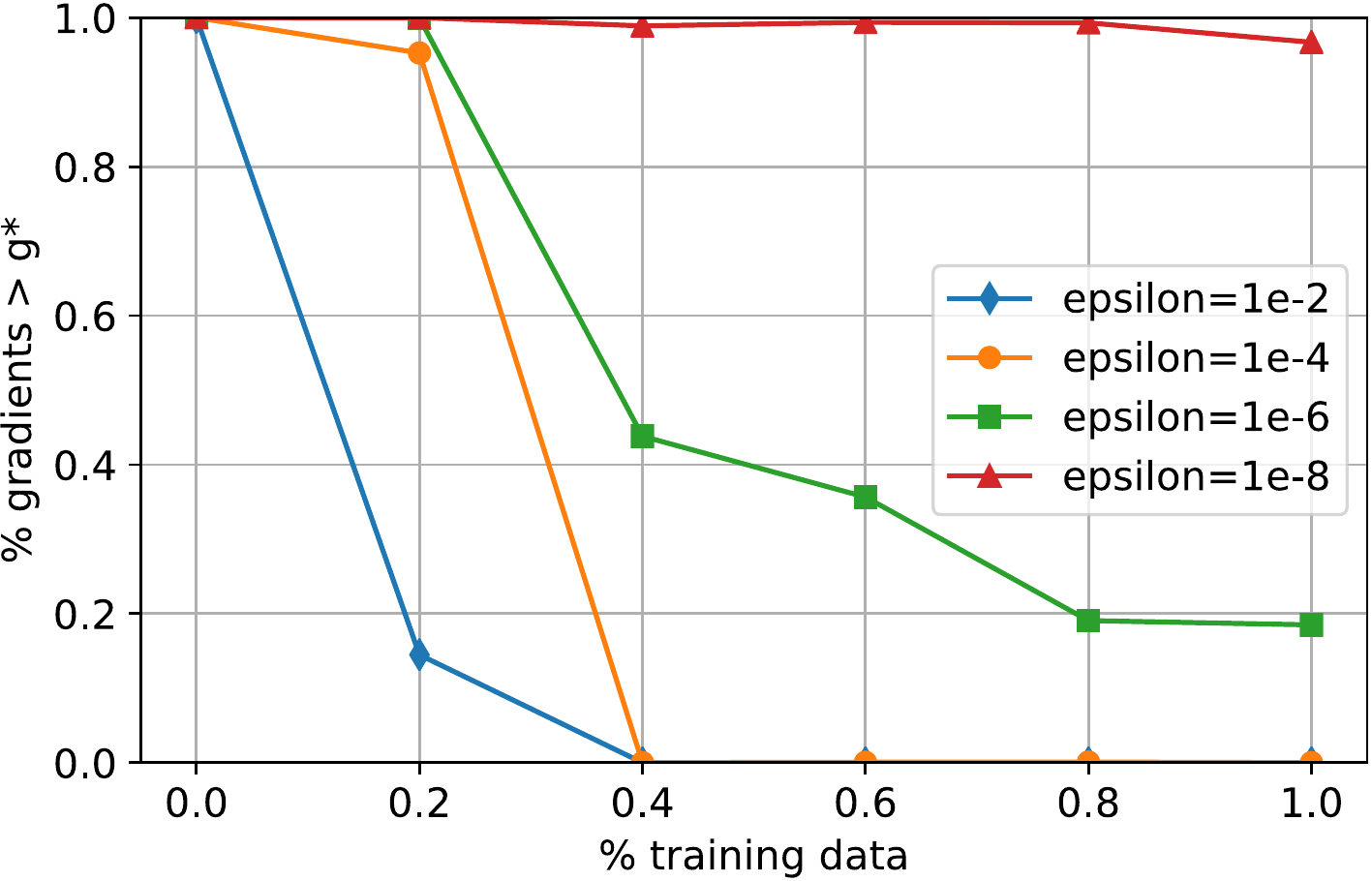}
	}
	\subfigure[Gradients $> 10g^*$]{
		\includegraphics[width=0.48\textwidth]{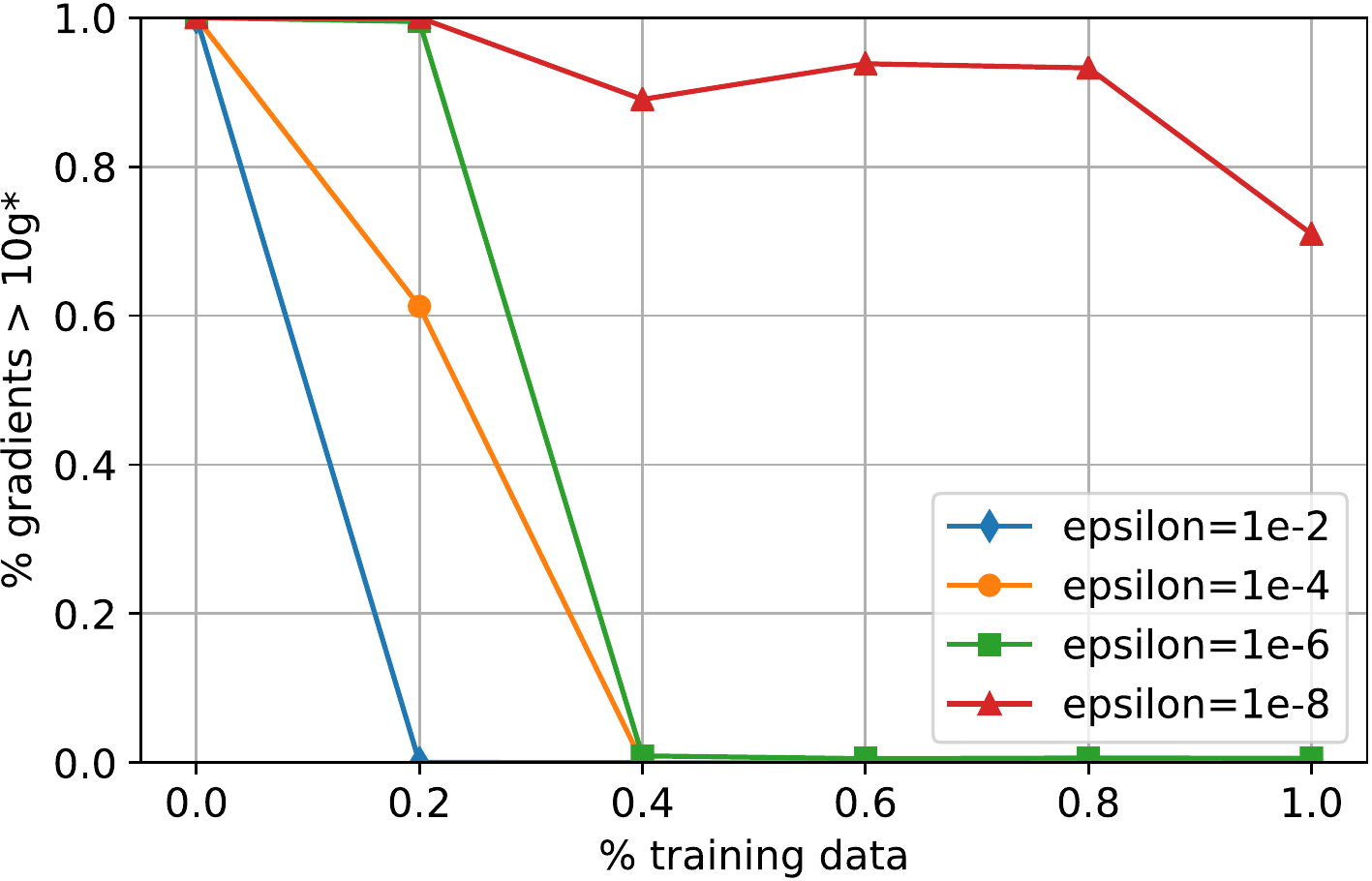}
	}
	\caption{Gradient decays with respect to $\epsilon$ of Adam. \emph{Note:} The x-axis means the portion of training examples fed in a model. The y-axis means the portion of gradients with magnitude larger than a threshold. $g^*$ refers to Eq.~\eqref{eq:gstar}, which influences the model convergence.}
	\label{fig:grad}
\end{figure}

Fig.~\ref{fig:grad} shows the gradient changes of FNN on iPinYou.
We collect gradients (absolute values, by default) of the first hidden layer in FNN at different training steps and calculate the ratio of the gradients greater than $g^*$ or $10g^*$.
From this figure, the gradients are always greater than the threshold when $\epsilon=10^{-8}$, but cross the threshold at different training steps when $\epsilon=10^{-2}, 10^{-4}, 10^{-6}$.
\revise{Thus we guess $\epsilon$ has a large impact on model convergence.}

\begin{table}[tbp]
	\small
	\centering
	\caption{Training Adam on iPinYou dataset with different $\epsilon$. \emph{Note:} ``-'' means model does not converge.}
	\label{tab:adam}
	\begin{tabular}{c|cccc|c|cccc}
		AUC & $\epsilon$=$10^{-2}$ & $\epsilon$=$10^{-4}$ & $\epsilon$=$10^{-6}$ & $\epsilon$=$10^{-8}$ & Log Loss & $\epsilon$=$10^{-2}$ & $\epsilon$=$10^{-4}$ & $\epsilon$=$10^{-6}$ & $\epsilon$=$10^{-8}$ \\ \hline
		LR & - & 76.38 & 75.31 & 74.84	& LR & - & 0.005691 & 0.005705 & 0.005719 \\
		FM & - & 75.28 & 77.17 & 75.69 & FM & - & 0.005674 & 0.005595 & 0.005655 \\
		FFM & - & 76.18 & 75.24 & 74.29 & FFM & - & 0.005695 & 0.005697 & 0.005709 \\
		CCPM & - & 77.37 & 77.53 & 76.83 & CCPM & - & 0.005584 & 0.005640 & 0.005622 \\
		FNN & 76.55 & 77.82 & 77.26 & 76.66 & FNN & 0.005597 & 0.005573 & 0.005568 & 0.005620 \\
		AFM & - & 75.30 & 77.71& 75.72& AFM & - &0.005658& \underline{0.005562} & 0.005654\\
		DeepFM & 74.96 & \underline{77.92} & 77.36 & 76.46 & DeepFM & 0.006189 & 0.005588 & 0.005581 & 0.005603 \\ \hline
		KFM & 62.62 & 72.77 & 76.90 &75.13 & KFM & 0.005997 & 0.005776 & 0.005630 & 0.005675 \\
		NIFM & 59.57 & 75.25 & 77.07 & 76.19 & NIFM & 0.005994 & 0.005664 & 0.005607 & 0.005622 \\
		IPNN & 75.80 & 78.17 & 77.12 & 75.90 & IPNN & 0.005703 & 0.005549 & 0.005577 & 0.005608 \\
		KPNN & 76.10 & 78.21 & 77.43 & 76.91 & KPNN & 0.005647 & 0.005563 & 0.005582 & 0.005611 \\
		\textbf{PIN} & 76.71 & \textbf{78.22} & 77.39 & 76.81 & PIN & 0.005602 & \textbf{0.005547} & 0.005587 & 0.005600
	\end{tabular}
\end{table}

We then test the parameter sensitivity of $\epsilon$ on iPinYou for this dataset is sparse and unbalanced. 
The results are shown in Table~\ref{tab:adam}. 
From this table, we conclude:
(i) Shallow models are more sensitive to $\epsilon$, and in some cases, these models do not converge after sufficient training steps.
(ii) An unbalanced dataset is sensitive to $\epsilon$, and sometimes empirical value $\epsilon=10^{-8}$ is not a good choice.
For these reasons, when presenting the overall performance in Table~\ref{tab:overall}, we use $\epsilon=10^{-8}$ on Criteo and Avazu, $\epsilon=10^{-5}$ on Huawei, and we use different $\epsilon$ (usually $10^{-4}$ or $10^{-6}$) on different models on iPinYou.

\subsubsection{Regularization}\label{sec:exp-prac-reg}
In Section~\ref{sec:issue-reg}, we discuss several regularization methods.
In this section, we study \change{L2 regularization, dropout, LN, and SELU}. \change{We choose FNN to perform the parameter study, since FNN is the most typical network structure.
}

\begin{figure}[htbp]
\subfigure[Criteo, Avazu, and iPinYou]{
\includegraphics[width=0.48\textwidth]{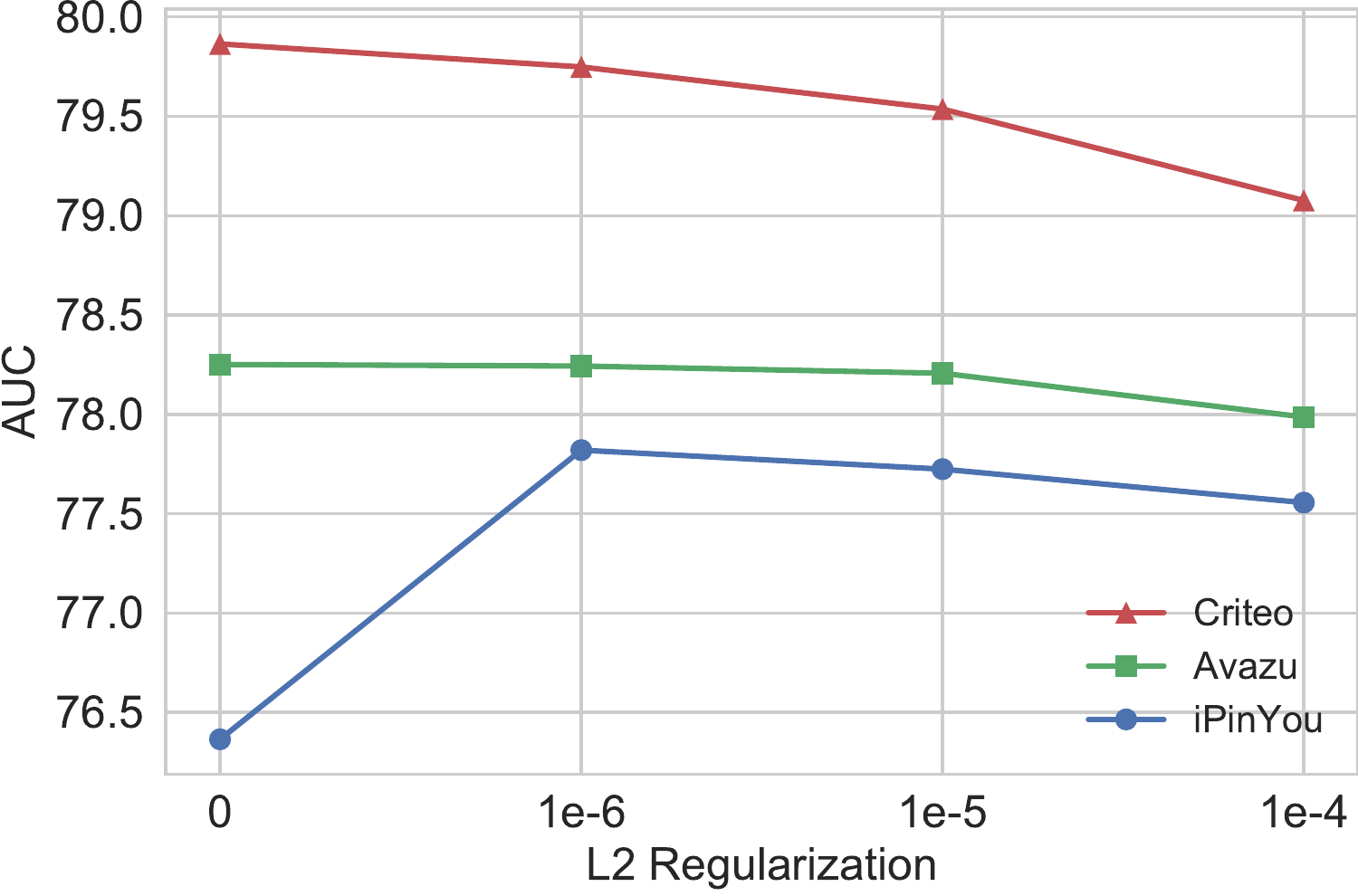}
}
\subfigure[Huawei]{
\includegraphics[width=0.48\textwidth]{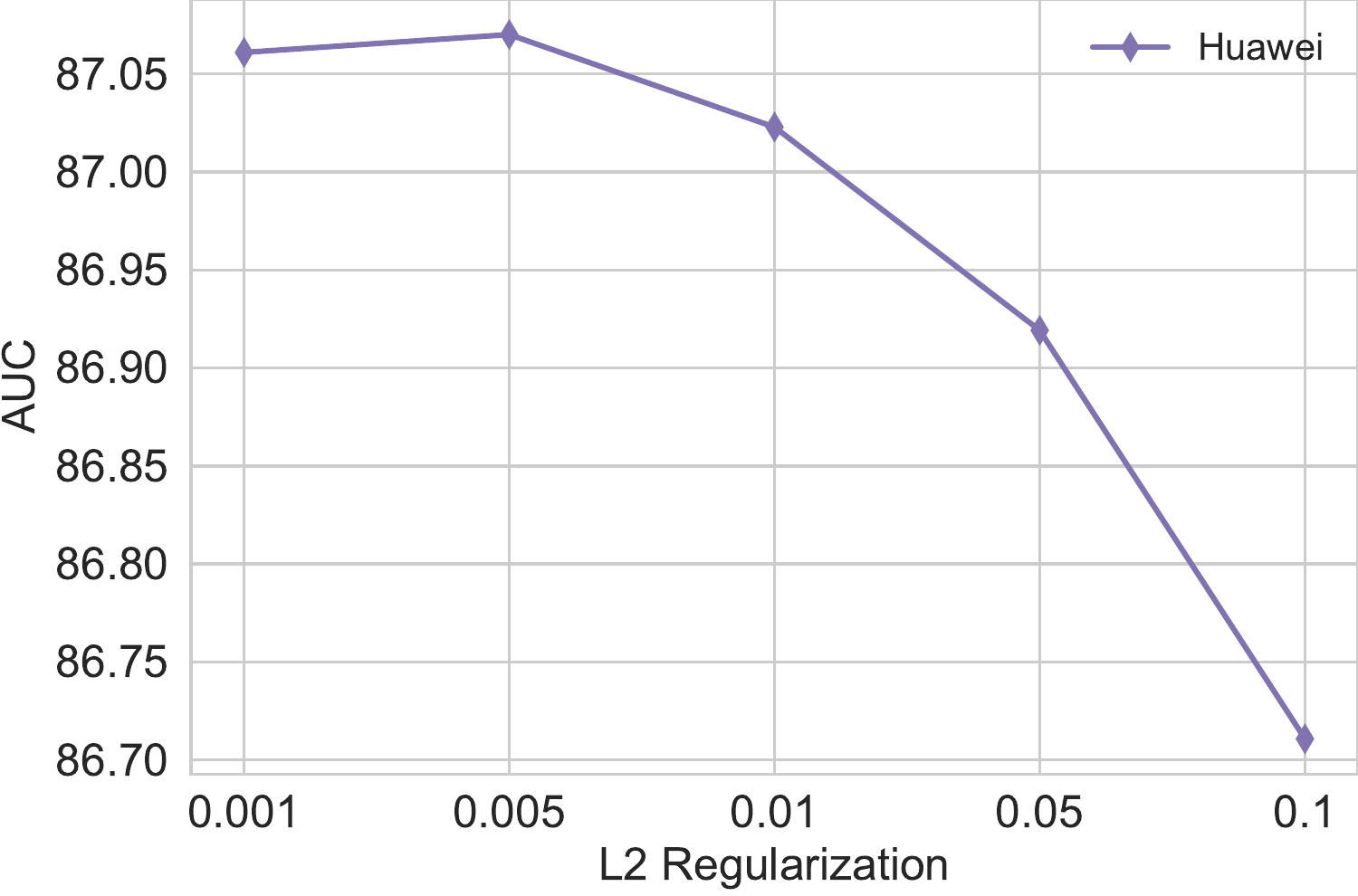}
}
\caption{FNN performance with respect to L2 regularization.}
\label{fig:l2}
\end{figure}

\change{Fig.~\ref{fig:l2} shows the L2 regularization results.
In Huawei, L2 regularization controls overfitting well, and keeps the embedding distribution stable during training.}
In Criteo, Avazu, and iPinYou, we use sparse L2 regularization and LN instead, because the input dimension is too large to apply L2 regularization.
\change{According to Fig.~\ref{fig:l2}, we use $L2=0, 0, 10^{-6}$, and $0.005$ on Criteo, Avazu, iPinYou and Huawei respectively.}

\begin{figure}[htbp]
	\subfigure[AUC]{
		\includegraphics[width=0.48\textwidth]{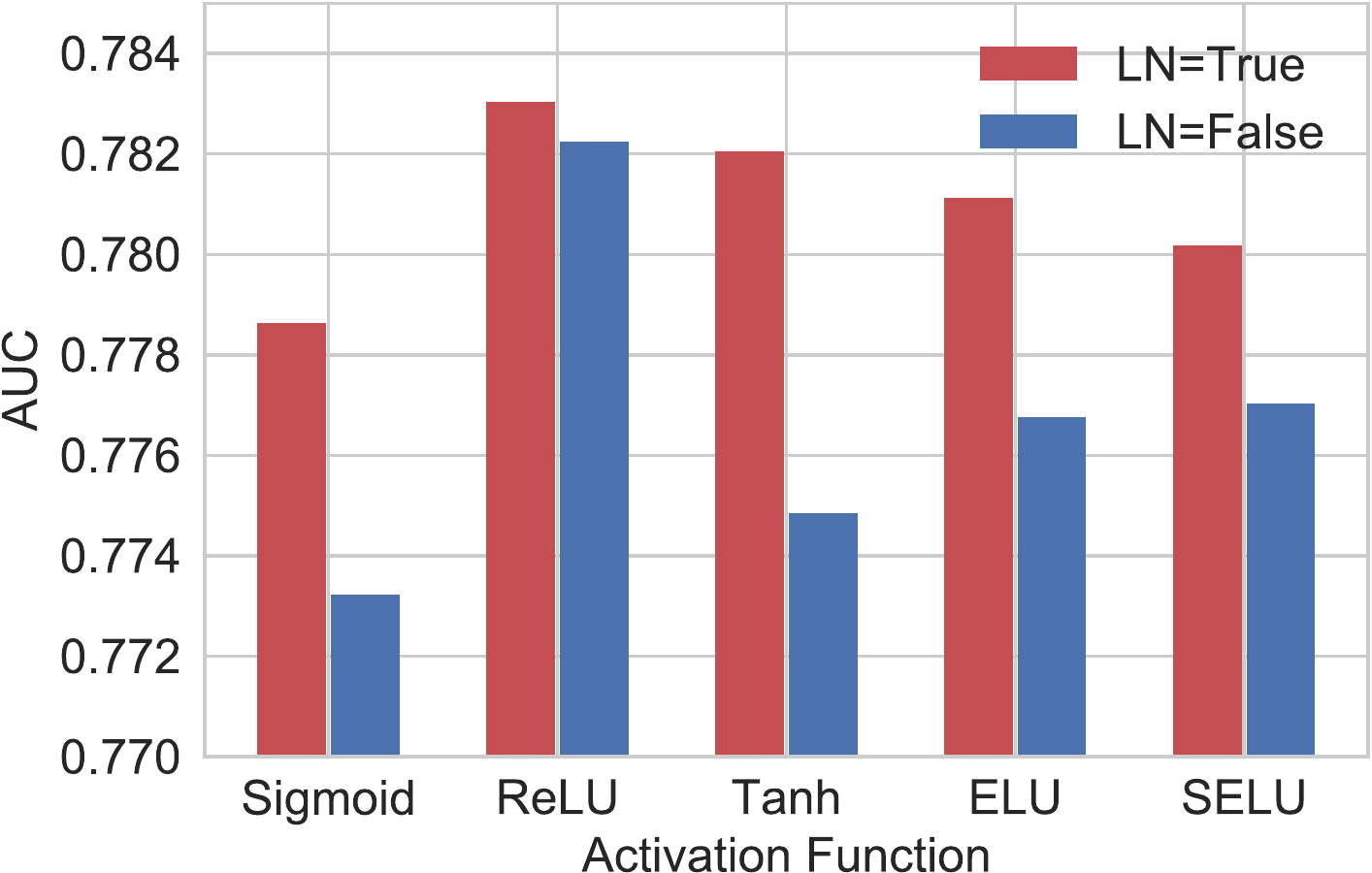}
	}
	\subfigure[Log Loss]{
		\includegraphics[width=0.48\textwidth]{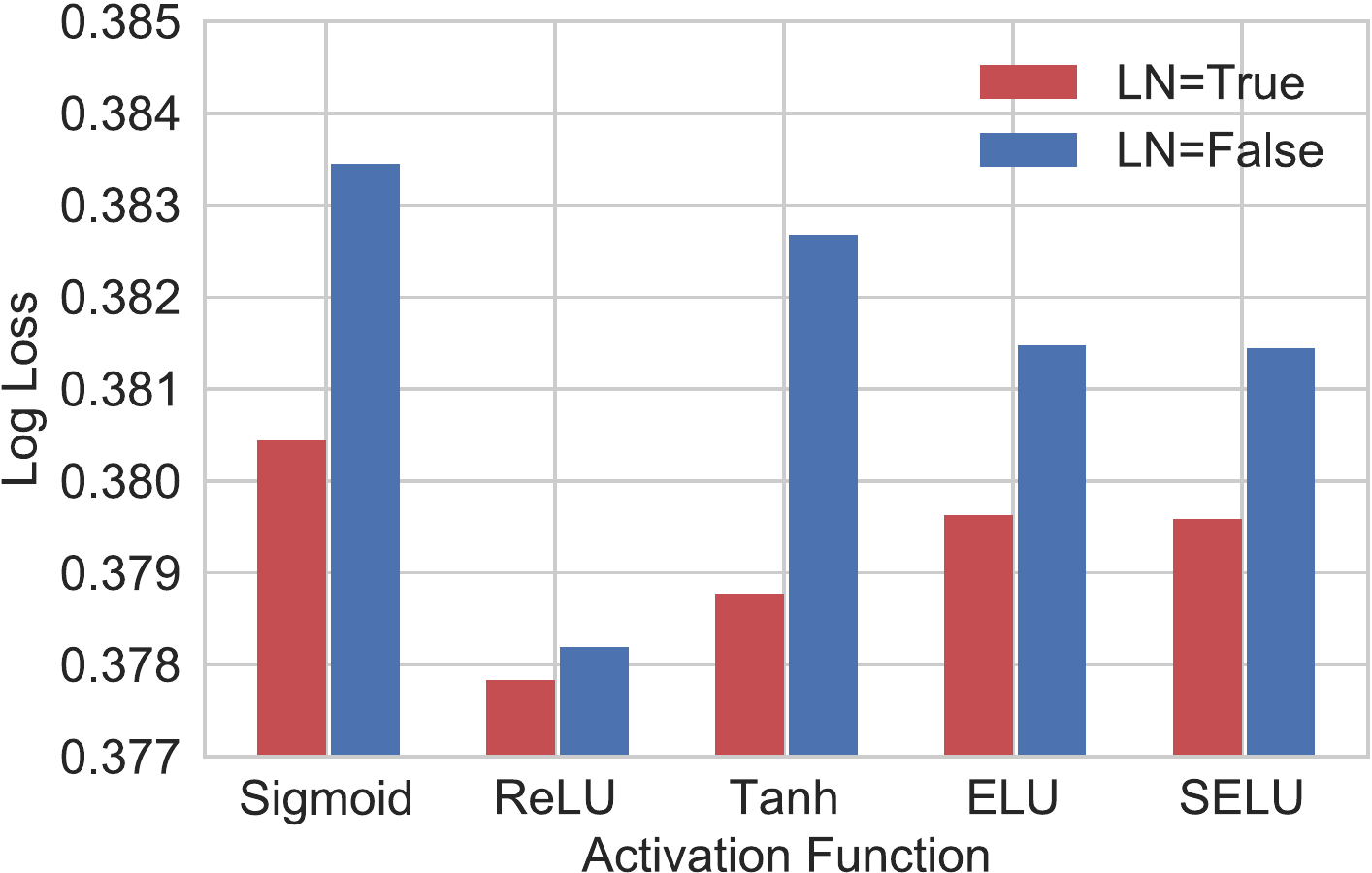}
	}
	\caption{FNN performance on Avazu with respect to different activation functions. \emph{Note:} LN means layer normalization.}
	\label{fig:act}
\end{figure}

We compare SELU with other activation functions on FNN, Avazu. 
The results are shown in Fig.~\ref{fig:act}.
We can observe that ReLU has the best performance, while SELU is similar to ELU. 
This may be because ReLU has more efficient gradient propagation.

\begin{figure}[tbp]
\subfigure[AUC]{
\includegraphics[width=0.48\textwidth]{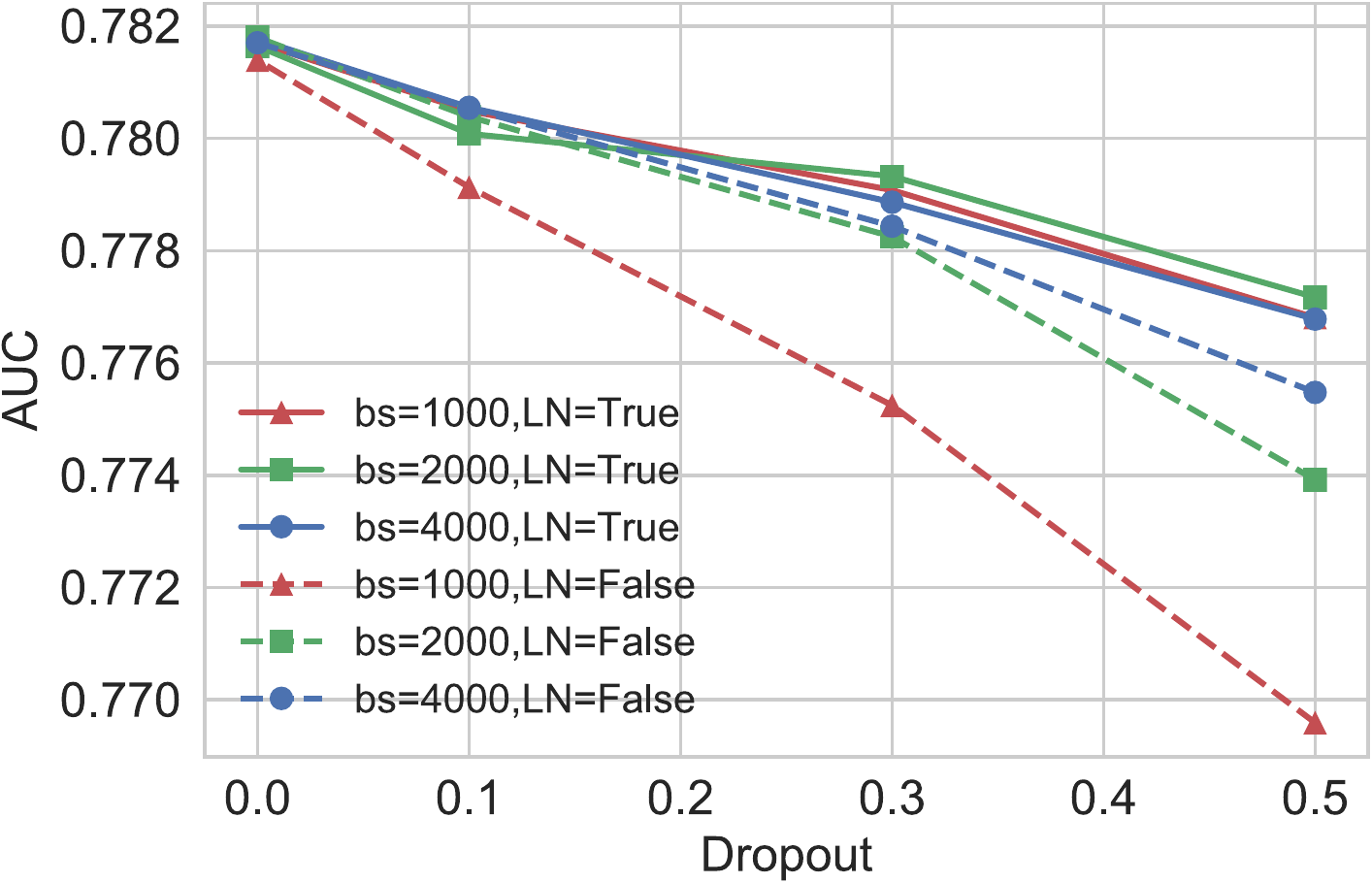}
}
\subfigure[Log Loss]{
\includegraphics[width=0.48\textwidth]{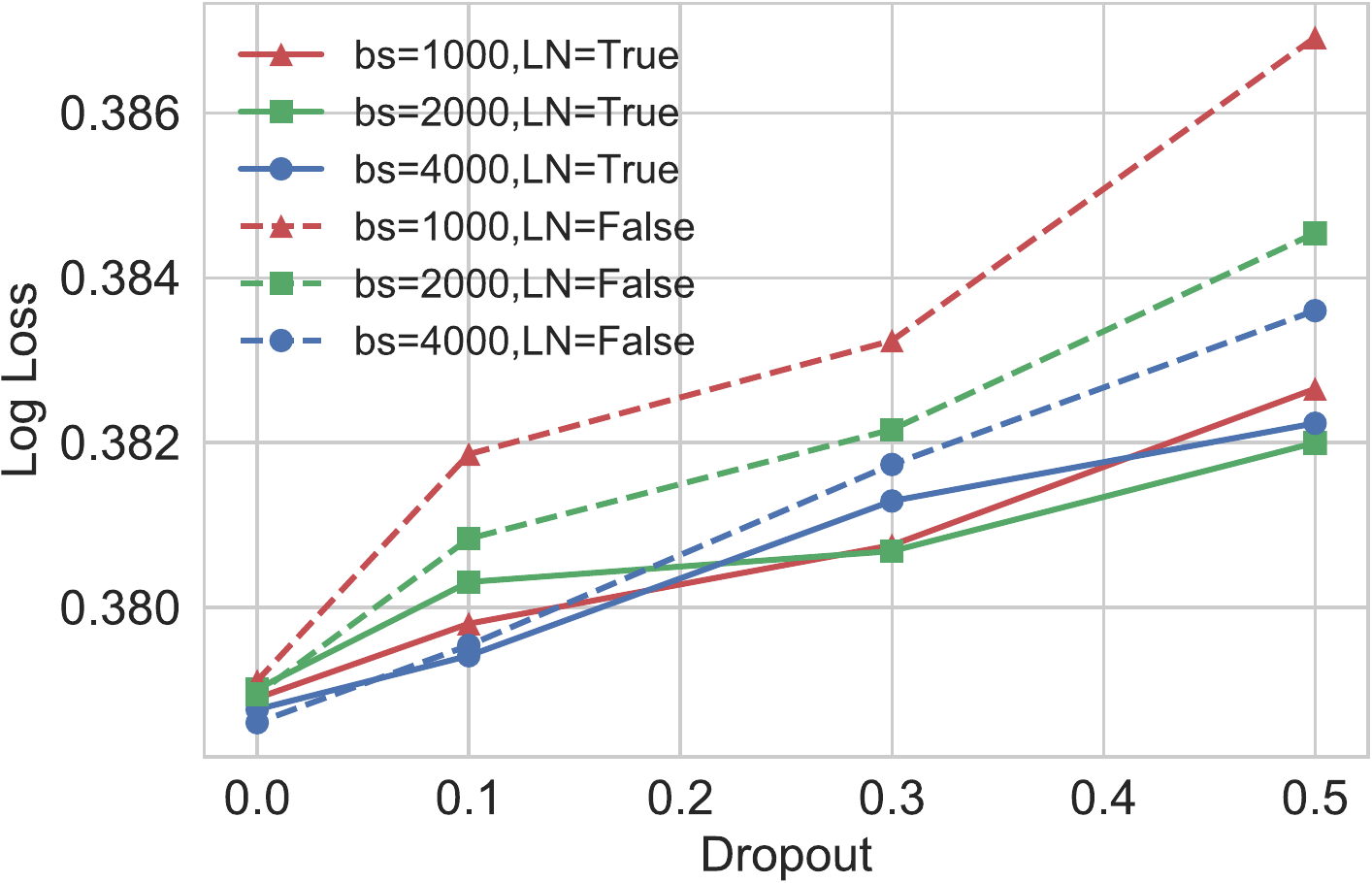}
}
\caption{FNN performance on Avazu with respect to dropout and layer normalization. \emph{Note:} dropout rate means the probability of a neuron being disabled when training a mini-batch.}
\label{fig:drop}
\end{figure}

We find the performance of dropout depends on data sparsity.
For sparse data, small mini-batch has a large bias, and dropout amplifies this bias, as shown in Fig.~\ref{fig:bias}.
We study dropout on FNN, Avazu.
From Fig.~\ref{fig:drop} we find: \revise{(i) Dropout decreases AUC. This becomes even worse when batch size is small. 
(ii) LN stabilizes dropout with different batch sizes.}
According to this result and the parameter study in \cite{guo2017deepfm}, we use LN on Criteo, Avazu, and iPinYou, and we use dropout value 0.1 (without LN) on Huawei.

\subsection{Feature Interaction Visualization}\label{sec:exp-fafi}
In Section~\ref{sec:method-fafi}, we analyze the weaknesses of FM and FFM, and refine feature interactions to field-aware feature interactions.
In this section, we propose a technique to visualize the learned feature interactions. 
Recall the cross term $\langle \bv_i, \bv_j \rangle$ in FM: (i) When the inner product is positive, this term pushes up the prediction to 1. (ii) When negative, this term pulls down the prediction to 0. (iii) When near 0, this term does not contribute to the prediction. 

\subsubsection{Mean Embedding}\label{exp-fafi-mean}

Different fields have different numbers of categories, ranging from tens (e.g., device) to millions (e.g., IP address), resulting in difficulties for commonly-used unsupervised methods (e.g., PCA, t-SNE \cite{maaten2008visualizing}).
Taking PCA as an example, the objective of PCA is defined as the summed reconstruction errors, $Err = \sum_{i} \sum_{j=1}^{N_i} err(\bv_i^j; \theta)$, where $\bv_i^j$ is the embedding vector of the $j$-th category of field $i$, $N_i$ is the field size of field $i$.
\revise{Recall the example in Table~\ref{tab:example}, there are 1009 embedding vectors, 7 for \ttt{WEEKDAY}, 2 for \ttt{GENDER}, and 1000 for \ttt{CITY}.
Thus the principal components will be dominated by \ttt{CITY}, and the interactions between \ttt{CITY} and \ttt{GENDER} are no longer mentained.}

Therefore, we turn to field-level analysis.
The simplest way is using mean embeddings.
A mean embedding is the center of the embedding vectors within a field, $\bar{\bv}_i = \sum_{j=1}^{N_i} \bv_i^j / N_i$. 

\begin{figure}[tbp]
	\subfigure[FM]{
		\includegraphics[width=0.3\linewidth]{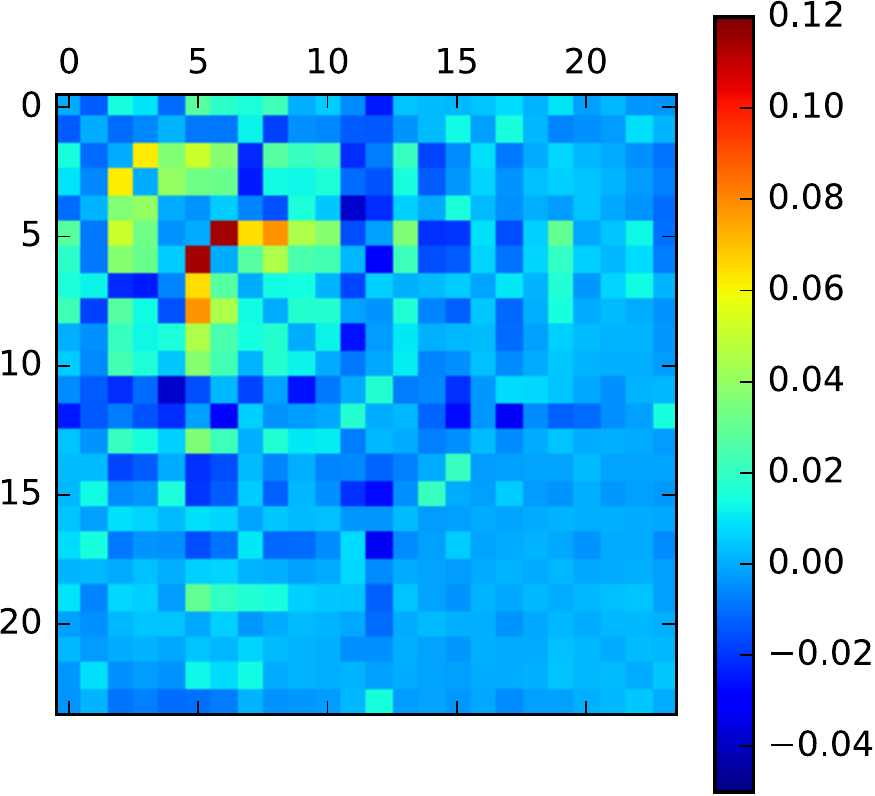}
		\label{fig:corr_1}	
	}
	\subfigure[FFM]{
		\includegraphics[width=0.3\linewidth]{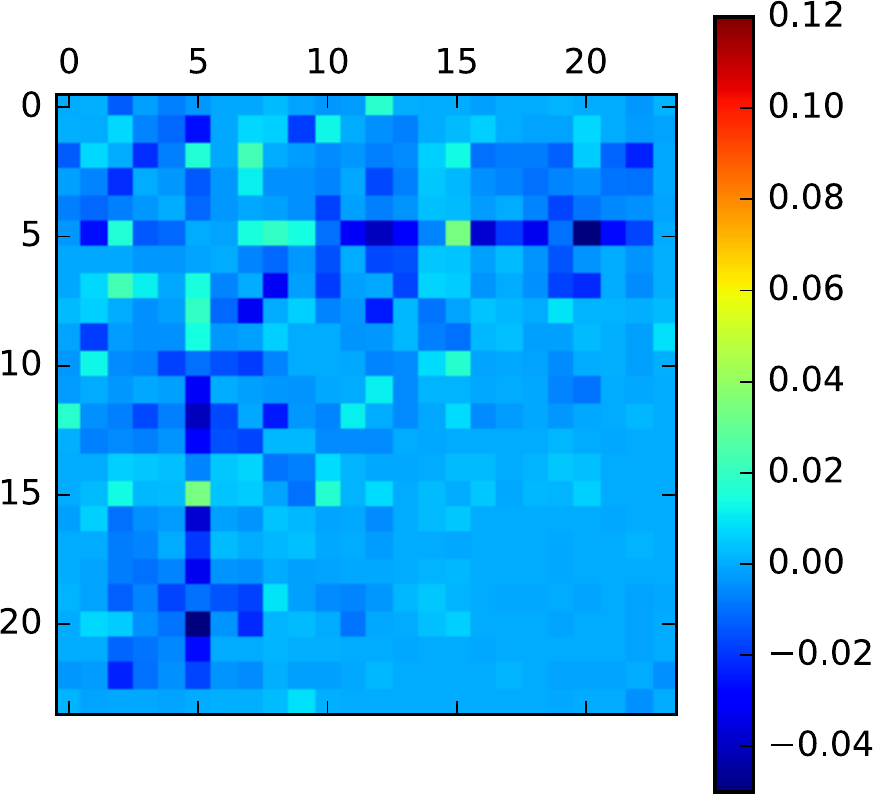}
		\label{fig:corr_2}	
	}
	\subfigure[KFM]{
		\includegraphics[width=0.3\linewidth]{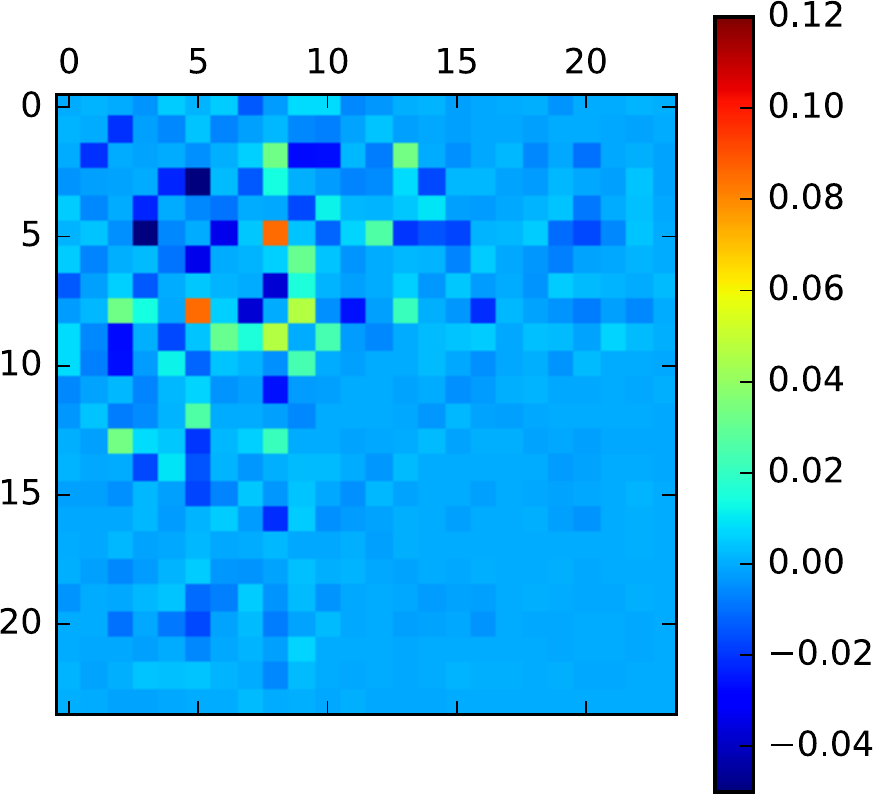}
		\label{fig:corr_4}	
	}
	\caption{Feature interaction heatmaps of FM, FFM, and KFM on Avazu. \emph{Note:} The x- and y-axis both represent fields in Avazu, these 24 fields are shown in Table~\ref{tab:fields}. 
		}
	\label{fig:corr}
\end{figure}

\subsubsection{Visualization}\label{sec:exp-fafi-vis}
We choose FM, FFM and KFM to visualize feature interactions.
The parameter settings follow Table~\ref{tab:param}, where the embedding size of FM and KFM is 40 and the embedding size of FFM is 4.
Fig.~\ref{fig:corr} shows the heatmaps generated from FM, FFM and KFM. The x- and y-axis of Fig.~\ref{fig:corr} both represent the 24 fields of Avazu, and the value of grid $(i,j)$ is the inner/kernel product of the mean embeddings of field $i$ and field $j$. 
The diagonal elements are set to 0, since a field does not interact with itself.

We assume feature interactions should be sparse.
(i) In category level, $m$-order feature combinations are roughly $C_n^m \bar{N_i}^m$, where $C_n^m$ denotes the combination number, $\bar{N_i}$ denotes the average field size.
Compared with the enormous feature combinations, feature interactions should be sparse in practice, due to the rare positive responses.
(ii) In field level, \revise{if a field has strong interactions with all other fields, this field provides little interactive information.
Thus inter-field interactions should also be sparse.}

According to the colorbar, we focus on bright (yellow or orange) and dark (dark blue) points because they have large absolute values, and we neglect light blue points because they have values near 0.
If the interaction between field $i$ and field $j$ is large, there will be a bright/dark point at grid $(i,j)$.
If the interactions between field $i$ and other fields are all large values, 
there will be a bright/dark bar in the $i$-th row/column.
With the sparse assumption of feature interactions, we expect isolated bright/dark points instead of bright/dark bars in a heatmap.

In Fig.~\ref{fig:corr}(a), the bright/dark bars in the left-top corner indicate these fields are strongly correlated. 
Therefore, we conclude FM has the coupled gradient issue, as discussed in Section~\ref{sec:method-fafi}.
In Fig.~\ref{fig:corr}(b), FFM learns clearer patterns than FM, since the bright/dark bars are shorter than those of FM. 
However, the short bars are still not expected, showing that FFM is limited by the small embedding size, as discussed in Section~\ref{sec:method-fafi}. 
In Fig.~\ref{fig:corr}(c), we find that most of the bars disappear, and clearly isolated points are displayed.
This figure proves that kernel product successfully solves the coupled gradient issue.

We find most useful feature interactions appear in the left-top corner (field 1 - field 13), which can guide feature engineering and model design. 
For example, we can use more complex models to capture these fields, or design higher-order cross features among these fields. 
For the right-bottom corner (field 14-24), these fields provide less interactive information. 
And a model can be compressed if we remove interactive parameters of these fields. 
The 24 Avazu fields are listed in Table.~\ref{tab:fields}.

\begin{table}[tbp]
	\centering
	\caption{Fields of Avazu.}
	\label{tab:fields}
	\begin{tabular}{c|c}
		Field 1-13 & Field 14-24 \\ \hline
		\tabincell{c}{C1, banner\_pos, site\_id, site\_domain, \\ site\_category, app\_id, app\_domain, \\ app\_category, device\_id, device\_ip, \\ device\_model, device\_type, \\ device\_conn\_type} & \tabincell{c}{C14, C15, C16, C17,\\ C18, C19, C20, C21,\\ day, hour, weekday}
	\end{tabular}
\end{table}

\subsection{Training Difficulty of Gradient-based DNN}\label{sec:exp-diff}
In Section~\ref{sec:method-diff}, we discuss the insensitive gradient issue of gradient-based DNN from a theoretical view.
In this section, we conduct a synthetic experiment to support the discussion.
This synthetic dataset is generated from a poly-2 function, where the bi-linear terms are analogous to interactions between categories. 
Based on this dataset, we investigate 
(i) the impact of data sparsity on DNN, and 
(ii) the ability of DNN in fitting poly-2 functions. 

The input $\bx$ of this dataset is randomly sampled from $N$ categories of $n$ fields, where each field size $N_i$ is randomly selected.
The output $y$ is binary labeled depending on the sum of linear terms and bi-linear terms 
\begin{align}
y & = \delta \Big(\sum_{i=1}^{n} w_i x_i + \sum_{i=1}^{n-1}\sum_{j=i+1}^{n} v_{i,j} x_i x_j + b + \epsilon \Big) \label{eq:poly2}\\
\delta(z) & = \begin{cases}
1, & \text{ if } z \ge \text{threshold}\\
0, & \text{ otherwise}
\end{cases} ~.
\end{align}

The data distribution $p(\bx)$ and $\bw, \bv, b$ are randomly sampled and fixed, and the data pairs $\{\bx, y\}$ are i.i.d. sampled to build the training and validation datasets.
We also add a small random noise $\epsilon$ to the sampled data.
We use DNNs with different depths and widths to fit the synthetic data.
We use AUC to evaluate these models on the validation dataset.

\begin{figure}[tbp]
	\subfigure[Sparsity]{
		\includegraphics[width=0.48\linewidth]{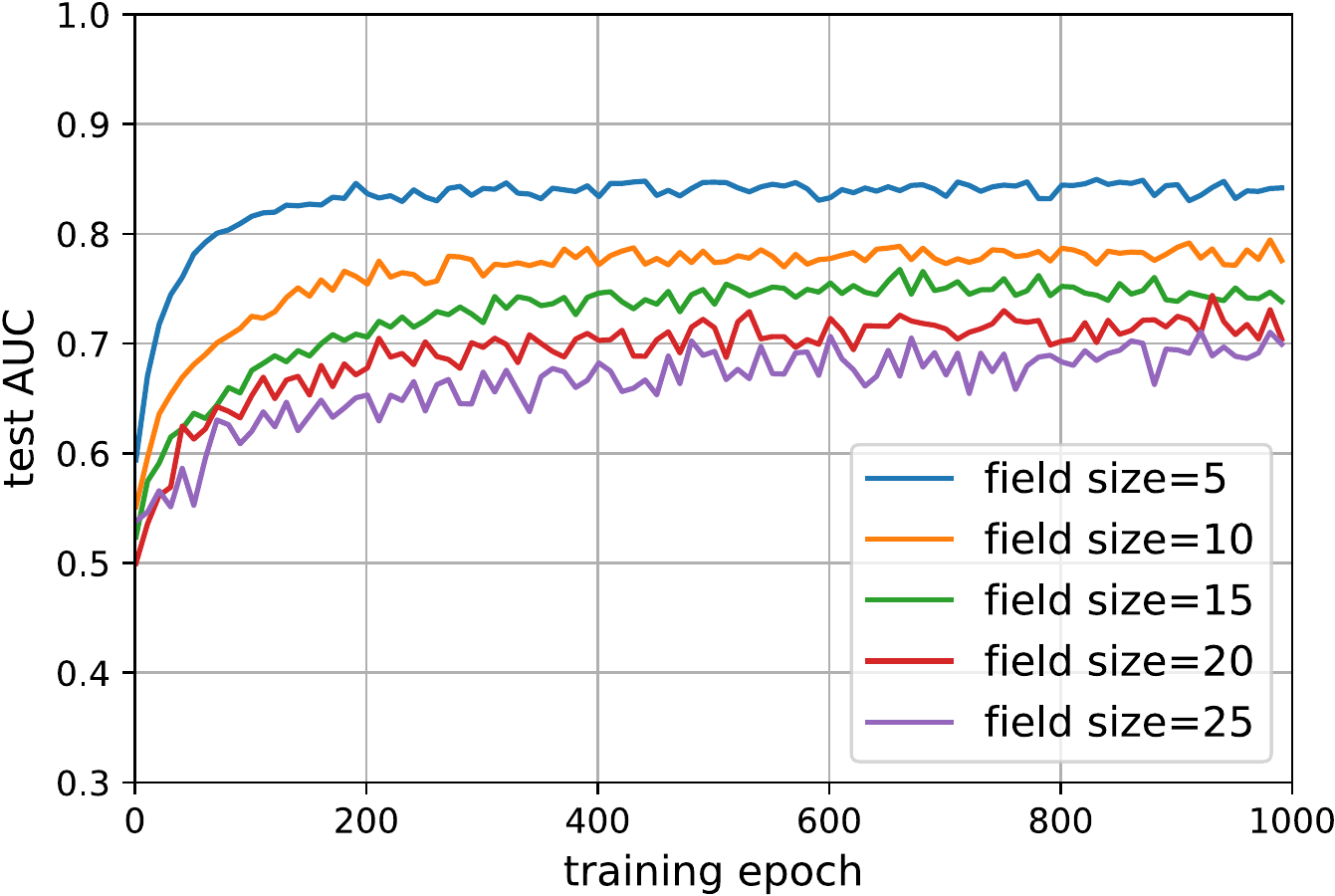}
		\label{fig:field_size}	
	}
	\subfigure[DNN convergence]{
		\includegraphics[width=0.48\linewidth]{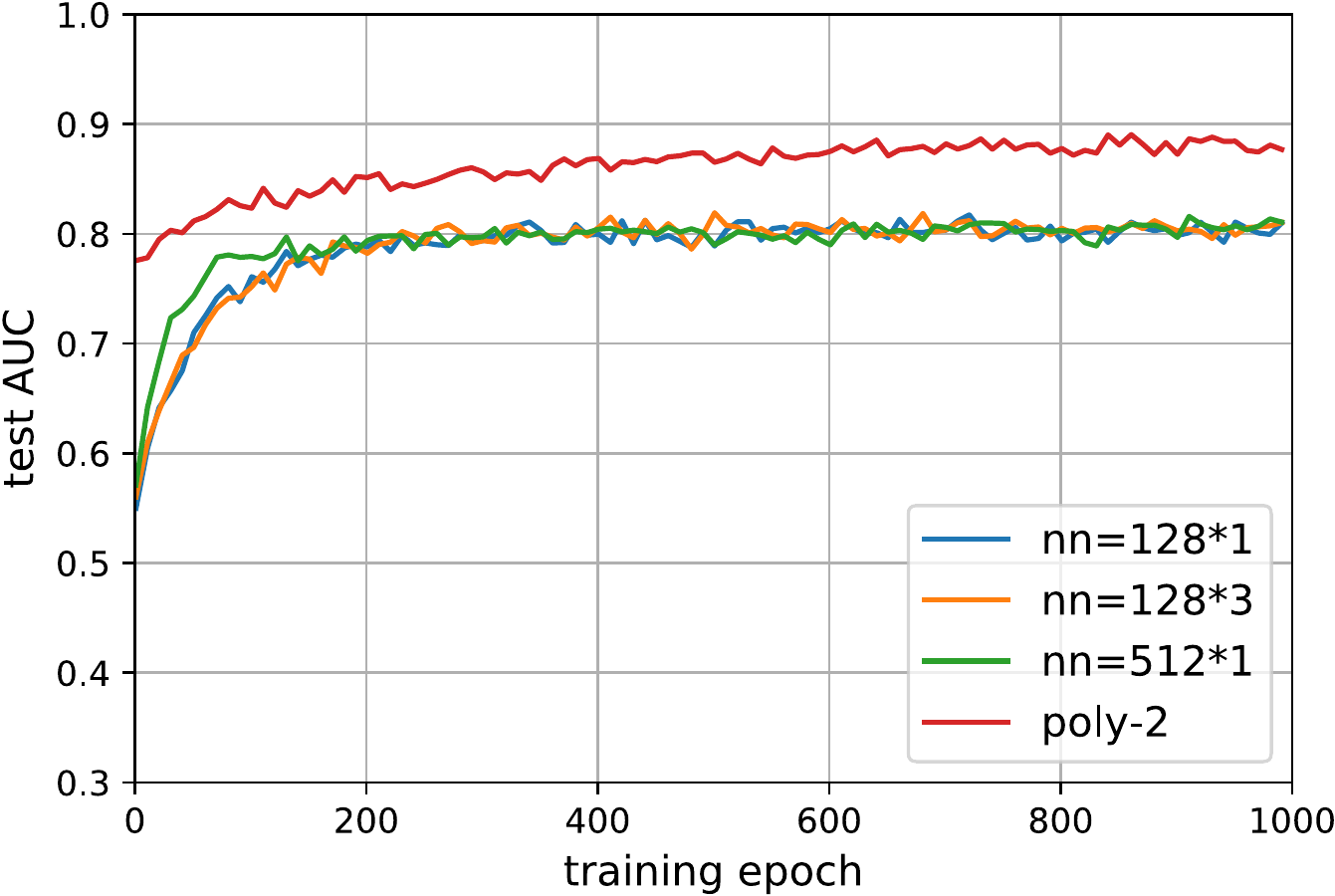}
		\label{fig:nn_kinds}	
	}
	
	\caption{DNN training curves of the synthetic experiments. \emph{Note:} Plot (a) verifies the influence of data sparsity. Plot (b) verifies DNN as not a perfect function approximator in such a problem.}
	\label{fig:synthetic}
\end{figure}

First, we study the impact of data sparsity by changing field sizes.
We sample a data instance from $p(\bx)$ and determine its label through Eq.~\eqref{eq:poly2}, with each field one-hot encoded.
It is a reciprocal relationship between field size and data sparsity.
The size of each field is a random number between $1$ and $2N/n$ (so that the average field size is $N/n$), where $N=\{200, 400, 600, 800, 1000\}$, $n=40$.
We use a DNN with 3 hidden layers of 128 neurons to fit the training dataset with different $N$ values, and the evaluation curves are shown in Fig.~\ref{fig:synthetic}(a).
From this figure, we conclude DNN training is more difficult if input data is sparser.

Second, we choose field size $=10$, $N=400$ to test DNN convergence.
Fig.~\ref{fig:synthetic}(b) compares three types of DNNs:  1 hidden layer of 128 neurons, 1 hidden layer of 512 neurons, and 3 hidden layers of 128 neurons.
We use Eq.~\eqref{eq:poly2} to fit this data, namely poly-2 regression (for short, poly-2).
Because the ground-truth is a poly-2 function, we treat poly-2 as the performance upper bound, shown as the red curve.
From this figure we can see, there is a consistent gap between DNNs and poly-2.
And unfortunately, increasing width or depth does not improve the performance of DNN.
In this experiment, we also try DNN with \{1, 3, 5\} hidden layers of \{128, 512, 1024\} neurons, but the performance changes slightly in different settings. 
This figure indicates that, in spite of universal approximation property, DNN cannot fit a simple poly-2 function perfectly through gradient descent, 
thus gradient-based DNN may not be a perfect function approximator for user response prediction.

Poly-2 and libFFM are also studied in \revise{\cite{juan2016field}, and FFM achieves promising results.
Thus, it is promising to add a feature extractor to explore interactive patterns in a DNN model.}
This synthetic experiment validates the discussion in Section~\ref{sec:method-diff}, and highlights the necessity of extracting feature interactions from the sparse input, as a complement of DNN classifiers.

\section{Conclusion}\label{sec:con}

In this paper, we study user response prediction over multi-field categorical data. 
\change{We find a coupled gradient issue of latent vector-based models and propose to learn field-aware feature interactions instead.
We propose kernel product methods to solve this problem as well as solving the memory bottleneck of FFM.}
\revise{We also find an insensitive gradient issue of DNN-based models, and we propose several PNN models to solve this problem.}
The improvement of PNNs over other DNN-based models proves the necessity of product layers. 
With both expressive feature extractors and powerful DNN classifiers, PNNs consistently outperform 8 baselines and achieve the state-of-the-art performance on 4 industrial datasets.
\revise{Besides, PIN makes great CTR improvements in online A/B test.
In future work, we will study different kernels and micro networks, 
and explore the generalization ability of DNN-based models in information systems.
}





  

\begin{acks}

The work is sponsored by Huawei Innovation Research Program. The corresponding author Weinan Zhang thanks the support of National
Natural Science Foundation of China (61632017, 61702327, 61772333), Shanghai Sailing Program (17YF1428200).

\end{acks}

\bibliographystyle{ACM-Reference-Format}
\bibliography{bibliography}

\end{document}